\def\Roman#1{\uppercase\expandafter{\romannumeral#1}}
\documentclass[a4paper,12pt]{article}
\usepackage{cite}
\usepackage{amssymb,amsfonts}
\usepackage{amsmath}
\usepackage[mathscr]{eucal}
\usepackage{mathrsfs}
\usepackage[utf8x]{inputenc}

\title{On the geometric representation of the path integral reduction Jacobian in the path integral for interacting systems:
The case of dependent coordinates in the description of reduced motion on the orbit space}

\author{S. N. Storchak\\
\small{ A. A. Logunov Institute for High Energy Physics}\\
\small{of NRC ``Kurchatov Institute'',}\\
\small{Protvino, 142281, Russian Federation,}\\
}

\begin{document}

  \maketitle

\begin{abstract}
A geometric representation of the previously obtained path integral reduction Jacobian  is found in the problem of the path integral quantization  of a mechanical system consisting of two interacting scalar particles moving on a product manifold  with the given free isometric action of the compact semisimple Lie group. The product manifold consists of a smooth compact finite-dimensional Riemannian manifold and a finite-dimensional vector space.
The reduction Jacobian we were dealing with was obtained for the case when, as in gauge theories, the local reduced motion is described using dependent coordinates. The result  is based on the scalar curvature formula for the original manifold which  is viewed as a total space of the principal fiber bundle. The calculation of the Christoffel symbols and the scalar curvature is done in a special nonholonomic basis, also known as the horizontal lift basis.
%\end{abstract}
%\vspace{1.5mm}
\begin{flushleft}
%PACS numbers: 03.65.Bz, 31.15.Kb 
\end{flushleft}

\end{abstract}

\section{Introduction}
In \cite{Storchak_2019, Storchak_2020} and \cite{Storchak_2021}, the method  developed earlier in \cite{Storchak_2, Storchak_2009, Storchak_1, Storchak_98}, by which it is possible to perform the reduction in the path integrals that are used for quantization of the mechanical systems with a symmetry was exteded to the the systems with interaction. The path integrals of these papers were  defined  in the same way as was done by Daletski\v{\i} and Belopol'skaya in \cite{Dalecky_1,Dalecky_2}. The measure of these path integrals are generated by stochastic processes given on a manifold. Using these path integrals, one can represent the solution of the backward Kolmogorov equation, thereby defining a global evolution semigroup acting in the space of functions on the manifold. It is important that this global evolution semigroup is  a limit of local evolution semigroups associated with local stochastic processes that are solutions of local stochastic differential equations given on charts of the manifold. This means that the behavior of the global semigroup is determined by these stochastic differential equations

The finite-dimensional mechanical systems with symmetry, which were considered in our works,  have geometric properties similar to those that we encounter when studying the reduction of gauge fields. In particular, mechanical
system from \cite{Storchak_2019,Storchak_2021} has geometric properties similar to those of field systems describing the interaction of Yang-Mills fields with scalar fields.

In all cases that were considered, it was shown that the measure of the path integral is not invariant under reduction.
As a consequence, this leads to the appearance of a reduction Jacobian, which gives an additional potential term to the reduced Hamiltonian. 
Of particular interest is the study
geometric representation of the resulting reduction Jacobian, since it can be assumed that its
 geometric structure can be the same as in the corresponding gauge systems.

In a pure Yang-Mills theory, the appearance of an additional term to the  potential as a result of the reduction  was demonstrated by J. Lott in \cite{Lott} when projecting the functional Schr\"{o}dinger equation onto the equation given on the orbit  space manifold. In his work, the dynamics on the base space  of the corresponding  principal fiber bundle (the orbit space manifold) was described using independent coordinates. Note that the geometric structure of the additional term to the potential in the reduced Hamiltonian coincides
with the reduction Jacobian obtained in \cite{Storchak_2021}, if in this article one neglects the motion in the vector space.

A similar approach, but using dependent variables, was considered by  
K. Gaw\c{e}dzki in \cite{Gawedzki}, where the geometric structure of the additional term  to the potential of the reduced Hamiltonian was also found. 
 This additional term looks different than in \cite{Lott}. But in fact, both additional terms must be representations of  the same geometrical structure. This could be verified by deriving the Jacobian of the Lott's  reduction from the expression  obtained by Gaw\c{e}dzki as a particular case.

Since in gauge theories the dynamics on the space of orbits of the gauge group is described in terms of dependent variables (gauge fields with additional restrictions) defined on the gauge surface, this leads to the need to study the case of using dependent coordinates in path integrals when considering reduction in finite-dimensional mechanical systems with symmetry.
It is from this point of view that we studied in \cite{Storchak_2021} the reduction of the path integral
for a mechanical system consisting of two interacting scalar particles moving on a product manifold (formed by a smooth compact finite-dimensional Riemannian manifold and a finite-dimensional vector space) on which a free isometric action of a compact  semisimple  Lie group is given.

The purpose of the present paper is to find a geometric representation of the reduction Jacobian obtained there for such an interacting system. To do this, we calculate the scalar curvature of the original manifold, which is the total space of the principal fiber bundle associated with the system. In the case when dependent coordinates are used as local coordinates on the total space of this bundle (together with coordinates defined on the vector space, as well as group coordinates), the calculation is carried out in a special nonholonomic coordinate basis.

The work is organized as follows. In Section 2, we  introduce the necessary definitions and briefly describe the result obtained in our previous paper. Section 3 with subsections is devoted to calculating the scalar curvature of the original manifold, where first the Christoffel symbols are calculated, then the components of the Ricci tensor, and finally the scalar curvature. In section 4, the integrand  in the obtained reduction Jacobian, is rewritten using the resulting scalar curvature formula.
Appendix A contains auxiliary  material consisting of definitions of the  projection operators together with their properties, the Killing relations for the horizontal metric, and some of the identities used in this article. 
Appendix B presents the results of calculating the scalar curvature terms. In Appendix C, we derive an identity that helps to get an explicit description of the scalar curvature.

\section{Definitions}
An original diffusion of two scalar particles on a smooth compact Riemannian manifold $ \tilde {\mathcal P} = \mathcal P \times \mathcal V $ is described by the backward Kolmogorov equation 
\begin{equation}
\left\{
\begin{array}{l}
\displaystyle
\left(
\frac \partial {\partial t_a}+\frac 12\mu ^2\kappa \bigl[\triangle
_{\cal P}(p_a)+\triangle
_{\cal V}(v_a)\bigr]+\frac
1{\mu ^2\kappa m}V(p_a,v_a)\right){\psi}_{t_b} (p_a,t_a)=0,\\
{\psi}_{t_b} (p_b,v_b,t_b)=\phi _0(p_b,v_b),
\qquad\qquad\qquad\qquad\qquad (t_{b}>t_{a}),
\end{array}\right.
\label{1}
\end{equation}
in which $\mu ^2=\frac \hbar m$ , $\kappa $  is a real positive
parameter,  $V(p,f)$ is the group-invariant potential term:
 $V(pg,g^{-1}v)=V(p,v)$, $g\in \mathcal G$, 
$\triangle _{\cal P}$ is the Laplace--Beltrami operator on the Riemannian
manifold $\cal P$, the scalar Laplacian $\triangle
_{\cal V}$ acts on the space of functions defined on  the vector space $\cal V$. In a chart $(U_{\cal P}\times U_{\cal V},{\varphi}^{\tilde{\mathcal P}} )$, ${\varphi}^{\tilde{\mathcal P}}=(\varphi^A,\varphi^a)$, from the atlas of the manifold $\tilde{\mathcal P}$, where the point $(p,v)$ has the  coordinates $(Q^A,f^a)$, $\triangle _{\cal P}$ is represented as\footnote{In  our formulas  we  assume that there is 
sum over the repeated indices. The indices denoted by the capital
letters  run from 1 to $n_{\cal P}=\rm{dim} \cal P$, and the indices of small Latin letters, except $i,j,k,l$, run from 1 to $n_{\cal V}=\dim \cal V$.}:
\begin{equation}
\triangle _{\cal P}(Q)=G^{-1/2}(Q)\frac \partial {\partial
Q^A}G^{AB}(Q)
G^{1/2}(Q)\frac\partial {\partial Q^B},
\label{2}
\end{equation}
where $G=\det (G_{AB})$, $G_{AB}(Q)=G(\frac{\partial}{\partial
Q^A},\frac{\partial}{\partial
Q^B})$
 is a metric tensor of the
 Riemannian manifold
 $\cal P$ in coordinate basis  $\{\frac{\partial}{\partial
Q^A}\}$.

The coordinate expression of  the operator $\triangle_{\cal V}$ is  
\[
 \triangle_{\cal V}(f)=G^{ab}\frac{\partial}{\partial f^a\partial f^b},
\]
where  $G^{ab}$ is an  inverse matrix to the positive definite, symmetrix matrix $G_{ab}$ representing the metric tensor of  $\mathcal V$
 in the coordinate basis $\{\frac{\partial}{\partial f^a}\}$.
We assume that  the matrix $G_{ab}$ consists of fixed constant elements. 
%It is also assumed that the matrix $G_{ab}$ may have nonzero off-diagonal entries.

The Schr\"odinger equation is obtained  from  the forward Kolmogorov equation associated with the equation (\ref{1}) in which  one must set $\kappa =i$.

With the definition of the path integral from \cite{Dalecky_1}, provided that the necessary smoothness requirements imposed on the terms of the equation are satisfied, the solution of the equation
equation (\ref{1}) can be written as follows
\begin{eqnarray}
{\psi}_{t_b} (p_a,v_a,t_a)&=&{\rm E}\Bigl[\phi _0(\eta_1 (t_b),\eta_2(t_b))\exp \{\frac
1{\mu
^2\kappa m}\int_{t_a}^{t_b}V(\eta_1(u),\eta_2(u))du\}\Bigr]\nonumber\\
&=&\int_{\Omega _{-}}d\mu ^\eta (\omega )\phi _0(\eta (t_b))\exp
\{\ldots 
\},
\label{orig_path_int}
\end{eqnarray}
where ${\eta}(t)=(\eta_1(t),\eta_2(t))$ is a global stochastic process on a manifold 
$\tilde{\cal P}=\cal P\times \cal V$, 
the measure  ${\mu}^{\eta}$ in 
the path space $\Omega _{-}=\{\omega (t)=(\omega^1(t),\omega^2(t)): \omega^{1,2} (t_a)=0,
\eta_1
(t)=p_a+\omega^1 (t), \eta_2(t)=v_a+\omega^2(t)\}$ is determined by the probability distribution of the stochastic  process $\eta(t)$.

 The equation (\ref{orig_path_int}) is  a symbolical notation of the global semigroup defined in \cite{Dalecky_1, Dalecky_2} as a limit of the superposition of the local evolution semigroups 
\begin{equation}
\!\psi _{t_b}(p_a,v_a,t_a)=U(t_a,t_b)\phi _0(p_a,v_a)=
{\lim}_q \Bigl[{\tilde U}_{\eta}(t_a,t_1)\cdot\ldots\cdot
{\tilde U}_{\eta}(t_{n-1},t_b)
\phi _0\Bigr](p_a,v_a),
\label{6}
\end{equation}
where  each 
${\tilde U}_{\eta}$ acting in the space of functions on the manifold $\tilde {\mathcal P}$ is
\begin{equation}
 {\tilde U}_{\eta}(s,t) \phi (p,v)={\rm E}_{s,p,v}[\phi (\eta_1
 (t),\eta_2(t))],\,\,\,\,\,\,
 s< t,\,\,\,\,\,\,\eta_1 (s)=p,\;\eta_2 (s)=v,
 \label{local_semigroup}
 \end{equation}
 where  ${\tilde U}_{\eta}(s,t) \phi (p,v)\equiv [{\tilde U}_{\eta}(s,t) \phi] (p,v)$.\footnote{Here the local evolution semigroups are temporarily written without an exponential with a potential term of the Hamiltonian operator. Because, in fact, the presence of this term is not essential in the transformations of path integrals we carried out.} 
These local evolution semigroups are also given by the corresponding path integrals.
 The local stochastic processes     $\{\eta_1^A(t),\eta_2^a(t)\}\in R^{n_{\tilde{\cal P}}}$  are  solutions of two  stochasic differential equations using It\^o's stochastic differentials:
\begin{align}
d\eta_1^A(t)&=\frac12\mu ^2\kappa G^{-1/2}\frac \partial {\partial
Q^B}(G^{1/2}G^{AB})dt+\mu \sqrt{\kappa }{\mathscr X}_{\bar{M}}^A(\eta_1
(t))dw^{
\bar{M}}(t),
\label{eta_1}\\
 d\eta_2^a(t)&=\mu \sqrt{\kappa }{\mathscr X}_{\bar{b}}^a
dw^{
\bar{b}}(t),
\label{eta_2}
\end{align}
where  ${\mathscr X}_{\bar{M}}^A$ and ${\mathscr X}_{\bar{b}}^a$ are  defined  by the local equalities
$\sum^{n_{\mathcal P}}_{\bar{\scriptscriptstyle K}\scriptscriptstyle =1}{\mathscr X
}_{\bar{K}}^A{\mathscr X}_{\bar{K}}^B=G^{AB}$ and  $\sum^{n_{\mathcal V}}_{\bar{\scriptscriptstyle b}\scriptscriptstyle =1}{\mathscr X
}_{\bar{b}}^a{\mathscr X}_{\bar{b}}^c=G^{ac}$, $dw^{\bar{M}}(t)$ and $dw^{\bar{b}}(t)$ are the independent Wiener processes.
(We  denote the Euclidean indices by over-barred indices.)

The left-hand side of (\ref{orig_path_int}) can also be  represented as follows
\[
\psi_{t_b} (p_a,v_a,t_a)=\int G_{\tilde{\cal P}}(p_b,v_b,t_b;p_a,v_a,t_a)\phi _0(p_b,v_b)dv_{\tilde{\cal P}}(p_b,v_b),
\]
where $G_{\tilde{\cal P}}$ is the  Green's function -- the kernel of the  semigroup $U(t_a,t_b)$,
$dv_{\tilde{\cal P}}(p,v)$ is a volume element of the  manifold $\tilde{\cal P}$.
In the path integral for the Green's function $ G_{\tilde{\cal P}}$, the integration is carried out in the space of paths with fixed both ends.

The configuration space of our original mechanical system is the Riemannian manifold $ \tilde{\mathcal P} = \mathcal P \times \mathcal V $ endowed with a smooth isometric free and proper action of the compact semisimple Lie group $\mathcal G$. 

 It is assumed that the  manifold $\tilde{\mathcal P}$  has  a  Riemannian metric, which in local coordinates for the  basis $(\frac{\partial}{\partial Q^A},\frac{\partial}{\partial f^a}$) is written as
\begin{equation}
 ds^2=G_{AB}(Q)dQ^AdQ^B+G_{ab}\,df^adf^b.
\label{metr_orig}
\end{equation}
This metric is invariant under the right action of the group  which acts as
 \[
 {\tilde Q}^A=F^A(Q,g),\;\;\;\;{\tilde f}^b=\bar D^b_a(g)f^a,
\]
where $\bar D^b_a(g)\equiv D^b_a(g^{-1})$,
 ($D^b_a(g)$ is the matrix of  the finite-dimensional representation of the group $\mathcal G$
acting in the vector space $\mathcal V$).

The  action of the group $ \mathcal G $ on the manifold $ \tilde {\mathcal P} $ leads to  the principal fiber bundle $ \pi ': \mathcal P \times \mathcal V \to \tilde {\mathcal M}=\mathcal P \times_{\mathcal G} \mathcal V $, so ${\tilde{\mathcal P}}$ can be regarded as the total space of this bundle. From this it follows that  the local coordinates $(Q^A, f^a)$ of the point $(p,v)\in \tilde{\mathcal P}$ can be expressed in terms of the  coordinates defined in the principal fiber bundle.

In gauge field theories, the calculations are performed in a special local coordinates known as  adapted coordinates.
A similar choice of coordinates was made  in the considered principal fiber bundle $\rm P(\tilde{\mathcal M},\mathcal G)$ in \cite{Storchak_2021}.
Note that to determine them, it is necessary to use such local submanifolds (local surfaces, or `gauge fixing surfaces' in gauge theories) of the total space $\tilde{\mathcal P}$ of the principal fiber bundle that intersect transversally with each of the orbits. Moreover, they must correspond to open neighborhoods $\tilde{U}_i$ from the atlas of the manifold defined on $\tilde {\mathcal M}$.

In our case, such local surfaces $\tilde{\Sigma}_i$ in the principal bundle $\rm P(\tilde {\mathcal M}, \mathcal G)$ are defined by local surfaces $\Sigma_i$ that have similar properties and are chosen in the total space $\mathcal P$ of the principal bundle $\rm P(\mathcal M, \mathcal G)$. From ${\Sigma}_i$ one  can first determine the local sections of this bundle, and then the local sections of the fiber bundle $\rm P(\tilde{\mathcal M},\mathcal G)$. The local surfaces $\tilde{\Sigma}_i$ in $\tilde{\mathcal P}$ are the images of maps given by the obtained  sections.

As in gauge theories, we assume that the local surfaces ${\Sigma}_i$ form the global surface ${\Sigma}$ in ${\mathcal P}$, which means that $\rm P({\mathcal M}, \mathcal G)$ is a trivial principal bundle. This implies that $\rm P(\tilde {\mathcal M}, \mathcal G)$ is also trivial, i.e.
$\tilde{\cal P}\sim\tilde{\mathcal M}\times \mathcal G$.

The  submanifold $\Sigma$ in $\mathcal P$ is given by the system of equations
$\chi^{\alpha}=0,\,\alpha=1,...,n_{\mathcal G}$, which means that the point with coordinates $Q^{\ast A}$ (provided that $ \chi^{\alpha}(Q^{\ast A})=0$) belongs to $\Sigma $. By this reason, the coordinates $Q^{\ast A}$ are called dependent coordinates.

The group coordinates $a^{\alpha}(Q)$ of a point $p\in \mathcal P$ are defined by the solution of the following equation
\[
 \chi^{\alpha}(F^A(Q,a^{-1}(Q)))=0.
\]
It means that
$
 Q^{\ast A}=F^A(Q,a^{-1}(Q)).
$
That is, the group element $g^{-1}(p)$ with the coordinates $a^{-1}(Q)$ carries the point $p$ to the submanifold  $\Sigma$. 

For a given (global) `gauge-fixing surface' $\Sigma$ in $\mathcal P$, the section ${\sigma}_{\Sigma}$ of the principal bundle $\pi:\cal P\to\cal M$ is defined as the inverse map to the projection
$\pi|_{\Sigma}:U_{\Sigma}\to U_{\cal M}$ (for local neighborhoods on $\Sigma$ and $\cal M$): 
$$\pi|_{\Sigma}\circ{\sigma}_{\Sigma}={\rm id}_{\pi|_{\Sigma}},\;\;\;\; {\sigma}_{\Sigma}([p])=p\,g^{-1}(p).$$
 The section ${\sigma}_{\tilde\Sigma}$ of the  principal fiber bundle $\rm P(\tilde {\mathcal M}, \mathcal G)$, ${\sigma}_{\tilde\Sigma}:\mathcal P \times_{\mathcal G} \mathcal V\to \mathcal P \times \mathcal V $, is given by 
 $${\sigma}_{\tilde\Sigma}([p,v])=({\sigma}_{\Sigma}([p]),g(p) v)=(p\,g^{-1}(p),g(p)v)=(p,v)g^{-1}(p).$$ 
 And the `gauge fixing surface' $\tilde{\Sigma}$ in $\mathcal P\times \mathcal V$ is the image of  ${\sigma}_{\tilde\Sigma}$. 
 
 Then it follows that using $\pi'_{(\tilde{\Sigma})}={\sigma}_{\tilde\Sigma}\circ\pi'$ we can define the principal fiber bundle $\pi'_{ ( \tilde{\Sigma})}:\mathcal P\times \mathcal V\to
{\tilde\Sigma}$, isomorphic to the principal bundle $\rm P(\tilde {\mathcal M}, \mathcal G)$. 

 The transition to adapted coordinates   carried out by the coordinate functions $\tilde \varphi^{-1}$:
\[
 \tilde \varphi^{-1} :(Q^A,f^b)\to (Q^{\ast}{}^A(Q),\tilde f^b(Q),a^{\alpha}(Q)\,),
\]
where
\[
\tilde f^b(Q) = D^b_c(a(Q))\,f^c,
\]       
($\bar D^b_c(a^{-1})\equiv D^b_c(a))$ and $a(Q)$ are the coordinates of the group element $g(p)$).

The inverse transformation is given by the coordinate functions $\tilde \varphi$: 
\[
 \tilde \varphi :(Q^{\ast}{}^B,\tilde f^b,a^{\alpha})\to (F^A(Q^{\ast},a), \bar D^c_b(a)\tilde f^b).
\]
These functions,  $\tilde \varphi$ and $\tilde \varphi^{-1} $, represent the bundle maps $\tilde \varphi : \tilde {\Sigma}\times\cal G \to {\mathcal P} \times \cal V$ and 
${\tilde \varphi^{-1}}: {\mathcal P} \times {\mathcal V} \to {\tilde\Sigma} \times {\mathcal G}$ for the isomorphic trivial principal bundles $\tilde {\Sigma}\times\mathcal G \to \tilde {\Sigma} $ and ${\rm P}(\tilde{\mathcal M},\mathcal G)$.

Thus,  the special  bundle coordinates in the total space of
  the principal fiber bundle 
$\pi':\mathcal P\times \mathcal V\to \mathcal P\times_{\mathcal G} \mathcal V$ are given by the following coordinate functions:
$$\tilde{\varphi}^{\tilde{\cal P}}(p,v)=(Q^{\ast}{}^A,\tilde f^b, a^{\alpha}),\;\;\rm {where}\;\; \tilde{\varphi}^{\tilde{\cal P}}:= \tilde \varphi^{-1}\circ \varphi^{\tilde{\cal P}}.$$

Replacement of the  coordinates $(Q^A,f^a)$ of a point $(p,v)\in \mathcal P\times V$  with adapted coordinates $(Q^{\ast}{}^B, \tilde f^b, a^{\alpha})$, 
\begin{equation}
Q^A=F^A(Q^{\ast}{}^B,a^{\alpha}),\;\;\;f^a=\bar D^a_b(a)\tilde f^b,
\label{transf_coord}
\end{equation}
 leads to the following transformations of the local coordinate vector fields: 
\begin{eqnarray}
\displaystyle 
&&\!\!\!\!\!\!\!\!\frac{\partial}{\partial f^a}=D^b_a(a)\frac{\partial}{\partial {\tilde f}^b},
\nonumber\\
&&\!\!\!\!\!\!\!\!\frac{\partial}{\partial Q^B}=\frac{\partial Q^{\ast}{}^A}{ \partial Q^B}\frac{\partial}{\partial Q^{\ast}{}^A}+\frac{\partial a^{\alpha}}{\partial Q^B}\frac{\partial}{\partial a^{\alpha}}+\frac{\partial {\tilde f}^b}{\partial Q^B}\frac{\partial}{\partial {\tilde f}^b}
\nonumber\\
&&\!\!\!\!\!\!\!\!\!\!\!\!\!=\check F^C_B\Biggl(N^A_C(Q^{\ast})\frac{\partial}{\partial Q^{\ast}{}^A}+{\chi}^{\mu}_C({\Phi}^{-1})^{\beta}_{\mu}\bar{v}^{\alpha}_{\beta}(a)\frac{\partial}{\partial a^{\alpha}}-{\chi}^{\mu}_C({\Phi}^{-1})^{\nu}_{\mu}(\bar J_{\nu})^b_p\tilde f^p\frac{\partial}{\partial {\tilde f}^b}\Biggr)\!.
\label{vectfield}
\end{eqnarray}
Here $\check F^C_B\equiv F^C_B(F(Q^{\ast},a),a^{-1})$ is an inverse matrix to the matrix $F^A_B(Q^{\ast},a)\equiv \frac{\partial F^A(Q,a)}{\partial Q^B}|_{Q=Q^{\ast}}$,
${\chi}^{\mu}_C\equiv \frac{\partial {\chi}^{\mu}(Q)}{\partial Q^C}|_{Q=Q^{\ast}}$, $({\Phi}^{-1})^{\beta}_{\mu}\equiv({\Phi}^{-1})^{\beta}_{\mu}(Q^{\ast})$ -- the matrix which is inverse to the Faddeev--Popov matrix:
\[
 ({\Phi})^{\beta}_{\mu}(Q)=K^A_{\mu}(Q)\frac{\partial {\chi}^{\beta}(Q)}{\partial Q^A},
\]
the matrix $\bar{v}^{\alpha}_{\beta}(a)$ is inverse of the matrix $\bar{u}^{\alpha}_{\beta}(a)$.\footnote{$\det \bar{u}^{\alpha}_{\beta}(a)$ is   the density of the right-invariant measure  given on the group $\mathcal G$.}

The operator $N^A_C$, defined  as
 $N^A_C(Q)=\delta^A_C-K^A_{\alpha}(Q)({\Phi}^{-1})^{\alpha}_{\mu}(Q){\chi}^{\mu}_C(Q)$, 
 is the projection operator ($N^A_BN^B_C=N^A_C$) onto planes perpendicular to the gauge orbits in $ {\rm P}(\mathcal M, \mathcal G)$, since $N^A_CK^C_{\alpha}=0$\cite{Creutz}.  $N^A_C(Q^{\ast})$ is the restriction of $N^A_C(Q)$ to the submanifold $ \Sigma $:
\[
 N^A_C(Q^{\ast})\equiv N^A_C(F(Q^{\ast},e))\;\;\;N^A_C(Q^{\ast})=F^B_C(Q^{\ast},a)N^M_B(F(Q^{\ast},a))\check F_M^A(Q^{\ast},a),
\]
where $e$ is the unity element of the group.

In the new coordinate basis $\displaystyle(\partial/\partial Q{}^{\ast A},\partial/\partial \tilde f^m,\partial/\partial a^{\alpha})$
the metric (\ref{metr_orig}) of the original manifold $\mathcal P \times \mathcal V$ is determined by the following metric tensor:
\begin{equation}
\displaystyle
{\tilde G}_{\cal A\cal B}(Q{}^{\ast},\tilde f,a)=
\left(
\begin{array}{ccc}
 G_{CD}(P_{\bot})^C_A (P_{\bot})^D_B & 0 & G_{CD}(P_{\bot})^C_AK^D_{\nu}\bar u^{\nu}_{\alpha}\\
 0 & G_{ab} & G_{ap}K^p_{\nu}\bar u^{\nu}_{\alpha}\\
G_{BC}K^C_{\mu}\bar u^{\mu}_{\beta} & G_{bp}K^p_{\nu}\bar u^{\nu}_{\beta} & d_{\mu\nu}\bar u^{\mu}_{\alpha}\bar u^{\nu}_{\beta}\\
\end{array}
\right)
\label{metric2c}
\end{equation}
where $G_{CD}(Q{}^{\ast})\equiv G_{CD}(F(Q{}^{\ast},e))$,
\[
 G_{CD}(Q{}^{\ast})=F^M_C(Q{}^{\ast},a)F^N_D(Q{}^{\ast},a)G_{MN}(F(Q{}^{\ast},a)), 
\]
the projection operator 
$(P_\bot)^A_B=\delta^A_B-\chi ^{\alpha}_{B}\,(\chi \chi ^{\top})^{-1}{}^{\beta}_{\alpha}\,(\chi ^{\top})^A_{\beta},$
and the components $K^A_{\mu}$ of the Killing vector fields  depend on $Q{}^{\ast}$,   $K^p_{\nu}=K^p_{\nu}(\tilde f)$, $\bar u^{\mu}_{\beta}=\bar u^{\mu}_{\beta}(a),$  $d_{\mu\nu}(Q{}^{\ast},\tilde f)$ 
is the tensor used to determine the metric given on the  
orbits of the group action. The components of $d_{\mu\nu}$  are  given by the following relation:
\begin{eqnarray*}
 d_{\mu\nu}(Q{}^{\ast},\tilde f)&=&K^A_{\mu}(Q{}^{\ast})G_{AB}(Q{}^{\ast})K^B_{\nu}(Q{}^{\ast})+K^a_{\mu}(\tilde f)G_{ab}K^b_{\nu}(\tilde f)
\nonumber\\
&\equiv&\gamma_{\mu \nu}(Q^{\ast})+\gamma'_{\mu \nu}(\tilde f).
\end{eqnarray*}

The pseudoinverse matrix ${\tilde G}^{\cal A\cal B}(Q{}^{\ast},\tilde f,a)$ to  the matrix (\ref{metric2c}) is
\begin{equation}
\displaystyle
\left(
\begin{array}{ccc}
{G}^{EF}N_E^AN_F^B & -G^{EF}N^A_E{\Lambda}^{\nu}_FK^a_{\nu} &  G^{EF}N^A_E{\Lambda}^{\beta}_F\bar v^{\alpha}_{\beta}\\
-G^{EF}N^B_F{\Lambda}^{\nu}_EK^b_{\nu} & G^{ba}+G^{EF}{\Lambda}^{\nu}_E{\Lambda}^{\mu}_FK^b_{\nu}K^a_{\mu} & -G^{EF}{\Lambda}^{\nu}_E{\Lambda}^{\mu}_FK^b_{\nu}{\bar v}^{\alpha}_{\mu}\\
G^{EF}N^B_F{\Lambda}^{\mu}_E\bar v_{\mu}^{\beta} & -G^{EF}{\Lambda}^{\nu}_E{\Lambda}^{\mu}_FK^a_{\mu}{\bar v}^{\beta}_{\nu} &G^{EF}{\Lambda}^{\nu}_E{\Lambda}^{\mu}_F{\bar v}^{\alpha}_{\nu}{\bar v}^{\beta}_{\mu}  \\
\end{array}
\right).
\label{metric2b}
\end{equation}
Here  ${\Lambda}^{\nu}_E\equiv({\Phi}^{-1})^{\nu}_{\mu}(Q{}^{\ast}){\chi}^{\mu}_E(Q{}^{\ast})$.

The pseudoinversion of ${\tilde G}_{\cal A\cal B}$ means that
\[
\displaystyle
{\tilde G}^{\tilde{\cal A}\tilde{\cal D}}{\tilde G}_{\tilde{\cal D}\tilde{\cal B}}=
\left(
\begin{array}{ccc}
  (P_{\bot})^A_B & 0 & 0\\
 0 & {\delta}^a_b & 0\\
0 & 0 & {\delta}^{\alpha}_{\beta}\\
\end{array}
\right).
%\label{orthogonal}
\]
  
The determinant of the matrix  (\ref{metric2c}) is 
$$\det {\tilde  G}_{{\cal A\cal B}}=(\det d_{\mu\nu})\;(\det {\bar u}^{\mu}_{\nu})^2\; H,$$
where
\begin{eqnarray}
 H= \displaystyle
\det \left(
\begin{array}{cc}
(P_{\bot})^{A'}_A\tilde G^{\rm H}_{A'B'}(P_{\bot})^{B'}_B & (P_{\bot})^{A'}_A\tilde G^{\rm H}_{A' a}\\
(P_{\bot})^{B'}_B\tilde G^{\rm H}_{b B'} & \tilde G^{\rm H}_{ab}\\
\end{array}
\right),
\label{det}
\end{eqnarray}
and
$\;\;\;{\tilde G}^{\rm H}_{AB}=G_{AB}-G_{AC}K^C_{\mu}d^{\mu\nu}K^D_{\nu}G_{DB},$
$\;\;\;\tilde G^{\rm H}_{Aa}=-G_{AB}K^B_{\mu}d^{\mu\nu}K^b_{\nu}G_{ba}$, $\tilde G^{\rm H}_{ba}=G_{ba}-G_{bc}K^c_{\mu}d^{\mu\nu}K_{\nu}^pG_{pa}$.

Note that $\det {\tilde  G}_{{\cal A\cal B}}$ does not vanish only on the surface $\tilde \Sigma $. On this surface $\det(P_{\bot})^{A}_B$ is equal to unity. 

The reduction of the path integral consists of several transformations of the original path integral. The first transformation of the path integral is related to the replacement of  integration variables in the original path integral. As new integration variables, we use such variables that are locally represented by adapted coordinates. The second transformation leads to factorization of the measure of the path integral, which makes it possible to separate the group variables from the variables describing the motion on the orbit space. This transformation is based on the use of the nonlinear filtering equation from the theory of the stochastic processes. We also aply the Girsanov transformation to avoid the appearance of stochastic integral as the argument of the exponential of the resulting Jacobian.
  
  As a result of all these transformations, we  obtained in \cite{Storchak_2021} an integral relation between the path integral given on $\tilde{\Sigma}$ and the path integral on $\tilde{\mathcal P}$. These path integrals were used to represent the integral relation between the corresponding Green's functions. For   reduction onto the level of zero momentum this integral relation is
\begin{eqnarray*}
 &&d_b^{-1/4}d_a^{-1/4}G_{\tilde{\Sigma}}(\pi'_{(\tilde{\Sigma})}(p_b,v_b),t_b;
 \pi'_{(\tilde{\Sigma})}(p_a,v_a),t_a)\nonumber\\
 &&\;\;\;\;\;\;\;\;\;\;\;\;\;\;\;\;\;\;\;\;\;\;\;\;\displaystyle=\int _{\cal G}G_{\tilde{\cal P}}(p_b\theta,v_b\theta,t_b;
p_a,v_a,t_a) d\mu(\theta) , 
%\label{green_funk_relat}
\end{eqnarray*}
where  $d_b$ and $d_a$ are the values of the $\det (d_{\alpha\beta})$ taken at the points $\pi'_{(\tilde{\Sigma})}(p_b,v_b)$ and $\pi'_{(\tilde{\Sigma})}(p_a,v_a)$.

The Green's function $G_{\tilde{\mathcal P}}(p_b,v_b,t_b;p_a,v_a,t_a)$ representing the kernal of the evolution semigroup ({\ref{orig_path_int}) acts in the Hilbert space of  functions with the scalar product $(\psi_1,\psi_2)=\int \psi_1\psi_2\,dv_{\tilde{\mathcal P}}$  ($dv_{\tilde{\mathcal P}}$ is a volume element of the manifold $\tilde{\mathcal P}$).

The global semigroup  determined by the Green's function $G_{\tilde{\Sigma}}$ acts in the Hilbert  space of the scalar functions on $\tilde{\Sigma}$ with the following scalar product $(\psi_1,\psi_2)=\int \psi_1 \psi_2\,dv_{\tilde{\Sigma}}.$ This 
 Green's function $G_{\tilde{\Sigma}}$ is given by the following path integral:  
\begin{eqnarray*}
 &&G_{\tilde{\Sigma}}(\pi'_{(\tilde{\Sigma})}(p_b,v_b),t_b;
 \pi'_{(\tilde{\Sigma})}(p_a,v_a),t_a)\\
 &&=\int\limits_{{\tilde{\xi}_{\tilde \Sigma} (t_a)=\pi'_{(\tilde\Sigma)} (p_a,v_a)}\atop  
{\tilde{\xi}_{\tilde \Sigma} (t_b)=\pi'_{(\tilde\Sigma)} (p_b,v_b)}} d{\mu}^{\tilde{\xi}_{\tilde \Sigma}}
\exp \Bigl\{\frac 1{\mu ^2\kappa
m}\int_{t_a}^{t_b}\Bigl[\tilde{V}(\tilde{\xi}_{\tilde \Sigma}(u))+J(\sigma (\tilde{\xi}_{\tilde \Sigma}(u)))\Bigr]du\Bigr\}.
\nonumber\\
\end{eqnarray*}
The   path integral measure ${\mu}^{\tilde{\xi}_{\tilde \Sigma}}$ is generated by the stochastic process 
$\tilde{\xi}_{\tilde \Sigma}(t)$, which
determined by the solutions of the local stochastic differential equations
\begin{eqnarray*}
 &&dQ^{\ast}{}^A(t)=\mu^2\kappa\Bigl(-\frac12h^{\tilde B\tilde M}\,{}^{ \mathrm  H}{\tilde \Gamma}^{ A}_{\tilde B\tilde M}+j^{ A}_{\Roman 1}\Bigr)dt
  +\mu\sqrt{\kappa}N^A_C\mathscr X^C_{\bar M}dw^{\bar M}_t,
 \label{sde_Q_ast_j}\,\,\rm{and}
\end{eqnarray*}
 \begin{eqnarray*}
  &&d\tilde f^a(t)=\mu^2\kappa\Bigl(-\frac12h^{\tilde B\tilde M}\,{}^{ \mathrm  H}{\tilde \Gamma}^{ a}_{\tilde B\tilde M}+j^{ a}_{\Roman 1}\Bigr)dt
    +\mu\sqrt{\kappa}\bigl(N^a_C\mathscr X^C_{\bar M}dw^{\bar M}_t+\mathscr X^a_{\bar b}dw^{\bar b}_t\bigr),
  \label{sde_f_j} 
 \end{eqnarray*}
where $ h^{\tilde B\tilde M}\equiv G^{EF}N^{\tilde B}_EN^{\tilde M}_F$,\footnote{The definition of the projection operator $N^{\tilde A}_{\tilde B}$ is given in Appendix A.} $j^{ A}_{\Roman 1}$ and $j^{ a}_{\Roman 1}$ are the components of the mean curvature normal of the orbit space. They are given by the following expressions:
\begin{align*}
 j^{ A}_{\Roman 1}&=\frac12 h^{BM}N^A_{B,M}+\frac12h^{\tilde B\tilde M}\,{}^{ \mathrm  H}{\tilde \Gamma}^{ A}_{\tilde B\tilde M}-\frac12h^{\tilde B\tilde M}N^A_C\,{}^{ \mathrm  H}{\tilde \Gamma}^{ C}_{\tilde B\tilde M},
\\
j^{ a}_{\Roman 1}&=-\frac12N^a_Ch^{BM}N^C_{B,M}-\frac12N^a_Ch^{\tilde B\tilde M}\,{}^{ \mathrm  H}{\tilde \Gamma}^{ C}_{\tilde B\tilde M}.
\end{align*}

In the above path integral, the integrand $J$ of the path integral reduction  Jacobian  
is given by
\begin{equation}
J=-\frac18\mu^2\kappa\bigl(\triangle_{\tilde{\Sigma}}^{{\scriptscriptstyle \rm H}}\sigma +\frac14<\partial\sigma,\partial \sigma>_{\tilde{\Sigma}}\bigr)=-\frac18\mu^2\kappa\, \tilde J,
\label{jacobian_reduct}
\end{equation}
where $$\triangle_{\tilde{\Sigma}}^{{\scriptscriptstyle \rm H}}\sigma=
h^{AB}\sigma_{AB}+2h^{Ab}\sigma_{Ab}
+h^{ab}\sigma_{ab}
-h^{\tilde B \tilde M}\,{}^{ \mathrm  H}{\tilde \Gamma}^{ A}_{\tilde B\tilde M}\sigma_A
-h^{\tilde B \tilde M}\,{}^{ \mathrm  H}{\tilde \Gamma}^{ a}_{\tilde B\tilde M}\sigma_a$$ 
 is the Laplacian on $\tilde{\Sigma}$, and 
$<\partial\sigma,\partial \sigma>_{\tilde{\Sigma}}$ is a quadratic form in first partial derivatives of $\sigma\equiv \ln d$ \footnote{These partial derivatives will be denoted as $\sigma_A=\partial \sigma /\partial Q^{\ast A}$, $\sigma_a=\partial \sigma /\partial \tilde f^a$, etc.}, defined by the formula
$$h^{AB}{\sigma}_A {\sigma}_B+2h^{aB}{\sigma}_a {\sigma}_B+h^{ab}{\sigma}_a {\sigma}_b.$$
%$$G^{CD}N^A_CN^B_D{\sigma}_A {\sigma}_B+2G^{CD}N^a_CN^B_D{\sigma}_a {\sigma}_B+(G^{CD}N^a_CN^b_D+G^{ab}){\sigma}_a %{\sigma}_b.$$
The expression for this quadratic form is obtained by using the pseudoinverse matrix to the matrix representing the ``horizontal metric''  on $\tilde{ \Sigma}$.

The Green's function $G_{\tilde{\Sigma}}$ satisfies the forward Kolmogorov equation with the operator
\begin{equation*}
\hat{H}_{\kappa}=
\frac{\hbar \kappa}{2m}\tilde\triangle _{\tilde{\Sigma}}-\frac{\hbar \kappa}{8m}\Bigl[\triangle_{\tilde{\Sigma}}^{{\scriptscriptstyle \rm H}}\sigma +\frac14<\partial\sigma,\partial \sigma>_{\tilde{\Sigma}}\Bigr]+\frac{1}{\hbar \kappa}\tilde V,
%\nonumber\\
\end{equation*}
where $$\tilde\triangle_{\tilde{\Sigma}} =\triangle_{\tilde{\Sigma}}^{{\scriptscriptstyle \rm H}}+2j^{ A}_{\Roman 1}\,\partial_A+2j^{ a}_{\Roman 1}\,\partial_a.$$
Note that at $\kappa =i$ the forward Kolmogorov equation becomes
the Schr\"odinger equation with the Hamilton operator 
$\hat H=-\frac{\hbar}{\kappa}{\hat H}_{\kappa}\bigl|_{\kappa =i}$.

\section{The scalar curvature  of the manifold  $\tilde {\mathcal P}$}
The scalar curvature of $\tilde {\mathcal P}$ will be calculated by using the special nonholonomic  basis, also known  as the horizontal lift basis. 
This basis consists of the horizontal vector fields $H_A$  and $H_p$ together with the left-invariant vector fields $L_{\alpha}=v^{\mu}_{\alpha}(a)\frac{\partial}{\partial a^{\mu}}$.
 In terms of the principal fiber bundle coordinates, the horizontal vector fields are 
\begin{eqnarray*}
 &&H_A(Q^{\ast},\tilde f,a)= \Bigl[N^T_A\Bigl(\frac{\partial}{\partial Q^{\ast T}}
-\tilde {\mathscr A}^{\alpha }_T L_{\alpha}\Bigl)+N^p_A\Bigl(\frac{\partial}{\partial {\tilde f}^p}-\tilde {\mathscr A}^{\alpha }_pL_{\alpha}\Bigl)\Bigr],\\
&&H_p(Q^{\ast},\tilde f,a)=\Bigl( \frac{\partial}{\partial {\tilde f}^p}-\tilde {\mathscr A}^{\alpha }_pL_{\alpha}\Bigl).\\
\end{eqnarray*}
These vector fields are defined with the help of the  components of the projection operator $N^{\tilde A}_{\tilde C}$ :
$N^{\tilde A}_{\tilde C}=(N^A_C,N^A_p,N^q_A,N^q_p).$ 

The commutation relations of the vector fields have the following form:
\begin{eqnarray*}
 &&[H_A,H_B]={\mathbb C}^T_{AB}\,H_T+{\mathbb C}^p_{AB}\,H_p+{\mathbb C}^{\alpha}_{AB}L_{\alpha}, \\
 &&[H_A,H_p]={\mathbb C}^q_{Ap}\,H_q+{\mathbb C}^{\alpha}_{Ap}L_{\alpha},\\
 &&[H_p,H_q]={\mathbb C}^{\alpha}_{pq}L_{\alpha},\\
&&[H_A,L_{\alpha}]=0,\;\;\;[H_p,L_{\alpha}]=0,\;\;\;[L_{\alpha},L_{\beta}]=c^{\gamma}_{\alpha \beta}L_{\gamma},\\
 \end{eqnarray*}
%\label{commrelat_AB}
where the structure constants of the nonholonimic basis  are 
\begin{eqnarray*}
&&{\mathbb C}^T_{AB}=({\Lambda}^{\gamma}_A N^R_B-{\Lambda}^{\gamma}_BN^R_A) K^{T}_{{\gamma}, R},\\
&&{\mathbb C}^p_{AB}=-N^D_AN^R_B({\Lambda}^{\alpha}_{R,D}-{\Lambda}^{\alpha}_{D,R}) K^p_{\alpha}  \;\;  -c^{\sigma}_{\alpha \beta}{\Lambda}^{\beta}_A{\Lambda}^{\alpha}_BK^p_{\sigma}, \\
&&{\mathbb C}^{\alpha}_{AB}=-N^S_AN^P_B\,\tilde{\mathscr F}^{\alpha}_{SP}-(N^E_AN^p_B-N^E_BN^p_A)\tilde{\mathscr F}^{\alpha}_{Ep}+N^a_AN^p_B\tilde{\mathscr F}^{\alpha}_{pa},\\
&&{\mathbb C}^q_{Ap}=({\bar J}_{\alpha})^q_p{\Lambda}^{\alpha}_A=K^q_{\alpha,p}{\Lambda}^{\alpha}_A,\;\;\;
{\mathbb C}^{\alpha}_{Ap}=-N^E_A\tilde{\mathscr F}^{\alpha}_{Ep}-N^q_A\tilde{\mathscr F}^{\alpha}_{qp},\;\;\;{\mathbb C}^{\alpha}_{pq}=-\tilde{\mathscr F}^{\alpha}_{pq}\,.\\
\end{eqnarray*}
The curvature tensor $\tilde{\mathscr F}^{\alpha}_{SP}$ of the mechanical connection ${\tilde{\mathscr A}^{\alpha}_P}=\bar{\rho}^{\alpha}_{\mu}{{\mathscr A}^{\mu}_P}$, with ${{\mathscr A}^{\mu}_P}=d^{\mu\nu}K^R_{\nu}G_{RP}$, is given by
\[
\tilde{\mathscr F}^{\alpha}_{SP}=\displaystyle\frac{\partial}{\partial Q^{\ast}{}^S}\,\tilde{\mathscr A}^{\alpha}_P- 
\frac{\partial}{\partial {Q^{\ast}}^P}\,\tilde{\mathscr A}^{\alpha}_S
+c^{\alpha}_{\nu\sigma}\, \tilde{\mathscr A}^{\nu}_S\,
\tilde{\mathscr A}^{\sigma}_P,\;\;(\tilde{\mathscr F}^{\alpha}_{SP}({Q^{\ast}},a)={\bar{\rho}}^{\alpha}_{\mu}(a)\,{\mathscr F}^{\mu}_{SP}(Q^{\ast})).  
\]
  The tensors $\tilde{\mathscr F}^{\alpha}_{Ep}$ and $\tilde{\mathscr F}^{\alpha}_{pa}$ are defined in a similar way. Note  also that ${\mathbb C}^q_{Ap}=-{\mathbb C}^q_{pA}$. 
  %It can also be shown that the first term on the right in ${\mathbb C}^p_{AB}$ is zero.

In our nonholonomic basis $(H_A,H_p,L_{\alpha})$, the metric tensor
(\ref{metric2c}) takes the  form
\begin{equation}
\displaystyle
{\check G}_{\cal A\cal B}(Q^{\ast},\tilde f,a)=
\left(
\begin{array}{ccc}
{\tilde G}^{\rm H}_{AB} & {\tilde G}^{\rm H}_{Ab} & 0\\
{\tilde G}^{\rm H}_{aB} & {\tilde G}^{\rm H}_{ab} & 0\\
0 & 0 &\tilde{d}_{\alpha \beta }  \\
\end{array}
\right)=\left( \begin{array}{cc}
{\tilde G}^{\rm H}_{\tilde A \tilde B}  & 0 \\
0 & \tilde{d}_{\alpha \beta }  \\
\end{array}
\right),
\label{metric2cc}
\end{equation}
with
\[
 {\check G}(H_A,H_B)\equiv{\tilde G}^{\rm H}_{AB},\;{\check G}(H_A,H_b)\equiv{\tilde G}^{\rm H}_{Ab},\;{\check G}(L_{\alpha},L_{\beta})\equiv\tilde{d}_{\alpha \beta }=\rho^{\alpha'}_{\alpha}\rho^{\beta'}_{\beta}d_{\alpha' \beta' }
\]

The metric tensor ${\tilde G}^{\rm H}_{\tilde{ A} \tilde{ B}}(Q^{\ast}{}^A,\tilde f ^p)$ of the ``horizontal metric'' has the following components:
%depending on $(Q^{\ast}{}^A,\tilde f ^m)$ 
\begin{eqnarray*}
&&{\tilde G}^{\rm H}_{AB}={{\tilde \Pi}}^{\tilde A}_{A}\,{{\tilde \Pi}}^{\tilde B}_B \,G_{\tilde A\tilde B}=G_{AB}-G_{AD}K^{D}_{\alpha}d^{\alpha \beta}K^R_{\beta}\,G_{RB},\\
%because of ${{\tilde \Pi}}^{\tilde C}_{A}\,{{\tilde \Pi}}^{\tilde D}_B \,G_{\tilde C\tilde D}= {{\tilde \Pi}}^{C}_{A}\,{{\tilde \Pi}}^{D}_B \,G_{CD}+
%{{\tilde \Pi}}^{q}_{A}\,{{\tilde \Pi}}^{p}_B \,G_{qp}$.
&&{\tilde G}^{\rm H}_{Ab}=-G_{AB}K^{B}_{\alpha}\,d^{\alpha \beta}K^p_{\beta}G_{pb},\;\;\;{\tilde G}^{\rm H}_{Ab}={\tilde G}^{\rm H}_{bA},\\
%Notice that ${\tilde G}^{\rm H}_{Am}$ is equal to
%$${\tilde G}^{\rm H}_{mA}=-G_{mq}K^{q}_{\mu}\,d^{\mu \nu}K^D_{\nu}G_{DA}.$$
&&{\tilde G}^{\rm H}_{ab}={{\tilde \Pi}}^r_aG_{rb}={{\tilde \Pi}}^{\tilde C}_{a}\,{{\tilde \Pi}}^{\tilde D}_b \,G_{\tilde C\tilde D}=G_{ab}-G_{ar}K^r_{\alpha}d^{\alpha \beta}K_{\beta}^pG_{pb}.\\
\end {eqnarray*}
 Note that  ${\tilde G}^{\rm H}_{\tilde{ A} \tilde{ B}}$  is defined on a local surface (submanifold) $\tilde \Sigma$.    In the case when the submanifold $\tilde \Sigma$ can be given parametrically, ${\tilde G}^{\rm H}_{\tilde{ A} \tilde{ B}}$  is  transformed into a metric tensor representing  the metric on the  orbit space $\mathcal P\times _\mathcal G V$.

The pseudoinverse matrix ${\check G}^{\cal A\cal B}$ to the matrix (\ref{metric2cc}) is represented as
\begin{equation*}
\displaystyle
{\check G}^{\cal A\cal B}=
\left(
\begin{array}{ccc}
{G}^{EF}N^A_EN^B_F & {G}^{EF}N^A_EN^q_F & 0\\
{G}^{EF}N^p_F N^B_E & {G}^{pq}+G^{AB}N^p_AN^q_B & 0\\
0 & 0 &\tilde{d}^{\alpha \beta }  \\
\end{array}
\right)\equiv
\left(
\begin{array}{ccc}
 h^{AB} & h^{Aq} & 0\\
 h^{pB} & h^{pq} & 0 \\
 0 & 0 & \tilde{d}^{\alpha \beta }\\
 \end{array}
 \right).
%\label{metric2bb}
\end{equation*}
This matrix is defined from the following orthogonality condition:
\[
\displaystyle
{\check G}^{\cal A\cal B}{\check G}_{\cal B\cal D}=
\left(
\begin{array}{ccc}
N^A_D& 0 & 0\\
 N^p_D & {\delta}^p_d & 0\\
0 & 0 & {\delta}^{\alpha}_{\beta} \\
\end{array}
\right)\equiv
\left(
\begin{array}{cc}
N^{\tilde A}_{\tilde D} & 0 \\
0 & {\delta}^{\alpha}_{\beta} \\
\end{array}
\right),
%\label{metric2c}
\]
where 
\[
\displaystyle
N^{\tilde A}_{\tilde D}=
\left(
\begin{array}{cc}
N^A_D& N^A_d\\
 N^p_D & N^p_d\\
\end{array}
\right)
%\label{NN}
\]
($N^A_d=0, N^p_d={\delta}^p_d$).

\subsection{The Christoffel symbols}
In the article, the computation of the Christoffel symbols ${\check{\rm\Gamma}}_{\mathscr A\mathscr B}^{\mathscr D}$ is carried out according to the following formula:
\begin{eqnarray}
&&2{\check{\rm\Gamma}}_{\mathscr A\mathscr B}^{\mathscr D}\check G(\partial_{\mathscr D},\partial_{\mathscr C})=\partial_{\mathscr A}\check G(\partial_{\mathscr B},\partial_{\mathscr C})+\partial_{\mathscr B}\check G(\partial_{\mathscr A},\partial_{\mathscr C})-\partial_{\mathscr C}\check G(\partial_{\mathscr A},\partial_{\mathscr B})\nonumber\\
&&\;\;\;\;\;-\check G(\partial_{\mathscr A},[\partial_{\mathscr B},\partial_{\mathscr C}])-\check G(\partial_{\mathscr B},[\partial_{\mathscr A},\partial_{\mathscr C}])+\check G(\partial_{\mathscr C},[\partial_{\mathscr A},\partial_{\mathscr B}]),
\label{christoff_main}
\end{eqnarray}
where by indices we mean generalized indices, such as, for example, $\mathscr A=(A,a,\alpha)$, and by $\partial_{\mathscr A}$ we denote one of the vector fields of our coordinate basis.
 We also assume that in (\ref{christoff_main}) there is a summation over repeated indices.

In the case when in (\ref{christoff_main}) \underline{ $\mathscr A=A, \mathscr B=B, \mathscr C=p$ } this formula is  written as
\begin{eqnarray*}
 &&2[{\check\Gamma}^D_{AB}{\tilde G}^{\rm H}_{Dp}+{\check\Gamma}^a_{AB}{\tilde G}^{\rm H}_{ap}]=
H_A{\tilde G}^{\rm H}_{Bp}+H_B{\tilde G}^{\rm H}_{Ap}-H_p{\tilde G}^{\rm H}_{AB}\nonumber\\
&&-\check G(H_A,[H_B,H_p])-\check G(H_B,[H_A,H_p])+\check G(H_p,[H_A,H_B]).
\end{eqnarray*}
This can be represented in the following form:
\begin{eqnarray*}
 &&2[{\check\Gamma}^D_{AB}{\tilde G}^{\rm H}_{Dp}+{\check\Gamma}^a_{AB}{\tilde G}^{\rm H}_{ap}]\\
&&\;\;\;=N^E_A{\tilde G}^{\rm H}_{Bp,E}+N^q_A{\tilde G}^{\rm H}_{Bp,q}+N^E_B{\tilde G}^{\rm H}_{Ap,E}+N^q_B{\tilde G}^{\rm H}_{Ap,q}-{\tilde G}^{\rm H}_{AB,p}
\nonumber\\
&&\;\;\;-\mathbb C^q_{Bp}{\tilde G}^{\rm H}_{Aq}-\mathbb C^q_{Ap}{\tilde G}^{\rm H}_{Bq}
+\mathbb C^T_{AB}{\tilde G}^{\rm H}_{pT}+\mathbb C^q_{AB}{\tilde G}^{\rm H}_{pq},
\end{eqnarray*}
or, explicitly, as 
\begin{eqnarray*}
 &&2[{\check\Gamma}^D_{AB}{\tilde G}^{\rm H}_{Dp}+{\check\Gamma}^a_{AB}{\tilde G}^{\rm H}_{ap}]=\\
&&N^E_A{\tilde G}^{\rm H}_{Bp,E}+N^q_A{\tilde G}^{\rm H}_{Bp,q}+N^E_B{\tilde G}^{\rm H}_{Ap,E}+N^q_B{\tilde G}^{\rm H}_{Ap,q}-{\tilde G}^{\rm H}_{AB,p}
\nonumber\\
&&-K^q_{\alpha,p}\Lambda^{\alpha}_B{\tilde G}^{\rm H}_{Aq}- K^q_{\alpha,p}\Lambda^{\alpha}_A{\tilde G}^{\rm H}_{Bq}
+(\Lambda^{\gamma}_AN^R_B-\Lambda^{\gamma}_BN^R_A)K^T_{\gamma, R}{\tilde G}^{\rm H}_{pT}
\nonumber\\
&&-c^{\sigma}_{\alpha \beta}\Lambda^{\beta}_A\Lambda^{\alpha}_BK^q_{\sigma}{\tilde G}^{\rm H}_{pq}.
\end{eqnarray*}

 When \underline{ $\mathscr A=A, \mathscr B=b, \mathscr C=C$}, ``$(A,b,C)$''-- case,  the formula (\ref{christoff_main}) leads to  the equation
\begin{eqnarray*}
 &&2[{\check\Gamma}^D_{Ab}{\tilde G}^{\rm H}_{DC}+{\check\Gamma}^a_{Ab}{\tilde G}^{\rm H}_{aC}]=
H_A{\tilde G}^{\rm H}_{bC}+H_B{\tilde G}^{\rm H}_{AC}-H_C{\tilde G}^{\rm H}_{Ab}\nonumber\\
&&-\check G(H_A,[H_b,H_C])-\check G(H_b,[H_A,H_C])+\check G(H_C,[H_A,H_b]).
\end{eqnarray*}

The right-hand side of this equation has the following terms:
\begin{eqnarray*}
 &&
H_A{\tilde G}^{\rm H}_{bC}+H_b{\tilde G}^{\rm H}_{AC}-H_C{\tilde G}^{\rm H}_{Ab}\nonumber\\
&&-\mathbb C^q_{bC}{\tilde G}^{\rm H}_{Aq}-\mathbb C^T_{AC}{\tilde G}^{\rm H}_{bT}-\mathbb C^q_{AC}{\tilde G}^{\rm H}_{qb}+\mathbb C^q_{Ab}{\tilde G}^{\rm H}_{Cq}.
\end{eqnarray*}
They can be rewritten as
\begin{eqnarray*}
 &&N^E_A{\tilde G}^{\rm H}_{bC,E}+N^q_A{\tilde G}^{\rm H}_{bC,q}+{\tilde G}^{\rm H}_{AC,b}-N^E_C{\tilde G}^{\rm H}_{Ab,E}-N^q_C{\tilde G}^{\rm H}_{Ab,q}
\nonumber\\
&&
-(\Lambda^{\gamma}_AN^R_C-\Lambda^{\gamma}_CN^R_A)K^T_{\gamma, R}{\tilde G}^{\rm H}_{bT}+
c^{\sigma}_{\alpha \beta}\Lambda^{\beta}_A\Lambda^{\alpha}_CK^q_{\sigma}{\tilde G}^{\rm H}_{qb}\nonumber\\  
&&+K^q_{\alpha,b}\Lambda^{\alpha}_C{\tilde G}^{\rm H}_{Aq}+ K^q_{\alpha,b}\Lambda^{\alpha}_A{\tilde G}^{\rm H}_{Cq}.
\end{eqnarray*}
The terms of this expression can be represented in a different form.
The first term is rewritten as
\[
 N^E_A({\tilde G}^{\rm H}_{bC,E}+{\tilde G}^{\rm H}_{EC,b}-{\tilde G}^{\rm H}_{bE,C})-N^E_A({\tilde G}^{\rm H}_{EC,b}-{\tilde G}^{\rm H}_{bE,C}).
\]
This leads to 
\[
 2N^E_A\,{}^{\mathrm  H}{\tilde \Gamma}_{bEC}-{\tilde G}^{\rm H}_{AC,b}+K^E_{\alpha}\Lambda^{\alpha}_A{\tilde G}^{\rm H}_{EC,b}+{\tilde G}^{\rm H}_{bA,C}-K^E_{\alpha}\Lambda^{\alpha}_A{\tilde G}^{\rm H}_{bE,C}.
\]

The sixth term is rewritten as
\[
 -[(\Lambda^{\gamma}_A(\delta ^R_C-K^R_{\varphi} \Lambda^{\varphi}_C)-(\Lambda^{\gamma}_C(\delta ^R_A-K^R_{\mu} \Lambda^{\mu}_A)]
K^T_{\gamma, R}{\tilde G}^{\rm H}_{bT}, 
\]
where, according to the identity $(\rm C)$ from Appendix A, one should use
\[
 K^T_{\gamma, R}{\tilde G}^{\rm H}_{bT}=-K^T_{\gamma}{\tilde G}^{\rm H}_{bT,R}-K^q_{\gamma}{\tilde G}^{\rm H}_{bq,R}.
\]
After that, we will first consider the transformation of  the $``\Lambda \Lambda$''-- terms which are on the right-hand side of the  equation. The transformed sixth term give
\begin{eqnarray*}
 \Lambda^{\gamma}_A\Lambda^{\varphi}_CK^R_{\varphi}K^T_{\gamma}{\tilde G}^{\rm H}_{bT,R}+\Lambda^{\gamma}_A\Lambda^{\varphi}_CK^R_{\varphi}K^q_{\gamma}{\tilde G}^{\rm H}_{bq,R}\\
-\Lambda^{\gamma}_C\Lambda^{\mu}_AK^R_{\mu}K^T_{\gamma}{\tilde G}^{\rm H}_{bT,R}-\Lambda^{\gamma}_C\Lambda^{\mu}_AK^R_{\mu}K^q_{\gamma}{\tilde G}^{\rm H}_{bq,R}.
\end{eqnarray*}
Using the identity $(\rm C)$ from Appendix A, we have
\[
 K^T_{\gamma}{\tilde G}^{\rm H}_{bT,R}+K^q_{\gamma}{\tilde G}^{\rm H}_{bq,R}=-{\tilde G}^{\rm H}_{bT}K^T_{\gamma, R}.
\]
So, we get 
\[
 -\Lambda^{\gamma}_C\Lambda^{\mu}_A{\tilde G}^{\rm H}_{bT}(K^R_{\mu}K^T_{\gamma,R}-K^R_{\gamma}K^T_{\mu,R})=-\Lambda^{\gamma}_C\Lambda^{\mu}_A{\tilde G}^{\rm H}_{bT}K^T_{\sigma} c^{\sigma}_{\mu \gamma}.
\]
Combining this with the seventh term of the original expression, we get
\[
 -\Lambda^{\mu}_A\Lambda^{\gamma}_C c^{\sigma}_{\mu \gamma}(K^T_{\sigma}{\tilde G}^{\rm H}_{bT}+K^q_{\sigma}{\tilde G}^{\rm H}_{qb}).
\]
But the expression in the bracket  is zero by the identity $(B)$ from Appendix A.
This means that there are no  $``\Lambda \Lambda$''-- terms in the final expression.

The remaining terms of the right-hand side of the considered equation are
\begin{enumerate}
 \item  
 $2N^E_A\,{}^{\mathrm  H}{\tilde \Gamma}_{bEC}-{\tilde G}^{\rm H}_{AC,b}+K^E_{\alpha}\Lambda^{\alpha}_A{\tilde G}^{\rm H}_{EC,b}+{\tilde G}^{\rm H}_{bA,C}-K^E_{\alpha}\Lambda^{\alpha}_A{\tilde G}^{\rm H}_{bE,C}$
\item $-K^q_{\alpha}\Lambda^{\alpha}_A{\tilde G}^{\rm H}_{bC,q}$
\item ${\tilde G}^{\rm H}_{AC,b}$
\item $-{\tilde G}^{\rm H}_{Ab,C}+K^E_{\alpha}\Lambda^{\alpha}_C{\tilde G}^{\rm H}_{Ab,E}$
\item $K^q_{\mu}\Lambda^{\mu}_C{\tilde G}^{\rm H}_{Ab,q}$
\item $\Lambda^{\gamma}_AK^T_{\gamma}{\tilde G}^{\rm H}_{bT,C}+\Lambda^{\gamma}_AK^q_{\gamma}{\tilde G}^{\rm H}_{bq,C}-\Lambda^{\gamma}_CK^T_{\gamma}{\tilde G}^{\rm H}_{bT,A}-\Lambda^{\gamma}_CK^q_{\gamma}{\tilde G}^{\rm H}_{bq,A}$
\item $0$
\item $\Lambda^{\alpha}_CK^q_{\alpha,b}{\tilde G}^{\rm H}_{Aq}=-\Lambda^{\alpha}_CK^q_{\alpha}{\tilde G}^{\rm H}_{Aq,b}-\Lambda^{\alpha}_CK^R_{\alpha}{\tilde G}^{\rm H}_{AR,b}$, by the identity $(D)$,
\item $\Lambda^{\alpha}_AK^q_{\alpha,b}{\tilde G}^{\rm H}_{Cq}=-\Lambda^{\alpha}_AK^q_{\alpha}{\tilde G}^{\rm H}_{Cq,b}-\Lambda^{\alpha}_AK^R_{\alpha}{\tilde G}^{\rm H}_{CR,b}$, by the identity $(D)$.
\end{enumerate}
 Note that there are  mutual cancellations between some terms: $(1.2)$\footnote{Notation (1.2) means the second term of the first row.} is cancelled with $(3.1)$,  $(1.3)$ with $(9.2)$, $(1.4)$ with $(4.1)$, $(1.5)$ with $(6.1)$.

The sum of  three terms $(6.3)+(4.2)+(8.2)$ is rewritten according to the formula
\[
\Lambda^{\gamma}_CK^T_{\gamma}(-{\tilde G}^{\rm H}_{bT,A}+{\tilde G}^{\rm H}_{Ab,T}-{\tilde G}^{\rm H}_{AT,b})=-2\Lambda^{\gamma}_CK^T_{\gamma}\,{}^{\mathrm  H}{\tilde \Gamma}_{AbT}.
\]

For the sum $(6.2)+(9.1)+(2.1)$ we have
\[
 \Lambda^{\gamma}_AK^q_{\gamma}({\tilde G}^{\rm H}_{bq,C}-{\tilde G}^{\rm H}_{Cq,b}-{\tilde G}^{\rm H}_{bC,q})
=-2\Lambda^{\gamma}_AK^q_{\gamma}\,{}^{\mathrm  H}{\tilde \Gamma}_{bqC}.
\]
And for $(6.4)+(5.1)+(8.1)$ --
\[
 \Lambda^{\gamma}_CK^q_{\gamma}(-{\tilde G}^{\rm H}_{bq,A}+{\tilde G}^{\rm H}_{Ab,q}-{\tilde G}^{\rm H}_{Aq,b})
=-2\Lambda^{\gamma}_CK^q_{\gamma}\,{}^{\mathrm  H}{\tilde \Gamma}_{Abq}.
\]
So we get
\[
 2(N^E_A\,{}^{\mathrm  H}{\tilde \Gamma}_{bEC}+N^q_A\,{}^{\mathrm  H}{\tilde \Gamma}_{bqC}-\Lambda^{\gamma}_CK^T_{\gamma}\,{}^{\mathrm  H}{\tilde \Gamma}_{AbT}-\Lambda^{\gamma}_CK^q_{\gamma}\,{}^{\mathrm  H}{\tilde \Gamma}_{Abq}).
\]
It can be shown that
\[
 K^T_{\gamma}\,{}^{\mathrm  H}{\tilde \Gamma}_{AbT}+K^q_{\gamma}\,{}^{\mathrm  H}{\tilde \Gamma}_{Abq}=0.
\]
This is done in the following way. We first rewrite the expression as
\[
 K^T_{\gamma}({\tilde G}^{\rm H}_{AT,b}+{\tilde G}^{\rm H}_{Tb,A}-{\tilde G}^{\rm H}_{Ab,T})+K^q_{\gamma}({\tilde G}^{\rm H}_{Aq,b}+{\tilde G}^{\rm H}_{qb,A}-{\tilde G}^{\rm H}_{Ab,q}).
\]
Making use of the identities $(D)$ and $(C)$ from Appendix A, we have the representations 
\[
 K^T_{\gamma}{\tilde G}^{\rm H}_{AT,b}=-K^q_{\gamma}{\tilde G}^{\rm H}_{Aq,b}-K^q_{\gamma,b}{\tilde G}^{\rm H}_{Aq}
\]
and
\[
 K^T_{\gamma}{\tilde G}^{\rm H}_{Tb,A}=-K^q_{\gamma}{\tilde G}^{\rm H}_{bq,A}-K^R_{\gamma,A}{\tilde G}^{\rm H}_{bR}.
\]
Taking them into account, we obtain the expression 
\[
 K^T_{\gamma}{\tilde G}^{\rm H}_{Ab,T}+K^q_{\gamma,b}{\tilde G}^{\rm H}_{Aq}+K^R_{\gamma,A}{\tilde G}^{\rm H}_{bR}+K^q_{\gamma}{\tilde G}^{\rm H}_{Ab,q},
\]
which is equal to zero by the Killing identity $\Roman{4}$ from Appendix A.
Therefore, as a result,  we have the following relation:
\[
 {\check\Gamma}^D_{Ab}{\tilde G}^{\rm H}_{DC}+{\check\Gamma}^a_{Ab}{\tilde G}^{\rm H}_{aC}=N^E_A\,{}^{\mathrm  H}{\tilde \Gamma}_{bEC}+N^q_A\,{}^{\mathrm  H}{\tilde \Gamma}_{bqC}, 
\]
which can be rewritten as 
\[
 {\check\Gamma}^{\tilde D}_{Ab}{\tilde G}^{\rm H}_{{\tilde D}C}=N^{\tilde E}_A\,{}^{\mathrm  H}{\tilde \Gamma}_{b{\tilde E}C}.
\]

In the case when \underline{ $\mathscr A=A, \mathscr B=b, \mathscr C=p$ }, the formula (\ref{christoff_main}) has the form
\begin{eqnarray*}
&&2[{\check\Gamma}^D_{Ab}{\tilde G}^{\rm H}_{Dp}+{\check\Gamma}^a_{Ab}{\tilde G}^{\rm H}_{ap}]=
H_A{\tilde G}^{\rm H}_{bp}+H_b{\tilde G}^{\rm H}_{Ap}-H_p{\tilde G}^{\rm H}_{Ab}\nonumber\\
&&-\check G(H_A,[H_b,H_p])-\check G(H_b,[H_A,H_p])+\check G(H_p,[H_A,H_b]).
\end{eqnarray*}

The right-hand side of this formula, 
\begin{eqnarray*}
 H_A{\tilde G}^{\rm H}_{bp}+H_b{\tilde G}^{\rm H}_{Ap}-H_p{\tilde G}^{\rm H}_{Ab}
-\mathbb C^q_{Ap}{\tilde G}^{\rm H}_{bq}+\mathbb C^q_{Ab}{\tilde G}^{\rm H}_{pq},
\end{eqnarray*}
is explicitly given by the expression 
\begin{eqnarray*}
 N^E_A{\tilde G}^{\rm H}_{bp,E}+N^q_A{\tilde G}^{\rm H}_{bp,q}+{\tilde G}^{\rm H}_{Ap,b}-{\tilde G}^{\rm H}_{Ab,p}
-K^q_{\alpha,p}\Lambda^{\alpha}_A{\tilde G}^{\rm H}_{bq}+ K^q_{\alpha,b}\Lambda^{\alpha}_A{\tilde G}^{\rm H}_{pq}.
\end{eqnarray*}
 The first term of the expression can be rewritten as follows
\[
  N^E_A({\tilde G}^{\rm H}_{bp,E}+{\tilde G}^{\rm H}_{pE,b}-{\tilde G}^{\rm H}_{bE,p})-N^E_A{\tilde G}^{\rm H}_{pE,b}+N^E_A{\tilde G}^{\rm H}_{bE,p}
\]
to get
\[
 2N^E_A\,{}^{\mathrm  H}{\tilde \Gamma}_{bEp}-N^E_A{\tilde G}^{\rm H}_{pE,b}+N^E_A{\tilde G}^{\rm H}_{bE,p}.
\]
Therefore, on the right side of the formula under consideration is the sum of the following terms:
\begin{enumerate}
 \item $2N^E_A\,{}^{\mathrm  H}{\tilde \Gamma}_{bEp}-{\tilde G}^{\rm H}_{pA,b}+K^E_{\alpha}\Lambda^{\alpha}_A{\tilde G}^{\rm H}_{pE,b}+{\tilde G}^{\rm H}_{bA,p}-K^E_{\alpha}\Lambda^{\alpha}_A{\tilde G}^{\rm H}_{bE,p}$
\item $-K^q_{\alpha,p}\Lambda^{\alpha}_A{\tilde G}^{\rm H}_{bp,q}$
\item ${\tilde G}^{\rm H}_{Ap,b}$
\item $-{\tilde G}^{\rm H}_{Ab,p}$
\item $-K^q_{\alpha,p}\Lambda^{\alpha}_A{\tilde G}^{\rm H}_{bq}=\Lambda^{\alpha}_A({\tilde G}^{\rm H}_{bq,p}K^q_{\alpha} +{\tilde G}^{\rm H}_{bE,p}K^E_{\alpha})$, by the identity $(B)$,
\item $K^q_{\alpha,b}\Lambda^{\alpha}_A{\tilde G}^{\rm H}_{pq}=-\Lambda^{\alpha}_A({\tilde G}^{\rm H}_{pq,b}K^q_{\alpha} +{\tilde G}^{\rm H}_{pE,p}K^E_{\alpha})$, by the identity $(B)$.
\end{enumerate}
There are some mutual cancellations between terms: (1.2) with (3.1), (1.3) with (6.2), (1.4) with (4.1), (1.5) with (5.2).
As a result, we get
\[
 2N^E_A\,{}^{\mathrm  H}{\tilde \Gamma}_{bEp}+K^q_{\alpha}\Lambda^{\alpha}_A(-{\tilde G}^{\rm H}_{bp,q} +{\tilde G}^{\rm H}_{bq,p}-{\tilde G}^{\rm H}_{pq,b}),
\]
or
\[
 2N^E_A\,{}^{\mathrm  H}{\tilde \Gamma}_{bEp}+2N^q_A\,{}^{\mathrm  H}{\tilde \Gamma}_{bqp}\equiv 2N^{\tilde E}_A\,{}^{\mathrm  H}{\tilde \Gamma}_{b{\tilde E}p}.
\]
The final relation of this subsection is 
\[
 {\check\Gamma}^{\tilde D}_{Ab}{\tilde G}^{\rm H}_{{\tilde D}p}=N^{\tilde E}_A\,{}^{\mathrm  H}{\tilde \Gamma}_{b{\tilde E}p}.
\]
This relation can be combined with what follows from $(A, b, C)$ -- case to get
\[
 {\check\Gamma}^{\tilde D}_{Ab}{\tilde G}^{\rm H}_{{\tilde D}\tilde C}=N^{\tilde E}_A\,{}^{\mathrm  H}{\tilde \Gamma}_{b{\tilde E}\tilde C}.
\]
To determine ${}^{\rm H}{\tilde \Gamma}_{b{\tilde E}}^{\tilde R}$, we use the following relation:
\[
 {\tilde \Gamma}_{b{\tilde E}\tilde C}={\tilde G}^{\rm H}_{\tilde R \tilde C}{}^{\rm H}{\tilde \Gamma}_{b{\tilde E}}^{\tilde R}.
\]
Note that  Christoffel symbols on the right-hand side of this equality are determined modulo terms $T$ satisfying ${\tilde G}^{\rm H}_{\tilde A \tilde M}T^{\tilde M}_{b\tilde C}=0$.
Using the above relation, we get
\[
 {\check\Gamma}^{\tilde D}_{Ab}{\tilde G}^{\rm H}_{{\tilde D}\tilde C}=N^{\tilde E}_A{\tilde G}^{\rm H}_{\tilde R \tilde C}{}^{\rm H}{\tilde \Gamma}_{b{\tilde E}}^{\tilde R}.
\]
Multiplying both sides of this equality by $G^{SF}N^T_SN^{\tilde C}_F$ and taking into account that
\[
 N^{\tilde C}_F{\tilde G}^{\rm H}_{{\tilde D}\tilde C}={\tilde G}^{\rm H}_{{\tilde D}F},\,\,\,\,G^{SF}{\tilde G}^{\rm H}_{{\tilde D}F}=\tilde {\Pi}^S_{\tilde D},\,\,\,\,\tilde {\Pi}^S_{\tilde D}N^T_S=N^T_{\tilde D},
\]
\[
 N^T_{\tilde D}{\check\Gamma}^{\tilde D}_{Ab}\equiv N^T_{ D}{\check\Gamma}^{D}_{Ab}+N^T_{r}{\check\Gamma}^{r}_{Ab}=N^T_{ D}{\check\Gamma}^{D}_{Ab},\;\; \rm{since}\;\; N^T_{r}=0,
\]
we come to
\[
 N^T_D{\check\Gamma}^{ D}_{Ab}=N^T_RN^{\tilde E}_A\,{}^{\rm H}{\tilde \Gamma}_{b{\tilde E}}^{R}.
\]
Hence it follows that
\[
 {\check\Gamma}^{ D}_{Ab}=N^{\tilde E}_A\,{}^{\rm H}{\tilde \Gamma}_{b{\tilde E}}^{D}
\]
(modulo the terms $X^D_{AbC}$ for which $N^T_DX^D_{AbC}=0$).

Similarly, one can obtain representations for other Christoffel symbols that do not include indices related to the group manifold:
\[
 {\check\Gamma}^{D}_{AB}=N^{\tilde E}_A\,{}^{\rm H}{\Gamma}^D_{B\tilde{E}}\;\;\;\;{\check\Gamma}^{D}_{ab}={}^{\rm H}{\Gamma}^D_{ab}
 \]
\[
 {\check\Gamma}^{a}_{AB}=N^{\tilde E}_A\,{}^{\rm H}{\Gamma}^a_{B\tilde{E}}\;\;\;\;{\check\Gamma}^{q}_{ab}={}^{\rm H}{\Gamma}^q_{ab}
 \]
\[
 \;\;{\check\Gamma}^{D}_{Ab}=N^{\tilde E}_A\,{}^{\rm H}{\Gamma}^D_{b\tilde{E}}\;\;\;\;\;\;\;{\check\Gamma}^{D}_{aB}={}^{\rm H}{\Gamma}^D_{aB}
\]
\[
 \;\;{\check\Gamma}^{p}_{Ab}=N^{\tilde E}_A\,{}^{\rm H}{\Gamma}^p_{b\tilde{E}}\;\;\;\;\;\;\;{\check\Gamma}^{D}_{aB}={}^{\rm H}{\Gamma}^D_{aB}
\]

As for the Christoffel symbols with indices related to the group manifold, they first of all arise from the commutator relations  of  the basis vector fields: 

\begin{eqnarray*}
 &&{\check\Gamma}^{\epsilon}_{ab}=-\frac12\tilde {\mathscr F}^{\epsilon}_{ab},\;\;\;\;{\check\Gamma}^{\epsilon}_{A\mu}=\frac12\tilde d^{\epsilon\nu}H_A\tilde d_{\mu\nu},\\
 &&{\check\Gamma}^{\epsilon}_{AB}=\frac12\mathbb{C}^{\epsilon}_{AB}=\frac12[-N^S_AN^R_B\tilde {\mathscr F}^{\epsilon}_{SR}-(N^E_AN^p_B-N^E_BN^p_A)\tilde {\mathscr F}^{\epsilon}_{Ep}-N^q_AN^p_B\tilde {\mathscr F}^{\epsilon}_{qp}],\\
 &&{\check\Gamma}^{\epsilon}_{a\alpha}=\frac12\tilde d^{\epsilon\mu}H_a\tilde d_{\alpha\mu},\;\;\;{\check\Gamma}^{\epsilon}_{\mu b}=\frac12\tilde d^{\epsilon\nu}H_b\tilde d_{\mu\nu},\;\;\;{\check\Gamma}^{\epsilon}_{\mu B}=\frac12\tilde d^{\epsilon\nu}H_B\tilde d_{\mu\nu},\\
&& {\check\Gamma}^{\nu}_{Ab}=\frac12(-N^E_A\tilde {\mathscr F}^{\nu}_{Eb}-N^q_A\tilde {\mathscr F}^{\nu}_{qb}),\;\;\;
{\check\Gamma}^{\nu}_{aB}=-\frac12(-N^E_B\tilde {\mathscr F}^{\nu}_{Ea}-N^q_B\tilde {\mathscr F}^{\nu}_{qa}),\\
&&( \mathbb C^{\alpha}_{Bp}=-\mathbb C^{\alpha}_{pB}).
\end{eqnarray*}

Other such Christoffel symbols are obtained from equations that are special cases of equation (\ref{christoff_main}):
\begin{enumerate}
%1 
\item 
${\check\Gamma}^{D}_{a \nu}{\tilde G}^{\rm H}_{DC}+{\check\Gamma}^{p}_{a \nu}{\tilde G}^{\rm H}_{pC}=-\frac12\mathbb{C}^{\mu}_{aC}\tilde d_{\nu\mu}$
%2
\item
${\check\Gamma}^{D}_{a \mu}{\tilde G}^{\rm H}_{Da}+{\check\Gamma}^{p}_{A\mu}{\tilde G}^{\rm H}_{pA}=-\frac12\mathbb{C}^{\epsilon}_{Aa}\tilde d_{\mu\epsilon}$
%3 
\item 
${\check\Gamma}^{D}_{a \nu}{\tilde G}^{\rm H}_{Db}+{\check\Gamma}^{p}_{a \nu}{\tilde G}^{\rm H}_{pb}=-\frac12\mathbb{C}^{\mu}_{ab}\tilde d_{\nu\mu}$
%4
\item
${\check\Gamma}^{D}_{\mu b}{\tilde G}^{\rm H}_{Da}+{\check\Gamma}^{p}_{\mu b}{\tilde G}^{\rm H}_{pa}=-\frac12\mathbb{C}^{\alpha}_{ba}\tilde d_{\mu\alpha}$
%5
\item
${\check\Gamma}^{D}_{\mu b}{\tilde G}^{\rm H}_{DC}+{\check\Gamma}^{p}_{\mu b}{\tilde G}^{\rm H}_{pC}=-\frac12\mathbb{C}^{\alpha}_{bC}\tilde d_{\mu\alpha}$
%6
\item
${\check\Gamma}^{D}_{\mu B}{\tilde G}^{\rm H}_{DC}+{\check\Gamma}^{p}_{\mu B}{\tilde G}^{\rm H}_{pC}=-\frac12\mathbb{C}^{\alpha}_{BC}\tilde d_{\mu\alpha}$
%7
\item
${\check\Gamma}^{D}_{\mu B}{\tilde G}^{\rm H}_{Db}+{\check\Gamma}^{p}_{\mu B}{\tilde G}^{\rm H}_{pb}=-\frac12\mathbb{C}^{\alpha}_{Bb}\tilde d_{\mu\alpha}$
%8
\item
${\check\Gamma}^{D}_{\mu \nu}{\tilde G}^{\rm H}_{Db}+{\check\Gamma}^{p}_{\mu \nu}{\tilde G}^{\rm H}_{pb}=-\frac12H_b\tilde d_{\mu\nu}$
%9
\item
${\check\Gamma}^{D}_{A \mu}{\tilde G}^{\rm H}_{DC}+{\check\Gamma}^{p}_{A\mu}{\tilde G}^{\rm H}_{pC}=-\frac12\mathbb{C}^{\epsilon}_{AC}\tilde d_{\mu\epsilon}$
%10
\item
${\check\Gamma}^{D}_{\mu \nu}{\tilde G}^{\rm H}_{DC}+{\check\Gamma}^{p}_{\mu \nu}{\tilde G}^{\rm H}_{pC}=-\frac12H_C\tilde d_{\mu\nu}$
\end{enumerate}
Combining the equations of the first and third rows of the above table, we rewrite these equations as
\[
 \left(\begin{matrix}{\tilde G}^H_{CD}&{\tilde G}^H_{Cp}\cr{\tilde G}^H_{bD}&{\tilde G}^H_{bp}\cr\end{matrix}\right)
\left(\begin{matrix}{\check\Gamma}^{D}_{a \nu}\cr {\check\Gamma}^{p}_{a \nu}\cr\end{matrix}\right)=-\frac12\left(\begin{matrix}
\mathbb{C}^{\mu}_{aC}\tilde d_{\nu\mu}\cr\mathbb{C}^{\mu}_{ab}\tilde d_{\nu\mu}\cr\end{matrix}\right),
\]
where
\[
 \mathbb{C}^{\mu}_{aC}=-\mathbb{C}^{\mu}_{Ca}=N^E_C\tilde{\mathscr F}^{\mu}_{Ea}+N^p_C\tilde{\mathscr F}^{\mu}_{pa}, \;\;\;\mathbb{C}^{\mu}_{ab}=-\tilde{\mathscr F}^{\mu}_{ab}.
\]
Multiplying the matrix equation (from the left) by the matrix
\[
 \left(\begin{matrix}{\tilde G}^{EF}N^A_EN^C_F&{\tilde G}^{EF}N^A_EN^b_F\cr {\tilde G}^{EF}N^q_FN^C_E & {\tilde G}^{bq}+{\tilde G}^{FE}N^b_FN^q_E\cr\end{matrix}\right),
\]
it can be shown that the left-hand side of the transformed equation will be equal to
\[
 \left(\begin{matrix} N^A_D&0\cr N^q_D&\delta^q_p\cr \end{matrix}\right)\left(\begin{matrix}{\check\Gamma}^{D}_{a \nu}\cr {\check\Gamma}^{p}_{a \nu}\cr\end{matrix}\right).
\]
On the right-hand side of the matrix equation, we have in the first row
\begin{eqnarray*}
&&-\frac12 \tilde d_{\nu\mu}[{\tilde G}^{EF}N^A_EN^C_F(N^{E'}_C\tilde{\mathscr F}^{\mu}_{E'a}+N^p_C\tilde{\mathscr F}^{\mu}_{pa})+{\tilde G}^{EF}N^A_EN^b_F\tilde{\mathscr F}^{\mu}_{ba}]
\nonumber\\
&&=-\frac12 \tilde d_{\nu\mu}[{\tilde G}^{EF}N^A_EN^T_F\tilde{\mathscr F}^{\mu}_{Ta}+{\tilde G}^{EF}N^A_EN^b_F\tilde{\mathscr F}^{\mu}_{ba}]
\nonumber\\
&&\equiv-\frac12 \tilde d_{\nu\mu}{\tilde G}^{EF}N^A_EN^{\tilde T}_F\tilde{\mathscr F}^{\mu}_{\tilde{T}a},\,\tilde{T}=(T,b).
\end{eqnarray*}
Here we have used the following properties of the projectors: $N^{\tilde A}_{\tilde B} N^{\tilde B}_{\tilde C}=N^{\tilde A}_{\tilde C}$.
In particular, this means that $N^p_CN^C_F=0$.

Thus, the first equation is
\begin{equation}
 N^A_D{\check\Gamma}^{D}_{a \nu}=-\frac12 \tilde d_{\nu\mu}{\tilde G}^{EF}N^A_EN^{\tilde T}_F\tilde{\mathscr F}^{\mu}_{\tilde{T}a}.
\label{eq_for_part_christoff}
\end{equation}
From this it follows that 
\begin{equation}
 {\check\Gamma}^{D}_{a \nu}=-\frac12 {\tilde G}^{DF}N^{\tilde T}_F\tilde{\mathscr F}^{\mu}_{\tilde{T}a}\tilde d_{\nu\mu}
\label{sol_part_christoff_eq}
\end{equation}
(modulo the terms for which $N^A_DX^D_{a\nu}=0$).

Then the second equation is
\[
 {\check\Gamma}^{q}_{a \nu}=-\frac12 {\tilde G}^{qb}\tilde{\mathscr F}^{\mu}_{ba}\tilde d_{\nu\mu}.
\]

In the same way, representations for the following Christoffel symbols can be obtained:
\begin{eqnarray*}
&&{\check\Gamma}^{D}_{\mu b}=-\frac12 {\tilde G}^{DF}N^{\tilde T}_F\tilde{\mathscr F}^{\sigma}_{\tilde{T}b}\tilde d_{\mu\sigma} \;\;\;\;\;\;\;
{\check\Gamma}^{q}_{\mu b}=\frac12 {\tilde G}^{qa}\tilde{\mathscr F}^{\sigma}_{ba}\tilde d_{\mu\sigma}\\
&&{\check\Gamma}^{D}_{A\mu }=\frac12 {\tilde G}^{DF}N^{\tilde R}_FN^{\tilde S}_A\tilde{\mathscr F}^{\sigma}_{\tilde{S}\tilde{R}}\tilde d_{\mu\sigma} \;\;\;
{\check\Gamma}^{q}_{A\mu }=\frac12 {\tilde G}^{qa}N^{\tilde E}_A\tilde{\mathscr F}^{\sigma}_{\tilde{E}a}\tilde d_{\mu\sigma}\\
&&{\check\Gamma}^{q}_{\mu B}=\frac12 {\tilde G}^{qb}N^{\tilde E}_B\tilde{\mathscr F}^{\sigma}_{\tilde{E}b }\tilde d_{\mu\sigma} \;\;\;\;\;\;\;\;\;\;
{\check\Gamma}^{D}_{\mu B}=\frac12 {\tilde G}^{DF}N^{\tilde R}_FN^{\tilde S}_B\tilde{\mathscr F}^{\sigma}_{\tilde{S}\tilde{R} }\tilde d_{\mu\sigma}\\
&&{\check\Gamma}^{D}_{\mu \nu}=-\frac12 {\tilde G}^{DF}N^{\tilde C}_F(H_{\tilde{C}}\,\tilde d_{\mu\nu}) \;\;\;\;\;
{\check\Gamma}^{q}_{\mu \nu}=-\frac12 {\tilde G}^{qb}(H_b\,\tilde d_{\mu\nu}).
\end{eqnarray*}
And also from (\ref{christoff_main}) it follows that
\begin{eqnarray*}
 {\check\Gamma}^{\sigma}_{\mu\nu}&=&\frac12{\tilde d}^{\sigma\gamma}(L_{\mu}{\tilde d}_{\nu\gamma}+L_{\nu}{\tilde d}_{\mu\gamma}-L_{\gamma}{\tilde d}_{\mu\nu}-c^{\varphi}_{\nu\gamma}{\tilde d}_{\mu\varphi}-c^{\varphi}_{\mu\gamma}{\tilde d}_{\nu\varphi}+c^{\varphi}_{\mu\nu}{\tilde d}_{\varphi\gamma})\\
 &=&\frac12{\tilde d}^{\sigma\gamma}(c^{\varphi}_{\mu\nu}{\tilde d}_{\varphi\gamma}-c^{\varphi}_{\gamma\nu}{\tilde d}_{\mu\varphi}-c^{\varphi}_{\gamma\mu}{\tilde d}_{\nu\varphi}).
\end{eqnarray*}

For the Christoffel symbols of the manifold $\tilde{\mathcal P}$, which have ``group" indices, there is another representation. It turns out that using such a representation, the calculation of the scalar curvature of the manifold can be done with less difficulty. Let us explain how these representations are obtained.

In our above calculation of the Christoffel symbol ${\check\Gamma}^{D}_{a \nu}$, we have obtained equation (\ref{eq_for_part_christoff})
 which has the ``solution'' (\ref{sol_part_christoff_eq}). Transforming the right-hand side  of the equation leads to
$$N^A_R{\check\Gamma}^{R}_{a \nu}=-\frac12 \tilde d_{\nu\mu}{\tilde G}^{DF}N^A_RN^R_DN^{\tilde T}_F\tilde{\mathscr F}^{\mu}_{\tilde{T}a}.$$
Then, as before, omitting $N^A_R$, we get the ``solution''
\begin{eqnarray*}
{\check\Gamma}^{R}_{a \nu}&=&-\frac12 \tilde d_{\nu\mu}{\tilde G}^{DF}N^R_DN^{\tilde T}_F\tilde{\mathscr F}^{\mu}_{\tilde{T}a}\\
&=&-\frac12 \tilde d_{\nu\mu}({\tilde G}^{DF}N^R_DN^{T}_F\tilde{\mathscr F}^{\mu}_{{T}a}+{\tilde G}^{DF}N^R_DN^{b}_F\tilde{\mathscr F}^{\mu}_{{b}a})\\
&=&-\frac12 \tilde d_{\nu\mu}(h^{RT}\tilde{\mathscr F}^{\mu}_{{T}a}+h^{Rb}\tilde{\mathscr F}^{\mu}_{{b}a}).
\end{eqnarray*}
The difference between the new ``solution'' and the previous one is equal to
$-\frac12 K^R_{\sigma}(G^{EF}\Lambda ^{\sigma}_EN^{\tilde T}_F\tilde{\mathscr F}^{\mu}_{\tilde{T}a}\tilde d_{\nu\mu})$. So,
$N^A_RK^R_{\sigma}(...)=0$. And two ``solutions'' belong to the same class.

The Christoffel symbols in the new representation are given by 
\begin{eqnarray*}
&&{\check\Gamma}^{D}_{AB}=N^{\tilde E}_A\,{}^{\rm H}{\Gamma}^D_{B\tilde{E}}\;\;\;\;{\check\Gamma}^{D}_{ab}={}^{\rm H}{\Gamma}^D_{ab}
\nonumber\\
&&{\check\Gamma}^{a}_{AB}=N^{\tilde E}_A\,{}^{\rm H}{\Gamma}^a_{B\tilde{E}}\;\;\;\;{\check\Gamma}^{q}_{ab}={}^{\rm H}{\Gamma}^q_{ab}
\nonumber\\
&&{\check\Gamma}^{D}_{Ab}=N^{\tilde E}_A\,{}^{\rm H}{\Gamma}^D_{b\tilde{E}}\;\;\;\;\;\;\;{\check\Gamma}^{D}_{aB}={}^{\rm H}{\Gamma}^D_{aB}
\nonumber\\
&&{\check\Gamma}^{p}_{Ab}=N^{\tilde E}_A\,{}^{\rm H}{\Gamma}^p_{b\tilde{E}}\;\;\;\;\;\;\;{\check\Gamma}^{p}_{aB}={}^{\rm H}{\Gamma}^p_{aB}
\nonumber\\
 &&{\check\Gamma}^{D}_{a\mu }=-\frac12( h^{DT}\tilde{\mathscr F}^{\sigma}_{{T}a}+h^{Db}\tilde{\mathscr F}^{\sigma}_{ba})\tilde d_{\mu\sigma},\;\;\;\;{\check\Gamma}^{D}_{\mu a}={\check\Gamma}^{D}_{a\mu}
 \nonumber\\
 &&{\check\Gamma}^{q}_{a\mu }=-\frac12( h^{qE}\tilde{\mathscr F}^{\sigma}_{Ea}+h^{qb}\tilde{\mathscr F}^{\sigma}_{ba})\tilde d_{\mu\sigma},\;\;\;\;\;\;{\check\Gamma}^{q}_{\mu a}={\check\Gamma}^{q}_{a\mu}
 \nonumber\\
 &&{\check\Gamma}^{D}_{A\mu }=\frac12[N^S_A( h^{DR}\tilde{\mathscr F}^{\sigma}_{SR}+h^{Dp}\tilde{\mathscr F}^{\sigma}_{Sp})+N^q_A(h^{DR}\tilde{\mathscr F}^{\sigma}_{qR}+h^{Dp}\tilde{\mathscr F}^{\sigma}_{qp})]
 \tilde d_{\mu\sigma}
 \nonumber\\
 &&%\;\;\;\;\;\;\;=\frac12( h^{DR}\tilde{\mathscr F}^{\sigma}_{AR}+h^{Dq}\tilde{\mathscr F}^{\sigma}_{Aq})\tilde %d_{\mu\sigma},\;\;\;\;
 {\check\Gamma}^{D}_{\mu A}={\check\Gamma}^{D}_{A\mu}
 \nonumber\\
&&{\check\Gamma}^{q}_{A\mu }=\frac12[N^S_A( h^{qR}\tilde{\mathscr F}^{\sigma}_{SR}+h^{qa}\tilde{\mathscr F}^{\sigma}_{Sa})+N^p_A(h^{qR}\tilde{\mathscr F}^{\sigma}_{pR}+h^{qa}\tilde{\mathscr F}^{\sigma}_{pa})]
 \tilde d_{\mu\sigma}
 \nonumber\\
 &&%\;\;\;\;\;\;\;=\frac12( h^{qR}\tilde{\mathscr F}^{\sigma}_{AR}+h^{qp}\tilde{\mathscr F}^{\sigma}_{Ap})\tilde %d_{\mu\sigma},\;\;\;\;\;
 {\check\Gamma}^{q}_{\mu A}={\check\Gamma}^{q}_{A\mu}
 \nonumber\\
 &&{\check\Gamma}^{\epsilon}_{AB}=\frac12\mathbb{C}^{\epsilon}_{AB}=\frac12N^E_A(N^D_B\tilde{\mathscr F}^{\epsilon}_{DE}+N^p_B\tilde{\mathscr F}^{\epsilon}_{pE})+\frac12N^p_A(N^E_B\tilde{\mathscr F}^{\epsilon}_{Ep}+N^q_B\tilde{\mathscr F}^{\epsilon}_{qp})
 \nonumber\\
 %&&\;\;\;\;\;\;\;\;=\frac12(N^E_A\tilde{\mathscr F}^{\epsilon}_{BE}+N^p_A\tilde{\mathscr F}^{\epsilon}_{Bp})
 %\nonumber\\
 && {\check\Gamma}^{\epsilon}_{Ab}=\frac12\mathbb{C}^{\epsilon}_{Ab}=\frac12(-N^E_A\tilde {\mathscr F}^{\epsilon}_{Eb}-N^q_A\tilde {\mathscr F}^{\epsilon}_{qb}),\;\;\;\;{\check\Gamma}^{\epsilon}_{bA}=-{\check\Gamma}^{\epsilon}_{Ab}
 \nonumber\\
 &&{\check\Gamma}^{\epsilon}_{aB}=-\frac12\mathbb{C}^{\epsilon}_{Ba}=\frac12(N^E_B\tilde {\mathscr F}^{\epsilon}_{Ea}+N^q_B\tilde {\mathscr F}^{\epsilon}_{qa}),\;\;\;\;{\check\Gamma}^{\epsilon}_{Ba}=-{\check\Gamma}^{\epsilon}_{aB}
 \nonumber\\
 &&{\check\Gamma}^{\epsilon}_{ab}=-\frac12\tilde {\mathscr F}^{\epsilon}_{ab},\;\;\;\;{\check\Gamma}^{\epsilon}_{A\mu}=\frac12\tilde d^{\epsilon\nu}H_A\tilde d_{\mu\nu},\;\;\;{\check\Gamma}^{\epsilon}_{a\mu}=\frac12\tilde d^{\epsilon\nu}H_a\tilde d_{\mu\nu},\;\;{\check\Gamma}^{\epsilon}_{\mu a}={\check\Gamma}^{\epsilon}_{a\mu}
 \nonumber\\
 &&{\check\Gamma}^{D}_{\mu\nu}=-\frac12(h^{DC}H_C\,\tilde d_{\mu\nu}+h^{Db}H_b\,\tilde d_{\mu\nu})
 \nonumber\\
 &&{\check\Gamma}^{q}_{\mu\nu}=-\frac12(h^{qC}H_C\,\tilde d_{\mu\nu}+h^{qb}H_b\,\tilde d_{\mu\nu})
 \nonumber\\
 &&{\check\Gamma}^{\alpha}_{\beta\gamma}=\frac12\tilde d^{\alpha\mu}(c^{\epsilon}_{\beta\gamma}\tilde d_{\epsilon\mu}-c^{\epsilon}_{\mu\gamma}\tilde d_{\epsilon\beta}-c^{\epsilon}_{\mu\beta}\tilde d_{\epsilon\gamma})
 \end{eqnarray*}

\subsection{The Ricci curvature tensor}
In the article, the Riemannian tensor $\check{R}_{\mathscr A\mathscr M\mathscr C \mathscr D}$  of the manifold $\tilde{\mathcal P}$ is defined by the Riemannian curvature operator $  \Omega(X,Y)=[\nabla_X,\nabla_Y]-\nabla_{[X,Y]}$ as follows
$$\check R(X,Y,Z,Z')=\check{G}_{\tilde{\mathcal P}}(\Omega(X,Y)Z,Z').$$

The components $\check R_{\mathscr A\mathscr C}=\check R_{\mathscr A\mathscr M\mathscr C}^{\;\;\;\;\;\;\;\;\;\;\mathscr M}$ of the Ricci curvature tensor of the metric (\ref{metric2cc}) will be evaluated in accordance with the following formula:
$$\check R_{\mathscr A\mathscr C}=\hat\partial_{\mathscr A}\check{\rm \Gamma}^{\mathscr P}_{\mathscr P\mathscr C}-\hat\partial_{\mathscr P}\check{\rm \Gamma}^{\mathscr P}_{\mathscr A\mathscr C}+\check{\rm \Gamma}^{\mathscr D}_{\mathscr P\mathscr C}\check{\rm \Gamma}^{\mathscr P}_{\mathscr A\mathscr D}-\check{\rm \Gamma}^{\mathscr E}_{\mathscr A\mathscr C}\check{\rm \Gamma}^{\mathscr P}_{\mathscr P\mathscr E}-\mathbb C^{\mathscr E}_{\mathscr A\mathscr P}\check{\rm \Gamma}^{\mathscr P}_{\mathscr E\mathscr C}.$$
In this formula, each  index given by a capital script letter denotes   three different indices, for example, $\mathscr A=(A,a,\alpha)$. This means that 
$\check R_{\mathscr A\mathscr C}$ is presented by the following components: $(\check R_{ A C}, \check R_{ A b}, \check R_{ b,A}, \check R_{ ab}, \check R_{ \alpha,\beta})$.

First,  consider the calculation   of the components  $\check R_{ A C}$ defined by 
$$\check R_{ A C}=\hat\partial_{A}\check{\rm \Gamma}^{\mathscr P}_{\mathscr PC}-\hat\partial_{\mathscr P}\check{\rm \Gamma}^{\mathscr P}_{AC}+\check{\rm \Gamma}^{\mathscr D}_{\mathscr PC}\check{\rm \Gamma}^{\mathscr P}_{A\mathscr D}-\check{\rm \Gamma}^{\mathscr E}_{AC}\check{\rm \Gamma}^{\mathscr P}_{\mathscr P\mathscr E}-\mathbb C^{\mathscr E}_{A\mathscr P}\check{\rm \Gamma}^{\mathscr P}_{\mathscr EC}.$$

They  are written as follows:
\begin{eqnarray}
 \check R_{ A C}&=&H_{A}\check{\rm \Gamma}^{ E}_{ EC}-H_{P}\check{\rm \Gamma}^{ P}_{ AC}
+H_{A}\check{\rm \Gamma}^{ p}_{ pC}-H_{q}\check{\rm \Gamma}^{q}_{ AC}
+\check{\rm \Gamma}^{ E}_{ DC}\check{\rm \Gamma}^{ D}_{A E}+\check{\rm \Gamma}^{ E}_{qC}\check{\rm \Gamma}^{q}_{ D C}
\nonumber\\
&&
+\check{\rm \Gamma}^{ q}_{ BC}\check{\rm \Gamma}^{ B}_{A q}+\check{\rm \Gamma}^{ q}_{ pC}\check{\rm \Gamma}^{ p}_{A q}
-\check{\rm \Gamma}^{ E}_{ AC}\check{\rm \Gamma}^{ P}_{ PE}-\check{\rm \Gamma}^{ E}_{ AC}\check{\rm \Gamma}^{ q}_{ qE}
-\check{\rm \Gamma}^{ q}_{ AC}\check{\rm \Gamma}^{ R}_{ Rq}-\check{\rm \Gamma}^{ q}_{ AC}\check{\rm \Gamma}^{ p}_{ pq}
\nonumber\\
&&-\mathbb C^{E}_{AR}\check{\rm \Gamma}^{R}_{ EC}-\mathbb C^{q}_{AR}\check{\rm \Gamma}^{R}_{qC}-\mathbb C^{q}_{Ap}\check{\rm \Gamma}^{p}_{qC}
\nonumber\\
&&+ H_{A}\check{\rm \Gamma}^{\alpha}_{ \alpha C}-L_{\alpha}\check{\rm \Gamma}^{\alpha}_{AC}+\check{\rm \Gamma}^{ E}_{ \mu C}\check{\rm \Gamma}^{\mu}_{ AE}+\check{\rm \Gamma}^{ q}_{\alpha C}\check{\rm \Gamma}^{\alpha}_{A q}
+\check{\rm \Gamma}^{\mu}_{ BC}\check{\rm \Gamma}^{ B}_{A \mu}+\check{\rm \Gamma}^{ \mu}_{ pC}\check{\rm \Gamma}^{ p}_{A \mu}
\nonumber\\
&&+\check{\rm \Gamma}^{ \mu}_{\nu C}\check{\rm \Gamma}^{\nu}_{A \mu}-\check{\rm \Gamma}^{ E}_{ AC}\check{\rm \Gamma}^{ \mu}_{\mu E}
-\check{\rm \Gamma}^{ q}_{ AC}\check{\rm \Gamma}^{ \mu}_{\mu q}
-\check{\rm \Gamma}^{ \alpha}_{ AC}\check{\rm \Gamma}^{ R}_{ R\alpha}
-\check{\rm \Gamma}^{ \alpha}_{ AC}\check{\rm \Gamma}^{ \mu}_{\mu \alpha}
\nonumber\\
&&-\mathbb C^{\mu}_{AR}\check{\rm \Gamma}^{R}_{\mu C}-\mathbb C^{\mu}_{Aq}\check{\rm \Gamma}^{q}_{\mu C}.
\label{R_AC}
\end{eqnarray}
Using the obtained Christoffel symbols $\check{\rm \Gamma}$ and the structure constants
$\mathbb C^{E}_{AR}$, $\mathbb C^{q}_{AR}$, $\mathbb C^{q}_{Ap}$, it  can be shown that the expression standing at the  first three lines of the right-hand side of (\ref{R_AC}) coinsides  with  
\[
 N^{\tilde S}_AN^{\tilde E}_{ M}\,{}^{\scriptscriptstyle \rm H}\!R_{\tilde S \tilde E C}{}^{ M},
\]
where 
\[
 {}^{\scriptscriptstyle \rm H}\!R_{\tilde S \tilde E C}{}^{M}=\partial_{\tilde S}{}^{\scriptscriptstyle \rm H}{\rm \Gamma}^{M}_{C\tilde E}-\partial_{\tilde E}{}^{\scriptscriptstyle \rm H}{ \rm \Gamma}^{M}_{C\tilde S}+{}^{\scriptscriptstyle \rm H}{ \rm \Gamma}^{\tilde K}_{C\tilde E}{}^{\scriptscriptstyle \rm H}{ \rm \Gamma}^{M}_{\tilde K\tilde S}-{}^{\scriptscriptstyle \rm H}{ \rm \Gamma}^{\tilde P}_{C\tilde S}{}^{\scriptscriptstyle \rm H}{ \rm \Gamma}^{M}_{\tilde P\tilde E}.
\]
Also note that the expression for ${}^{\scriptscriptstyle \rm H}\!R_{\tilde S \tilde E C}{}^{M}$ is similar to the expression used for the Riemannian curvature tensor of the Riemannian manifold. 
%Note, however, that in our case the metric of the manifold is degenerate. 

Therefore,  $\check R_{ A C}$ is represented as
 $\check R_{AC}=N^{\tilde S}_AN^{\tilde E}_{ M}\,{}^{\scriptscriptstyle \rm H}\!R_{\tilde S\tilde E C}{}^{ M}+\check R'_{AC}$. 

It can be verified that the same is true for other components  of the Ricci tensor $\check R_{\mathscr A\mathscr C}$.  For   $\check R_{ A b}$  defined as 
$$\check R_{ A b}=\hat\partial_{A}\check{\rm \Gamma}^{\mathscr P}_{\mathscr Pb}-\hat\partial_{\mathscr P}\check{\rm \Gamma}^{\mathscr P}_{Ab}+\check{\rm \Gamma}^{\mathscr D}_{\mathscr Pb}\check{\rm \Gamma}^{\mathscr P}_{A\mathscr D}-\check{\rm \Gamma}^{\mathscr E}_{Ab}\check{\rm \Gamma}^{\mathscr P}_{\mathscr P\mathscr E}-\mathbb{C}^{\mathscr E}_{A\mathscr P}\check{\rm \Gamma}^{\mathscr P}_{\mathscr Eb}$$
and given by the following expression
\begin{eqnarray}
 \check R_{ A b}&=&H_{A}\check{\rm \Gamma}^{ B}_{ Bb}-H_{B}\check{\rm \Gamma}^{ B}_{ Ab}
+H_{A}\check{\rm \Gamma}^{ p}_{ pb}-H_{p}\check{\rm \Gamma}^{p}_{ Ab}
+\check{\rm \Gamma}^{ B}_{ Cb}\check{\rm \Gamma}^{ C}_{A B}+\check{\rm \Gamma}^{ B}_{pb}\check{\rm \Gamma}^{p}_{AB}
\nonumber\\
&&
+\check{\rm \Gamma}^{ q}_{ Cb}\check{\rm \Gamma}^{ C}_{A q}+\check{\rm \Gamma}^{ q}_{ pb}\check{\rm \Gamma}^{ p}_{A q}
-\check{\rm \Gamma}^{ E}_{ Ab}\check{\rm \Gamma}^{ B}_{ BE}-\check{\rm \Gamma}^{ E}_{ AC}\check{\rm \Gamma}^{ p}_{ pE}
-\check{\rm \Gamma}^{ q}_{ Ab}\check{\rm \Gamma}^{ B}_{ Bq}-\check{\rm \Gamma}^{ q}_{ Ab}\check{\rm \Gamma}^{ p}_{ pq}
\nonumber\\
&&-\mathbb C^{E}_{AB}\check{\rm \Gamma}^{B}_{ Eb}-\mathbb C^{q}_{AB}\check{\rm \Gamma}^{B}_{qB}-\mathbb C^{q}_{Ap}\check{\rm \Gamma}^{p}_{qb}
\nonumber\\
&&+ H_{A}\check{\rm \Gamma}^{\alpha}_{ \alpha b}-L_{\alpha}\check{\rm \Gamma}^{\alpha}_{Ab}+\check{\rm \Gamma}^{ B}_{ \alpha b}\check{\rm \Gamma}^{\alpha}_{ AB}+\check{\rm \Gamma}^{ q}_{\alpha b}\check{\rm \Gamma}^{\alpha}_{A q}
+\check{\rm \Gamma}^{\alpha}_{Cb}\check{\rm \Gamma}^{ C}_{A \alpha}+\check{\rm \Gamma}^{ \alpha}_{ pb}\check{\rm \Gamma}^{ p}_{A \alpha}
\nonumber\\
&&+\check{\rm \Gamma}^{ \alpha}_{\mu b}\check{\rm \Gamma}^{\mu}_{A \alpha}-\check{\rm \Gamma}^{ E}_{ Ab}\check{\rm \Gamma}^{ \alpha}_{\alpha E}
-\check{\rm \Gamma}^{ q}_{ Ab}\check{\rm \Gamma}^{ \alpha}_{\alpha q}
-\check{\rm \Gamma}^{ \alpha}_{ Ab}\check{\rm \Gamma}^{ B}_{ B\alpha}-\check{\rm \Gamma}^{ \alpha}_{ Ab}\check{\rm \Gamma}^{ p}_{ p\alpha}
-\check{\rm \Gamma}^{ \alpha}_{ Ab}\check{\rm \Gamma}^{ \mu}_{\mu \alpha}
\nonumber\\
&&-\mathbb C^{\alpha}_{AB}\check{\rm \Gamma}^{B}_{\alpha b}-\mathbb C^{\alpha}_{Ap}\check{\rm \Gamma}^{p}_{\alpha b},
\label{R_Ab}
\end{eqnarray}
  the  first three lines on the right-hand side of (\ref{R_Ab}) are  the same as in\\  
$N^{\tilde S}_AN^{\tilde E}_{ M}\,{}^{\rm H}\!R_{\tilde S \tilde E b}{}^{ M}$.
 Therefore, we can represent $\check R_{ A b}$ as
\[
 \check R_{Ab}=N^{\tilde S}_AN^{\tilde E}_{ M}\,{}^{\scriptscriptstyle \rm H}\!R_{\tilde S\tilde E b}{}^{ M}+\check R'_{Ab}. 
\]

The components $\check R_{ bA}$ are defined as
\begin{eqnarray}
 \check R_{ bA}&=&H_{b}\check{\rm \Gamma}^{ B}_{ BA}-H_{B}\check{\rm \Gamma}^{ B}_{ bA}
+H_{b}\check{\rm \Gamma}^{ p}_{ pA}-H_{p}\check{\rm \Gamma}^{p}_{b A}
+\check{\rm \Gamma}^{ B}_{ CA}\check{\rm \Gamma}^{ C}_{b B}+\check{\rm \Gamma}^{ B}_{pA}\check{\rm \Gamma}^{p}_{bB}
\nonumber\\
&&
+\check{\rm \Gamma}^{ q}_{ CA}\check{\rm \Gamma}^{ C}_{b q}+\check{\rm \Gamma}^{ q}_{ pA}\check{\rm \Gamma}^{ p}_{b q}
-\check{\rm \Gamma}^{ E}_{ bA}\check{\rm \Gamma}^{ B}_{ BE}-\check{\rm \Gamma}^{ E}_{ bA}\check{\rm \Gamma}^{ p}_{ pE}
-\check{\rm \Gamma}^{ q}_{ bA}\check{\rm \Gamma}^{ B}_{ Bq}-\check{\rm \Gamma}^{ q}_{ bA}\check{\rm \Gamma}^{ p}_{ pq}
\nonumber\\
&&-\mathbb C^{q}_{bB}\check{\rm \Gamma}^{B}_{qA}
\nonumber\\
&&+ H_{b}\check{\rm \Gamma}^{\alpha}_{ \alpha A}-L_{\alpha}\check{\rm \Gamma}^{\alpha}_{bA}+\check{\rm \Gamma}^{ B}_{ \alpha A}\check{\rm \Gamma}^{\alpha}_{ bB}+\check{\rm \Gamma}^{ q}_{\alpha A}\check{\rm \Gamma}^{\alpha}_{b q}
+\check{\rm \Gamma}^{\alpha}_{CA}\check{\rm \Gamma}^{ C}_{b \alpha}+\check{\rm \Gamma}^{ \alpha}_{ pA}\check{\rm \Gamma}^{ p}_{b \alpha}
\nonumber\\
&&+\check{\rm \Gamma}^{ \alpha}_{\mu A}\check{\rm \Gamma}^{\mu}_{b \alpha}-\check{\rm \Gamma}^{ E}_{ bA}\check{\rm \Gamma}^{ \alpha}_{\alpha E}
-\check{\rm \Gamma}^{ q}_{ bA}\check{\rm \Gamma}^{ \alpha}_{\alpha q}
-\check{\rm \Gamma}^{ \alpha}_{ bA}\check{\rm \Gamma}^{ B}_{ B\alpha}-\check{\rm \Gamma}^{ \alpha}_{ bA}\check{\rm \Gamma}^{ p}_{ p\alpha}
-\check{\rm \Gamma}^{ \alpha}_{ bA}\check{\rm \Gamma}^{ \mu}_{\mu \alpha}
\nonumber\\
&&-\mathbb C^{\alpha}_{bB}\check{\rm \Gamma}^{B}_{\alpha A}-\mathbb C^{\alpha}_{bp}\check{\rm \Gamma}^{p}_{\alpha A},
\label{R_bA}
\end{eqnarray}
and we have
\[
 \check R_{bA}=N^{\tilde E}_{ M}\,{}^{\scriptscriptstyle \rm H}\!R_{b\tilde E A}{}^{ M}+\check R'_{bA}. 
\]

The components $\check R_{ ab}$ are defined as
\begin{eqnarray}
 \check R_{ ab}&=&H_{a}\check{\rm \Gamma}^{ B}_{ Bb}-H_{B}\check{\rm \Gamma}^{ B}_{ ab}
+H_{a}\check{\rm \Gamma}^{ p}_{ pb}-H_{q}\check{\rm \Gamma}^{q}_{ab }
+\check{\rm \Gamma}^{ B}_{ Db}\check{\rm \Gamma}^{ D}_{aB }+\check{\rm \Gamma}^{ B}_{qb}\check{\rm \Gamma}^{q}_{aB}
\nonumber\\
&&
+\check{\rm \Gamma}^{ p}_{ Bb}\check{\rm \Gamma}^{B}_{ap}+\check{\rm \Gamma}^{ p}_{ qb}\check{\rm \Gamma}^{ q}_{ap}
-\check{\rm \Gamma}^{ R}_{ ab}\check{\rm \Gamma}^{ B}_{ BR}-\check{\rm \Gamma}^{ R}_{ ab}\check{\rm \Gamma}^{ q}_{ qR}
-\check{\rm \Gamma}^{ p}_{ ab}\check{\rm \Gamma}^{ R}_{ Rp}-\check{\rm \Gamma}^{ p}_{ ab}\check{\rm \Gamma}^{q}_{ qp}
\nonumber\\
&&-\mathbb C^{p}_{aD}\check{\rm \Gamma}^{D}_{pb}
\nonumber\\
&&+ H_{a}\check{\rm \Gamma}^{\alpha}_{ \alpha b}-L_{\alpha}\check{\rm \Gamma}^{\alpha}_{ab}+\check{\rm \Gamma}^{ B}_{ \alpha b}\check{\rm \Gamma}^{\alpha}_{ aB}+\check{\rm \Gamma}^{ p}_{\alpha b}\check{\rm \Gamma}^{\alpha}_{ap}
+\check{\rm \Gamma}^{\alpha}_{Bb}\check{\rm \Gamma}^{ B}_{a \alpha}+\check{\rm \Gamma}^{ \alpha}_{ pb}\check{\rm \Gamma}^{ p}_{a \alpha}
\nonumber\\
&&+\check{\rm \Gamma}^{ \alpha}_{ \mu b}\check{\rm \Gamma}^{\mu}_{a \alpha}-\check{\rm \Gamma}^{ R}_{ ab}\check{\rm \Gamma}^{ \alpha}_{\alpha R}-\check{\rm \Gamma}^{ p}_{ ab}\check{\rm \Gamma}^{ \alpha}_{\alpha p}-\check{\rm \Gamma}^{ \alpha}_{ ab}\check{\rm \Gamma}^{ R}_{R \alpha}-\check{\rm \Gamma}^{ \alpha}_{ ab}\check{\rm \Gamma}^{ p}_{p \alpha}
%&&h^{ab}\check{\rm \Gamma}^{\alpha}_{B b}\check{\rm \Gamma}^{B}_{a\alpha}=\frac14h^{ab}(h^{ET}\tilde{\mathscr F}^{\alpha}_{Eb}\tilde{\mathscr F}^{\sigma}_{Ta}%%%+h^{Ep}\tilde{\mathscr F}^{\alpha}_{Eb}\tilde{\mathscr F}^{\sigma}_{pa})\tilde d_{\alpha\sigma}
 -\check{\rm \Gamma}^{ \alpha}_{ ab}\check{\rm \Gamma}^{ \mu}_{\mu \alpha}
\nonumber\\
&&-\mathbb C^{\alpha}_{aB}\check{\rm \Gamma}^{B}_{\alpha b}-\mathbb C^{\alpha}_{ap}\check{\rm \Gamma}^{p}_{\alpha b},
\label{R_ab}
\end{eqnarray}
and in this case
\[
 \check R_{ab}=N^{\tilde E}_{ M}\,{}^{\scriptscriptstyle \rm H}\!R_{a\tilde E b}{}^{ M}+\check R'_{ab}. 
\]

The components  $\check R_{\alpha \beta}$ of the Ricci curvature tensor  are  defined by the following formula
$$\check R_{\alpha \beta}=L_{\alpha}\check{\rm \Gamma}^{\mathscr P}_{\mathscr P\beta}-\hat\partial_{\mathscr P}\check{\rm \Gamma}^{\mathscr P}_{\alpha\beta}+\check{\rm \Gamma}^{\mathscr D}_{\mathscr P\beta}\check{\rm \Gamma}^{\mathscr P}_{\alpha\mathscr D}-\check{\rm \Gamma}^{\mathscr E}_{\alpha\beta}\check{\rm \Gamma}^{\mathscr P}_{\mathscr P\mathscr E}-\mathbb C^{\mathscr E}_{\alpha\mathscr P}\check{\rm \Gamma}^{\mathscr P}_{\mathscr E\beta},$$
which means that
\begin{eqnarray}
 \check R_{ \alpha\beta}&=&L_{\alpha}\check{\rm \Gamma}^{\mu}_{\mu\beta}-L_{\mu}\check{\rm \Gamma}^{ \mu}_{\alpha\beta}+\check{\rm \Gamma}^{ \mu}_{\nu\beta}\check{\rm \Gamma}^{\nu}_{\alpha\mu}-\check{\rm \Gamma}^{ \mu}_{\alpha\beta}\check{\rm \Gamma}^{ \nu}_{\nu\mu}-c^{\mu}_{\alpha\nu}\check{\rm \Gamma}^{ \nu}_{\mu\beta}
 \nonumber\\
&&+L_{\alpha}\check{\rm \Gamma}^{B}_{B\beta}-H_{B}\check{\rm \Gamma}^{ B}_{ \alpha\beta}+L_{\alpha}\check{\rm \Gamma}^{q}_{q\beta}-H_{q}\check{\rm \Gamma}^{ q}_{ \alpha\beta}+
+\check{\rm \Gamma}^{ B}_{ Db}\check{\rm \Gamma}^{ D}_{\alpha\beta }+\check{\rm \Gamma}^{ B}_{qb}\check{\rm \Gamma}^{q}_{\alpha B}
\nonumber\\
&&+\check{\rm \Gamma}^{ B}_{\mu\beta}\check{\rm \Gamma}^{\mu}_{\alpha B}+\check{\rm \Gamma}^{ q}_{R\beta }\check{\rm \Gamma}^{R}_{\alpha q}+\check{\rm \Gamma}^{ q}_{p\beta }\check{\rm \Gamma}^{p}_{\alpha q}+\check{\rm \Gamma}^{ q}_{\mu\beta }\check{\rm \Gamma}^{\mu}_{\alpha q}+\check{\rm \Gamma}^{\mu}_{B\beta }\check{\rm \Gamma}^{B}_{\alpha \mu}+\check{\rm \Gamma}^{\mu}_{q\beta }\check{\rm \Gamma}^{q}_{\alpha \mu}
\nonumber\\
&&-\check{\rm \Gamma}^{ E}_{ \alpha\beta}(\check{\rm \Gamma}^{ R}_{ RE}+\check{\rm \Gamma}^{ q}_{ qE}+\check{\rm \Gamma}^{ \mu}_{ \mu E})
-\check{\rm \Gamma}^{ q}_{ \alpha\beta}(\check{\rm \Gamma}^{ R}_{ Rq}+\check{\rm \Gamma}^{ p}_{ pq}+\check{\rm \Gamma}^{ \mu}_{ \mu q})
\nonumber\\
&&-\check{\rm \Gamma}^{\mu}_{ \alpha\beta}(\check{\rm \Gamma}^{ R}_{ R\mu}+\check{\rm \Gamma}^{ p}_{ p\mu}).
\label{R_alphbet}
\end{eqnarray}
From this it follows that
$\check R_{\alpha \beta}$ can also be represented as
$$\check R_{\alpha \beta}=R_{\alpha \beta}+\check R'_{\alpha \beta},$$
where
\[
 R_{\alpha \beta}=L_{\alpha}\check{\rm \Gamma}^{\mu}_{\mu\beta}-L_{\mu}\check{\rm \Gamma}^{\mu}_{\alpha\beta}+\check{\rm \Gamma}^{\mu}_{\nu\beta}\check{\rm \Gamma}^{\nu}_{\alpha\mu}-\check{\rm \Gamma}^{\mu}_{\alpha\beta}\check{\rm \Gamma}^{\nu}_{\nu\mu}-c^{\mu}_{\alpha\nu}\check{\rm \Gamma}^{\nu}_{\mu\beta}
\]
are  the components of the Ricci curvature tensor  for a Lie group $\mathcal G$.
The scalar curvature of the orbit 
\[
 R_{\mathcal G}\equiv\tilde{d}^{\alpha\beta}R_{\alpha \beta}=\frac12{d}^{\mu\nu}c^{\sigma}_{\mu\alpha}c^{\alpha}_{\nu\sigma}+\frac14d_{\mu\sigma}d^{\alpha\beta}d^{\epsilon\nu}c^{\mu}_{\epsilon\alpha}c^{\sigma}_{\nu\beta}.
\]
\subsection{The calculation of the scalar curvature}
In the horizontal lift basis, the scalar curvature of the original manifold $\tilde{\mathcal P} $ is defined  by the following formula:
\[
\check R_{\tilde{\mathcal P}}=h^{AC}\check R_{AC}+h^{Ab}\check R_{Ab}+h^{bA}\check R_{bA}+h^{ab}\check R_{ab}+\tilde{d}^{\alpha\beta}\check R_{\alpha\beta}.
\]
Since the scalar curvature of the manifold ${\tilde{\Sigma}}$ with the metric tensor ${\tilde G}^{\rm H}_{\tilde{ A} \tilde{ B}} $ is 
\begin{equation}
{}^{\scriptscriptstyle \rm H}\!R_{\tilde{\Sigma}}=h^{SC}N^{\tilde E}_{ M}\,{}^{\scriptscriptstyle \rm H}\!R_{ S\tilde E C}{}^{ M}+h^{Sb}N^{\tilde E}_{ M}\,{}^{\scriptscriptstyle \rm H}\!R_{\tilde S\tilde E b}{}^{ M}+h^{bA}N^{\tilde E}_{ M}\,{}^{\scriptscriptstyle \rm H}\!R_{b\tilde E A}{}^{ M}+h^{ab}N^{\tilde E}_{ M}\,{}^{\scriptscriptstyle \rm H}\!R_{a\tilde E b}{}^{ M}
\label{scal_curv_sigma}
\end{equation}
and $\tilde{d}^{\alpha\beta}R_{\alpha \beta}=R_{\mathcal G}$, it remains to determine
\[
 h^{AC}\check R'_{AC}+h^{Ab}\check R'_{Ab}+h^{bA}\check R'_{bA}+h^{ab}\check R'_{ab}+
 \tilde{d}^{\alpha\beta}\check R'_{\alpha \beta}.
\]
An explicit representation of the terms of this sum is given in Appendix B.

In order to find the expression for the scalar curvature in the form we need, the terms related to the tensors $h^{AC}\check R'_{ A C},h^{Ab}\check R'_{ A b},h^{bA}\check R'_{b A}$ and $h^{ab}\check R'_{ab}$ should be slightly reexpressed and combined in a certain way.
Consider the terms of these
tensors, which are in the first lines of Appendix B.

 In  $h^{AC}\check R'_{AC}$, the first line is represented by the following equality
\begin{eqnarray*}
&h^{AC}H_{A}\check{\rm \Gamma}^{\alpha}_{ \alpha C}&=\frac12h^{AC}H_A(\tilde d^{\alpha\nu}H_C\tilde d_{\alpha\nu})
=\frac12h^{AC}N^{\tilde E}_A\partial_{\tilde E}( d^{\alpha\nu}N^{\tilde F}_C\partial_{\tilde F} d_{\alpha\nu}).
\end{eqnarray*}
The right side of this equality (without the factor $h^{AC}$) can be rewritten as  
\begin{eqnarray*}
&&\frac12[N^E_AN^B_C\partial_E(d^{\alpha\nu}\partial_Bd_{\alpha\nu})+N^E_AN^p_C\partial_E(d^{\alpha\nu}\partial_pd_{\alpha\nu})
\nonumber\\
 &&+N^q_AN^B_C\partial_q(d^{\alpha\nu}\partial_Bd_{\alpha\nu})+N^q_AN^p_C\partial_q(d^{\alpha\nu}\partial_pd_{\alpha\nu})
 \nonumber\\
 &&+N^E_AN^B_{C,E}(d^{\alpha\nu}\partial_Bd_{\alpha\nu})+     N^E_AN^p_{C,E}(d^{\alpha\nu}\partial_pd_{\alpha\nu})+N^q_AN^p_{C,q}(d^{\alpha\nu}\partial_pd_{\alpha\nu})].
\end{eqnarray*}
Multiplying the previous expression by $h^{AC}$, we get
\[
 \frac12[h^{EB}\partial_E(d^{\alpha\nu}\partial_Bd_{\alpha\nu})+h^{EC}N^B_{C,E}(d^{\alpha\nu}\partial_Bd_{\alpha\nu})+h^{EC}N^p_{C,E}(d^{\alpha\nu}\partial_pd_{\alpha\nu})].
\]
The resulting expression is  the contribution to $\check R_{\tilde{\mathcal P}}$ from the terms given in the first line of $h^{AC}\check R'_{AC}$.

In $h^{Ab}\check R'_{Ab}$, in the first line, we have 
$$h^{Ab}H_A \check{\rm \Gamma}^{\alpha}_{ \alpha b}=\frac12h^{Ab}H_A(\tilde d^{\alpha\nu} H_b\tilde d_{\nu \alpha})=\frac12h^{Eb}\partial_E(d^{\alpha\nu} \partial_b\, d_{\nu \alpha}).$$

The next contribution comes from $h^{bA}\check R'_{bA}$, where we have
\begin{align*}
h^{bA}H_b \check{\rm \Gamma}^{\alpha}_{ \alpha A}&=\frac12h^{bA}H_b(\tilde d^{\alpha\nu} H_A\tilde d_{\nu \alpha})\nonumber\\
&=\frac12[h^{bE}\partial_b(d^{\alpha\nu} \partial_E d_{\nu \alpha})+h^{bA}\partial_b(N^p_A)(d^{\alpha\nu} \partial_p d_{\nu \alpha})]\nonumber\\
&=\frac12h^{bE}\partial_b(d^{\alpha\nu} \partial_E d_{\nu \alpha}).
\end{align*}
The last equality, because $h^{bA}N^p_{A,b}=-G^{LM}N^b_LN^A_{M,B}N^p_A=0.$

And from the first line of $h^{ab}\check R'_{ab}$  we get 
\[
h^{ab}H_a \check{\rm \Gamma}^{\alpha}_{ \alpha b}=\frac12h^{ab}H_a(\tilde d^{\alpha\nu} H_b\tilde d_{\nu \alpha})=\frac12h^{ab}\partial_a( d^{\alpha\nu} \partial_b d_{\nu \alpha}).
\]

Consider now also what contribution to $\check R_{\tilde{\mathcal P}}$ will be made by similar terms presented in ${\tilde d}^{\alpha\beta}\check R'_{\alpha\beta}$. They are given by terms derived from the second and third lines of ${\tilde d}^{\alpha\beta}\check R'_{\alpha\beta}$: 
\begin{eqnarray*}
&&-\tilde d^{\alpha\beta}H_B\check{\rm \Gamma}^{B}_{\alpha\beta}=-\tilde d^{\alpha\beta}H_B(-\frac12h^{BC}H_C\tilde d_{\alpha\beta}-\frac12h^{Bb}H_b\tilde d_{\alpha\beta}),
 \\
 &&-\tilde d^{\alpha\beta}H_q\check{\rm \Gamma}^{q}_{\alpha\beta}=-\tilde d^{\alpha\beta}H_q(-\frac12h^{qC}H_C\tilde d_{\alpha\beta}-\frac12h^{qb}H_b\tilde d_{\alpha\beta}).
 \end{eqnarray*}

The right side of the first equality without the factor ${\tilde d}^{\alpha\beta}$ (the terms from the second line in ${\tilde d}^{\alpha\beta}\check R'_{\alpha\beta}$) can be rewritten as follow:
\begin{align*}
 &\frac12H_B(h^{BC}H_C\tilde d_{\alpha\beta}+h^{Bb}H_b\tilde d_{\alpha\beta})=\frac12H_B(h^{BC})(H_C\tilde d_{\alpha\beta})+\frac12H_B(h^{Bb})(H_b\tilde d_{\alpha\beta})
 \nonumber\\
 &+\frac12h^{BC}(H_BH_C\tilde d_{\alpha\beta})+\frac12h^{Bb}(H_BH_b\tilde d_{\alpha\beta})
\nonumber\\
 &=\frac12H_B(h^{BC})(N^E_C\tilde {\mathscr D}_E\tilde d_{\alpha\beta}+N^p_C\tilde {\mathscr D}_p\tilde d_{\alpha\beta})+\frac12H_B(h^{Bb})(\tilde {\mathscr D}_b\tilde d_{\alpha\beta})
 \nonumber\\
 &+\frac12h^{BC}(N^R_B\tilde {\mathscr D}_R(H_C\tilde d_{\alpha\beta})+N^q_B\tilde {\mathscr D}_q(H_C\tilde d_{\alpha\beta}))\\
 &+\frac12h^{Bb}(N^R_B\tilde {\mathscr D}_R(\tilde {\mathscr D}_b\tilde d_{\alpha\beta})+N^q_B\tilde {\mathscr D}_q(\tilde {\mathscr D}_b\tilde d_{\alpha\beta})).
 \end{align*}
In these formulas by $\tilde {\mathscr D}_E\tilde d_{\alpha\beta}$ we denote the covariant derivative defined as
\[
 \tilde {\mathscr D}_E\tilde d_{\alpha\beta}=\partial_E\tilde d_{\alpha\beta}-c^{\sigma}_{\mu\alpha}\tilde{\mathscr A}^{\mu}_E\tilde d_{\sigma\beta}-c^{\sigma}_{\mu\beta}\tilde{\mathscr A}^{\mu}_E\tilde d_{\sigma\alpha},
\]
$\tilde{\mathscr A}^{\mu}_E=\bar{\rho}^{\mu}_{\alpha}{\mathscr A}^{\alpha}_E(Q^{\ast},\tilde f)$. Also, $\tilde {\mathscr D}_E\tilde d_{\alpha\beta}={\rho}^{\mu}_{\alpha}{\rho}^{\nu}_{\beta}{\mathscr D}_Ed_{\mu\nu}$.
  A similar definition is used for the covariant derivative $\tilde {\mathscr D}_p\tilde d_{\alpha\beta}$.
 
 Since $h^{BC}N^q_B=0$ and $h^{Bb}N^q_B=0$, and also $h^{BC}N^R_B=h^{RC}$, the obtained  expression can be rewritten as
\begin{align*}
 &\frac12H_B(h^{BC})(N^E_C\tilde {\mathscr D}_E\tilde d_{\alpha\beta}+N^p_C\tilde {\mathscr D}_p\tilde d_{\alpha\beta})+\frac12H_B(h^{Bb})(\tilde {\mathscr D}_b\tilde d_{\alpha\beta})\\
 &+\frac12h^{RC}N^E_{C,R}\tilde {\mathscr D}_E\tilde d_{\alpha\beta}+\frac12h^{RC}N^p_{C,R}\tilde {\mathscr D}_p\tilde d_{\alpha\beta}\\
 &+\frac12h^{RE}\tilde {\mathscr D}_R\tilde {\mathscr D}_E\tilde d_{\alpha\beta}+\frac12h^{Rb}\tilde {\mathscr D}_R\tilde {\mathscr D}_b\tilde d_{\alpha\beta}.
\end{align*}

The terms from the third line in ${\tilde d}^{\alpha\beta}\check R'_{\alpha\beta}$ (also without the first factor ${\tilde d}^{\alpha\beta}$) can be rewritten as
\begin{align*}
 \frac12(H_q(h^{qE})\tilde {\mathscr D}_E\tilde d_{\alpha\beta}+H_q(h^{qb})\tilde {\mathscr D}_b\tilde d_{\alpha\beta}+h^{qE}\tilde {\mathscr D}_q\tilde {\mathscr D}_E\tilde d_{\alpha\beta}+h^{qb}\tilde {\mathscr D}_q\tilde {\mathscr D}_b\tilde d_{\alpha\beta}).
\end{align*}
To obtain such an expression, we used the following properties: $H_q(h^{qC})N^E_C=H_q(h^{qE})$, since $N^E_C$ depends only on $Q^{\ast}$, and $H_q(h^{qC})N^p_C=0$.

Note also that
$$\tilde d^{\alpha\beta}\tilde {\mathscr D}_E\tilde d_{\alpha\beta}=d^{\alpha\beta}\partial_Ed_{\alpha\beta}=\partial_E(\ln \det(d_{\alpha\beta}))\equiv\sigma_E$$
for a semisimple Lie groups, and a similar equality holds for $\tilde d^{\alpha\beta}\tilde {\mathscr D}_p\tilde d_{\alpha\beta}$.

Hence after taking into account the factor $\tilde d^{\alpha\beta}$,   we will use a new representation for the following expressions:
$$\tilde d^{\alpha\beta}(N^E_C\tilde {\mathscr D}_E\tilde d_{\alpha\beta}+N^p_C\tilde {\mathscr D}_p\tilde d_{\alpha\beta})=N^E_C\sigma_E+N^p_C\sigma_p=\sigma_C.$$
The last equality due to  the identity derived in Appendix C.

Further transformations of the terms of the obtained expressions will be carried out using the following identity:
\[
 \tilde d^{\alpha\beta}(\tilde {\mathscr D}_E\tilde {\mathscr D}_C\tilde d_{\alpha\beta})
 =\partial_E(d^{\alpha\beta}\partial_Cd_{\alpha\beta})+d^{\alpha'\epsilon'}d^{\nu'\beta'}({\mathscr D}_Ed_{\epsilon'\nu'})({\mathscr D}_Cd_{\alpha'\beta'}).
\]
As a result, we get that $-\tilde d^{\alpha\beta}H_B\check{\rm \Gamma}^{B}_{\alpha\beta}$ is equal to 
\begin{align*}
 &\frac12h^{RE}(\partial _R\partial_E\sigma+d^{\alpha'\epsilon'}d^{\nu'\beta'}({\mathscr D}_Rd_{\epsilon'\nu'})({\mathscr D}_Ed_{\alpha'\beta'}))\\
 &+\frac12h^{Rb}(\partial _R\partial_b\sigma+d^{\alpha'\epsilon'}d^{\nu'\beta'}({\mathscr D}_Rd_{\epsilon'\nu'})({\mathscr D}_bd_{\alpha'\beta'}))\\
 &+\frac12h^{RC}(N^E_{C,R}\sigma_E+N^p_{C,R}\sigma_p)
 +\frac12(H_B(h^{BC})\sigma_C+H_B(h^{Bb})\sigma_b).
 \end{align*}

And for  $-\tilde d^{\alpha\beta}H_q\check{\rm \Gamma}^{q}_{\alpha\beta}$ we get the following expression:
\begin{align*}
 &\frac12h^{qE}(\partial _q\partial_E\sigma+d^{\alpha'\epsilon'}d^{\nu'\beta'}({\mathscr D}_qd_{\epsilon'\nu'})({\mathscr D}_Ed_{\alpha'\beta'}))\\
 &+\frac12h^{qb}(\partial _q\partial_b\sigma+d^{\alpha'\epsilon'}d^{\nu'\beta'}({\mathscr D}_qd_{\epsilon'\nu'})({\mathscr D}_bd_{\alpha'\beta'}))\\
 & +\frac12(\partial_q(h^{qE})\sigma_E+\partial_q(h^{qb})\sigma_b).
 \end{align*}

As a further transformation, we need formulas representing partial derivatives of $h^{\tilde D\tilde E}$.  They can be derived from the general formula by which
\[
 \partial_{\tilde A}h^{\tilde D\tilde E}=-h^{\tilde B\tilde D}\,{}^{\rm H}{\rm \Gamma}^{\tilde E}_{\tilde B\tilde A}-h^{\tilde C\tilde E}\,{}^{\rm H}{\rm \Gamma}^{\tilde D}_{\tilde C\tilde A}.
\]
For particular indices, this formula gives the following representations:
\[
 \partial_{E}h^{EC}=-h^{EB}\,{}^{\rm H}{\rm \Gamma}^{C}_{BE}-h^{Eb}\,{}^{\rm H}{\rm \Gamma}^{C}_{bE}-h^{BC}\,{}^{\rm H}{\rm \Gamma}^{E}_{BE}-h^{bC}\,{}^{\rm H}{\rm \Gamma}^{E}_{bE},
\]
\begin{align*}
 \partial_{E}h^{Eb}&=-h^{EA}\,{}^{\rm H}{\rm \Gamma}^{b}_{AE}-h^{Ea}\,{}^{\rm H}{\rm \Gamma}^{b}_{aE}-h^{Ab}\,{}^{\rm H}{\rm \Gamma}^{E}_{AE}-h^{ab}\,{}^{\rm H}{\rm \Gamma}^{E}_{aE},\\
 \partial_{q}h^{qC}&=-h^{qA}\,{}^{\rm H}{\rm \Gamma}^{C}_{Aq}-h^{qa}\,{}^{\rm H}{\rm \Gamma}^{C}_{aq}-h^{AC}\,{}^{\rm H}{\rm \Gamma}^{q}_{Aq}-h^{aC}\,{}^{\rm H}{\rm \Gamma}^{q}_{aq},\\
 \partial_{q}h^{qp}&=-h^{qA}\,{}^{\rm H}{\rm \Gamma}^{p}_{Aq}-h^{qa}\,{}^{\rm H}{\rm \Gamma}^{p}_{aq}-h^{Ap}\,{}^{\rm H}{\rm \Gamma}^{q}_{Aq}-h^{ap}\,{}^{\rm H}{\rm \Gamma}^{q}_{aq}.
\end{align*}
 The terms of $-\tilde d^{\alpha\beta}H_B\check{\rm \Gamma}^{B}_{\alpha\beta}$  with partial derivatives of   $h^{\tilde A\tilde B}$, 
 \[
  \frac12[N^E_B\partial_E(h^{BM})\sigma_M+N^p_B\partial_p(h^{BM})\sigma_M+N^E_B\partial_E(h^{Bb})\sigma_b+N^p_B\partial_p(h^{Bb})\sigma_b],
 \]
 must be re-expressed  using the following substitutions:
%\begin{eqnarray}
 %1.\;&&\!\!\!\!\!
 %N^E_B\partial_E(h^{BM})=-h^{ED}\,{}^{\rm H}\Gamma^M_{DE}-h^{Ep}\,{}^{\rm H}\Gamma^M_{pE}-h^{TM}N^E_B\,{}^{\rm H}\Gamma^B_{TE}\\
 %&&\;\;\;\;\;\;\;\;\;\;\;\;\;\;\;\;\;-h^{pM}N^E_B\,{}^{\rm H}\Gamma^B_{pE}\;\;(\times \sigma_M)
 %\\2.\;
 %&&\!\!\!\!\! N^p_B\partial_p(h^{BM})=-h^{TM}N^p_B\,{}^{\rm H}\Gamma^B_{Tp}-h^{qM}N^p_B\,{}^{\rm H}\Gamma^B_{qp}\;\;(\times \sigma_M)
%\\3.\;
%&&\!\!\!\!\!N^E_B\partial_E(h^{Bb})=-h^{ED}\,{}^{\rm H}\Gamma^b_{DE}-h^{Ep}\,{}^{\rm H}\Gamma^b_{pE}-h^{Tb}N^E_B\,{}^{\rm H}%\Gamma^B_{TE}\\
%&&\;\;\;\;\;\;\;\;\;\;\;\;\;\;\;\;\;-h^{pb}N^E_B\,{}^{\rm H}\Gamma^B_{pE}\,\,(\times \sigma_b)
%\\4.\;
%&&\!\!\!\!\!N^p_B\partial_p(h^{Bb})=-h^{Tb}N^p_B\,{}^{\rm H}\Gamma^B_{Tp}-h^{qb}N^p_B\,{}^{\rm H}\Gamma^B_{qp}\,\,(\times \sigma_b)
 % \label{hh^{BM}}
 % \end{eqnarray}
 \begin{align}
N^E_B\partial_E(h^{BM})&=-h^{ED}\,{}^{\rm H}\Gamma^M_{DE}-h^{Ep}\,{}^{\rm H}\Gamma^M_{pE}-h^{TM}N^E_B\,{}^{\rm H}\Gamma^B_{TE}
 -h^{pM}N^E_B\,{}^{\rm H}\Gamma^B_{pE},
 %(\times \sigma_M)
 \nonumber\\
 N^p_B\partial_p(h^{BM})&=-h^{TM}N^p_B\,{}^{\rm H}\Gamma^B_{Tp}-h^{qM}N^p_B\,{}^{\rm H}\Gamma^B_{qp},
 %\;\;(\times \sigma_M)
 \nonumber\\
N^E_B\partial_E(h^{Bb})&=-h^{ED}\,{}^{\rm H}\Gamma^b_{DE}-h^{Ep}\,{}^{\rm H}\Gamma^b_{pE}-h^{Tb}N^E_B\,{}^{\rm H}\Gamma^B_{TE}
-h^{pb}N^E_B\,{}^{\rm H}\Gamma^B_{pE},
%\,\,(\times \sigma_b)
\nonumber\\
N^p_B\partial_p(h^{Bb})&=-h^{Tb}N^p_B\,{}^{\rm H}\Gamma^B_{Tp}-h^{qb}N^p_B\,{}^{\rm H}\Gamma^B_{qp}.
%\,\,(\times \sigma_b)
  \label{dif_h^{BM}}
  \end{align}

 Those  terms of
$-\tilde d^{\alpha\beta}H_q\check{\rm \Gamma}^{q}_{\alpha\beta}$ 
that include  partial derivatives of $h^{\tilde A\tilde B}$ are equal to
\[
  \frac12 (\partial_q h^{qE})\sigma_E+\frac12(\partial_q h^{qb})\sigma_b,
 \]
where
 \begin{align}
 %1.\;&&
 %\!\!\!\!\!
 \partial_q(h^{qE})&=-h^{qD}\,{}^{\rm H}\Gamma^E_{Dq}-h^{qp}\,{}^{\rm H}\Gamma^E_{pq}-h^{TE}\,{}^{\rm H}\Gamma^q_{Tq}-h^{pE}\,{}^{\rm H}\Gamma^q_{pq},\nonumber\\
%\\2.\;
 %&&\!\!\!\!\!
 \partial_q(h^{qb})&=-h^{qD}\,{}^{\rm H}\Gamma^b_{Dq}-h^{qp}\,{}^{\rm H}\Gamma^b_{pq}-h^{Tb}\,{}^{\rm H}\Gamma^q_{Tq}-h^{pb}\,{}^{\rm H}\Gamma^q_{pq}.
\label{dif_h^{qb}} 
 \end{align}
 
Thus, we see that the expressions we have considered contain terms of certain types. In these expressions, there are terms with second partial derivatives of $\sigma$, terms with Christoffel symbols and ``$\mathscr D\mathscr D$''-terms.  

The summation of terms with second partial derivatives of $\sigma$ coming from all Ricci tensors gives the initial part of the Laplace-Beltrami operator on the manifold $\tilde {\Sigma}$, which is applied to $\sigma$.
To obtain the second part of this Laplacian, consisting of terms with Christoffel coefficients, it is necessary to take into account the contributions to $\check R_{\tilde{\mathcal P}}$ from other terms of the Ricci tensors. In particularly, in $h^{AC}\check R'_{ A C}$ we have  two such terms:
\begin{align*}
 -h^{AC}\check \Gamma^E_{AC}\check \Gamma^{\mu}_{\mu E}&=-\frac12h^{AC}N^{\tilde R}_A\,{}^{\rm H}\Gamma ^E_{C\tilde R}(N^D_E\sigma_D+N^p_E\sigma_p)=-\frac12h^{AC}N^{\tilde R}_A\,{}^{\rm H}\Gamma ^E_{C\tilde R}\sigma_E\\
 &=-\frac12h^{AC}N^{ R}_A\,{}^{\rm H}\Gamma ^E_{CR}\sigma_E\;\;\;\;\;\rm{and}\\
%\end{align*}
%\begin{align*}
 -h^{AC}\check \Gamma^q_{AC}\check \Gamma^{\mu}_{\mu q}&=-\frac12h^{AC}N^{\tilde E}_A\,{}^{\rm H}\Gamma ^q_{C\tilde E}\sigma_q
 =-\frac12h^{AC}N^{ E}_A\,{}^{\rm H}\Gamma ^q_{CE}\sigma_q.
\end{align*}
The additional contributions from $h^{Ab}\check R'_{ A b}$ consist of 
\begin{align*}
 -h^{Ab}\check \Gamma^E_{Ab}\check \Gamma^{\alpha}_{\alpha E}=-\frac12h^{Db}\,{}^{\rm H}\Gamma ^E_{bD}\sigma_E\;\;\;\;\rm {and}\;\;\;
 -h^{Ab}\check \Gamma^q_{Ab}\check \Gamma^{\alpha}_{\alpha q}=-\frac12h^{Eb}\,{}^{\rm H}\Gamma ^E_{bE}\sigma_q.
\end{align*}
The same contributions will be from $h^{bA}\check R'_{b A }$, that is,  from $-h^{bA}\check \Gamma^E_{bA}\check \Gamma^{\alpha}_{\alpha E}$ and $-h^{bA}\check \Gamma^q_{bA}\check \Gamma^{\alpha}_{\alpha q}$. 

The additional contributions from $h^{ab}\check R'_{a b}$ are represented by the following terms: 
$$-h^{ab}\check \Gamma^R_{ab}\check \Gamma^{\alpha}_{\alpha R}=-\frac12h^{ab}\,{}^{\rm H}\Gamma ^R_{ab}\sigma_R,\;\;\;\;-h^{ab}\check \Gamma^p_{ab}\check \Gamma^{\alpha}_{\alpha p}=-\frac12h^{ab}\,{}^{\rm H}\Gamma ^p_{ab}\sigma_p.$$

As for the Christoffel terms of $d^{\alpha\beta}\check R'_{\alpha\beta}$ obtained after replacements (\ref{dif_h^{BM}}) and (\ref{dif_h^{qb}}), note that all ``undesirable'' terms such as  $h^{TM}N^E_B\,{}^{\rm H}\Gamma^B_{TE}\sigma_M$  are cancelled with  the similar  terms derived from $-\tilde d^{\alpha\beta}\check \Gamma^E_{\alpha\beta}\check \Gamma^{R}_{RE}$, $ -\tilde d^{\alpha\beta}\check \Gamma^E_{\alpha\beta}\check \Gamma^{q}_{qE}$, $-\tilde d^{\alpha\beta}\check \Gamma^q_{\alpha\beta}\check \Gamma^{R}_{Rq}$ and $-\tilde d^{\alpha\beta}\check \Gamma^q_{\alpha\beta}\check \Gamma^{p}_{pq}$.

Therefore, by collecting all the terms with the second partial derivatives of sigma, as well as all the aforementioned terms with the Christoffel coefficients at the first partial derivatives of sigma, we get  $\triangle_{\tilde{\Sigma}}^{\scriptscriptstyle\rm H}\sigma$, where $\triangle_{\tilde{\Sigma}}^{\scriptscriptstyle\rm H}$ is the Laplace-Beltrami  operator  for the manifold with the ``horizontal metric'' defined by the metric tensor ${\tilde G}^{\scriptscriptstyle\rm H}_{\tilde A \tilde B}(Q^{\ast}{}^A,\tilde f ^p)$ in (\ref{metric2cc}).
%, and $\sigma =\ln \det d_{\alpha\beta}$.

Together with $\triangle_{\tilde{\Sigma}}^{\scriptscriptstyle\rm H}\sigma$, we get a term with derivatives of the projection operators
\[
 h^{RC}(N^E_{C,R}\sigma_E+N^p_{C,R}\sigma_p).
\]
But from the identity obtained  in Appendix C, it follows that this term is equal to zero.

Next consider the ``$\mathscr D\mathscr D$''-terms. They are present in  all components of the Ricci tensor $\check R^{'}_{\mathscr A\mathscr C}$. But ``$\mathscr D\mathscr D$''-terms in $\tilde d^{\alpha\beta}\check R'_{\alpha\beta} $ are mutually cancelled.
The contribution of all other terms to $\check R_{\tilde{\mathcal P}}$ is given by the following expression: 
$$||j||^2=\frac14h^{A'B'}d^{\alpha\epsilon}d^{\nu\beta}({\mathscr D}_{A'}d_{\epsilon\nu})({\mathscr D}_{B'}d_{\alpha\beta}),$$ which, as can be shown, is the ``square'' of the second fundamental form of the orbit.
Note that the primed capital letters used here as indices  mean, in fact, two indices. For example, $A'=(A,a)$.

In addition to the above contributions, there are two more types of terms that we must account for as contributions to the scalar curvature $\check R_{\tilde{\mathcal P}}$. One of them represents a quadratic form in  the first partial derivatives of $\sigma=\ln d$:
\[
 \frac14(h^{AB}\sigma_A\sigma_B+h^{Ab}\sigma_A\sigma_b+h^{bA}\sigma_b\sigma_A+h^{ab}\sigma_a\sigma_b)\equiv\frac14
 <\partial \sigma,\partial \sigma>_{\tilde{\Sigma}}.
\]
This contribution  comes only from  $\tilde d^{\alpha\beta}\check R'_{\alpha\beta} $. 

The last type of terms whose contribution to the scalar curvature
 must be taken into account are ``$\mathscr F\mathscr F $''- terms. 

As a result of our calculation, we get the following representation for the scalar curvature $\check R_{\tilde{\mathcal P}}$
\begin{eqnarray}
{\check R}_{\tilde{\mathcal P}}={}^{\scriptscriptstyle\rm H}\!{ R}_{\tilde{\Sigma}}+{ R}_{\mathcal G}
+\frac14\mathscr F^2
%+\frac14  h^{A'B'} h^{C'D'}d_{\mu\nu}{\mathscr F}^{\mu}_{A'C'}{\mathscr F}^{\nu}_{B'D'}
+||j||^2
%\nonumber\\
+\triangle_{\scriptscriptstyle \tilde {\Sigma}}^{\scriptscriptstyle\rm H}\ln d+\frac14<\partial \sigma,\partial \sigma>_{\tilde{\Sigma}}\,, 
\label{curvat_itog}
\end{eqnarray}
where ${}^{\scriptscriptstyle\rm H}\!{R}_{\tilde{\Sigma}}$ is from (\ref{scal_curv_sigma}), and  we have used the evident  symbolical notation for
$$\mathscr F^2\equiv h^{A'B'} h^{C'D'}d_{\mu\nu}{\mathscr F}^{\mu}_{A'C'}{\mathscr F}^{\nu}_{B'D'}\,.$$

\section{The geometrical representation of $\tilde {J}$}
Having obtained the expression (\ref{curvat_itog}) for the scalar curvature ${\check R}_{\tilde{\mathcal P}}$, we can  compare  its terms with the terms of $\tilde {J}$ in the integrand of the reduction Jacobian (\ref{jacobian_reduct}).  Since both expressions have similar terms,
this allows us to rewrite the formula (\ref{curvat_itog}) as follows
 \[
  {\check R}_{\tilde{\mathcal P}}={}^{\scriptscriptstyle\rm H}\!{ R}_{\tilde{\Sigma}}+{ R}_{\mathcal G}
+\frac14\mathscr F^2
%+\frac14  h^{A'B'} h^{C'D'}d_{\mu\nu}{\mathscr F}^{\mu}_{A'C'}{\mathscr F}^{\nu}_{B'D'}
+||j||^2+\tilde{ J}.
 \]
In turn, this means that the Hamilton operator of the Schrödinger equation on the manifold $\tilde{\Sigma}$ can be represented in the following form:
\[
 \hat H_{\tilde {\Sigma}}=-\frac{\hbar^2}{2m}\tilde{\triangle}_{\tilde \Sigma}+\frac{\hbar ^2}{8m}\Bigl[{\check R}_{\tilde{\mathcal P}}-{}^{\scriptscriptstyle\rm H}\!{ R}_{\tilde{\Sigma}}-{ R}_{\mathcal G}
-\frac14\mathscr F^2
%+\frac14  h^{A'B'} h^{C'D'}d_{\mu\nu}{\mathscr F}^{\mu}_{A'C'}{\mathscr F}^{\nu}_{B'D'}
-||j||^2\Bigr]+\tilde{ V}.
\]
In conclusion, it is worth noting that ${\check R}_{\tilde{\mathcal P}}={ R}_{{\mathcal P}}$, since in our case we  are dealing with the product manifold ${\tilde{\mathcal P}}= {\mathcal P}\times \mathcal V$,  and the scalar curvature $R_{\mathcal V}=0$.

\section*{Acknowledgment}
 The author is grateful to V.O. Soloviev for helpful discussions.

\section*{Appendix A}
\section*{Projectors, their properties,\\ some identities, Killing relations}
\setcounter{equation}{0}
\def\theequation{A.\arabic{equation}}
\subsection*{The horizontal projector ${\tilde \Pi}^{\tilde A}_{\tilde B}$}

By definition
$${\tilde \Pi}^{\tilde A}_{\tilde B}={\delta}^{\tilde A}_{\tilde B}-K^{\tilde A}_{\alpha}d^{\alpha \beta}K^{\tilde D}_{\beta}G_{\tilde D \tilde B}.$$
Its components are
$${\tilde \Pi}^{\tilde A}_{\tilde B}\equiv({\tilde \Pi}^A_B,{\tilde \Pi}^A_b,{\tilde \Pi}^a_B,{\tilde \Pi}^a_b),$$

$${{\tilde \Pi}}^{A}_{B}={\delta}^A_B-K^{A}_{\alpha}d^{\alpha \beta}K^{D}_{\beta}G_{D B},\;\;\;\;\;
{{\tilde \Pi}}^{A}_{b}=-K^{A}_{\mu}d^{\mu \nu}K^{p}_{\nu}G_{pb},$$

$${{\tilde \Pi}}^{a}_{B}=-K^{a}_{\mu}d^{\mu \nu}K^{D}_{\nu}G_{D B},\;\;\;\;
{{\tilde \Pi}}^{a}_{b}={\delta}^a_b-K^{a}_{\mu}d^{\mu \nu}K^{r}_{\nu}G_{rb}.$$

The main properties:
\[
 {\tilde \Pi}^{\tilde A}_{\tilde B}{\tilde \Pi}^{\tilde B}_{\tilde C}={\tilde \Pi}^{\tilde A}_{\tilde C}, \;\;\;
{{\tilde \Pi}}^{\tilde L}_{\tilde B}N^{\tilde A}_{\tilde L}=N^{\tilde A}_{\tilde B},\;\;\;
{{\tilde \Pi}}_{\tilde L}^{\tilde A}N_{\tilde C}^{\tilde L}={{\tilde \Pi}}_{\tilde C}^{\tilde A},\;\;\;
{{\tilde \Pi}}^{\tilde A}_{\tilde E}K^{\tilde E}_{\alpha}=0.
\]

\subsection*{The projector $N^{\tilde A}_{\tilde B}$}

%\[
% N^{\tilde A}_{\tilde B}=\delta ^{\tilde A}_{\tilde B}-K^{\tilde %A}_{\alpha}%({\Phi}^{-1})^{\mu}_{\nu}\,{\chi}^{\nu}_{\tilde B}, \;\;\;({\chi}={\chi}(Q^{\ast}))
%\]
This projector has the following components: 
$$N^{\tilde A}_{\tilde C}\equiv(N^A_B,N^A_b,N^a_B,N^a_b),$$ 

$$N^A_B={\delta}^A_B-K^A_{\mu}({\Phi}^{-1})^{\mu}_{\nu}\,{\chi}^{\nu}_B,\;\;
N^A_b=0,\;\;  N^a_B=-K^a_{\alpha}({\Phi}^{-1})^{\alpha}_{\mu}\,{\chi}^{\mu}_B,
%=-K^m_{\alpha}{\Lambda}^{\alpha}_B,
\;\;  N^a_b={\delta}^a_b.$$
The main properties of the projector are
$$N^{\tilde A}_{\tilde B} N^{\tilde B}_{\tilde C}=N^{\tilde A}_{\tilde C},\;\;\;
(P_{\bot})^{\tilde L}_{\tilde B}\,N^{\tilde C}_{\tilde L}=(P_{\bot})^{\tilde C}_{\tilde B},\;\;\;      
N^{\tilde A}_{\tilde B}\,(P_{\bot})^{\tilde C}_{\tilde A}=N^{\tilde C}_{\tilde B}.
$$

\subsection*{The projector $(P_{\bot})^{\tilde A}_{\tilde B}$}

The projector is defined as
\[
 (P_{\bot})^{\tilde A}_{\tilde B}\equiv(\,(P_{\bot})^A_B,\, (P_{\bot})^A_b,\, (P_{\bot})^a_B,\,(P_{\bot})^a_b\,),
\]
\[
 (P_\bot)^A_B=\delta^A_B-\chi ^{\alpha}_{B}\,(\chi \chi ^{\top})^{-1}{}^{\beta}_{\alpha}\,(\chi ^{\top})^A_{\beta},\;(\chi ^{\top})^A_{\mu}=G^{AB}{\gamma}_{\mu\nu}\chi ^{\nu}_B,\;{\gamma}_{\mu\nu}=K^A_{\mu}G_{AB}K^B_{\nu}, 
\]
 $(P_{\bot})^{A}_{b}=0,\;\;(P_{\bot})^{a}_{B}=0,\;\;(P_{\bot})^{a}_{b}=\delta ^a_b.$

\subsection*{Some identities derived from $K^{\tilde R}_{\gamma}{\tilde G}^H_{{\tilde R\tilde A} }=0$}

(1) $\tilde A\to A$
\[
 K^R_{\gamma}{\tilde G}^H_{RA}+K^p_{\gamma}{\tilde G}^H_{pA}=0\;\;\;\;\;{\rm or}\;\;\;K^{\tilde R}_{\gamma}{\tilde G}^H_{{\tilde R}A}=0
\]
%The result of differentiation w.r.t $Q^{D}$:
\[
(A)\;\;\;\;K^R_{\gamma, D}{\tilde G}^H_{RA}+K^R_{\gamma}{\tilde G}^H_{RA,D}+K^p_{\gamma}{\tilde G}^H_{pA,D}=0
\]
%The result of differentiation w.r.t  $f^q$:
\[
\!\!\!\!(D)\;\;\;\;{\tilde G}^H_{AR,q}K^R_{\gamma}+{\tilde G}^H_{Ap,q}K^p_{\gamma}+{\tilde G}^H_{Ap}K^p_{\gamma,q}=0 
\]

$\!\!\!\!\!\!\!(2)$ $\tilde A\to p$
\[
 {\tilde G}^H_{pq}K^q_{\mu}+{\tilde G}^H_{pA}K^A_{\mu}=0\;\;\;\;\;{\rm or}\;\;\;{\tilde G}^H_{p{\tilde R}}K^{\tilde R}_{\mu}=0
\]
%The result of differentiation w.r.t  $f^n$:
\[
\!\!\!\!(B)\;\;\;\;{\tilde G}^H_{pR,n}K^R_{\mu}+{\tilde G}^H_{pr,n}K^r_{\mu}+{\tilde G}^H_{pr}K^r_{\mu,n}=0
\]
%The result of differentiation w.r.t $Q^{D}$:
\[
\!\!\!(C)\;\;\;\;{\tilde G}^H_{pR,D}K^R_{\mu}+{\tilde G}^H_{pr,D}K^r_{\mu}+{\tilde G}^H_{pR}K^R_{\mu,n}=0 
\]
These relations are obtained as a result of the differentiations.

\subsection*{Killing relations for the horizontal metric $G^H_{\tilde A \tilde B}$}

\[
 {\tilde G}^H_{\tilde A \tilde B,\tilde D}K^{\tilde D}_{\alpha}+{\tilde G}^H_{\tilde R \tilde B}K^{\tilde R}_{\alpha,\tilde A}+{\tilde G}^H_{\tilde A \tilde R}K^{\tilde R}_{\alpha ,\tilde B}=0
\]

\begin{flushleft}
         $\;\;\;\;\;\;{\tilde A}\to A,\;\;\;{\tilde B} \to B$
\end{flushleft}

\begin{flushleft}
         $\rm {{\Roman 1}}\,.$ $\;\;\;\;{\tilde G}^H_{A B,D}K^{D}_{\alpha}+{\tilde G}^H_{A B,p}K^{p}_{\alpha}+{\tilde G}^H_{R B}K^{R}_{\alpha,A}+{\tilde G}^H_{AR }K^{R}_{\alpha,B}=0$
\end{flushleft}

\begin{flushleft}
         $\;\;\;\;\;\;{\tilde A}\to p,\;\;\;{\tilde B} \to q$
\end{flushleft}

 \begin{flushleft}
         $\rm {{\Roman 2}}$\,. $\;\;\;\; {\tilde G}^H_{pq,D}K^{D}_{\alpha}+{\tilde G}^H_{pq,r}K^{r}_{\alpha}+{\tilde G}^H_{rq}K^{r}_{\alpha,p}+{\tilde G}^H_{pr }K^{R}_{\alpha,q}=0$
\end{flushleft}

\begin{flushleft}
         $\;\;\;\;\;\;{\tilde A}\to p,\;\;\;{\tilde B} \to B$
\end{flushleft}

 \begin{flushleft}
         $\rm {{\Roman 3}}$\,. $\;\;\;\;{\tilde G}^H_{pB,D}K^{D}_{\alpha}+{\tilde G}^H_{pB,r}K^{r}_{\alpha}+{\tilde G}^H_{rB}K^{r}_{\alpha,p}+
{\tilde G}^H_{pR}K^{R}_{\alpha,B}=0$
\end{flushleft}

\begin{flushleft}
         $\;\;\;\;\;\;{\tilde A}\to B,\;\;\;{\tilde B} \to p$
\end{flushleft}

\begin{flushleft}
         $\rm {{\Roman 4}}$\,. $\;\;\;\;{\tilde G}^H_{Bp,D}K^{D}_{\alpha}+{\tilde G}^H_{Bp,r}K^{r}_{\alpha}+{\tilde G}^H_{Rp}K^{R}_{\alpha,B}+{\tilde G}^H_{Br}K^{r}_{\alpha,p}=0$
\end{flushleft}

$\rm{\Roman 4}=\rm{\Roman 3}$

\appendix
\section*{Appendix B}
\section*{$h^{\tilde A \tilde B}\check R'_{\tilde A \tilde B}$-terms of the scalar curvature $\check R_{\tilde{\mathcal P}}$}
%$h^{AC}\check R'_{AC},h^{Ab}\check R'_{Ab}, h^{bA}\check R'_{bA}, h^{ab}\check R'_{ab}, 
 %\tilde{d}^{\alpha\beta}\check R'_{\alpha \beta}$}
\setcounter{equation}{0}
\def\theequation{A.\arabic{equation}}
\subsection*{\underline{$h^{AC}\check R'_{AC}$}}
To get $h^{AC}\check R'_{AC}$, the following terms presented in this subsection must be multiplied by $h^{AC}$:
\begin{eqnarray*}
 %&&\hat\partial_{\mathscr A}
 &&H_{A}\check{\rm \Gamma}^{\alpha}_{ \alpha C}=\frac12H_A(\tilde d^{\alpha\nu}H_C\tilde d_{\alpha\nu})=\frac12N^{\tilde E}_A\partial_{\tilde E}( d^{\alpha\nu}N^{\tilde F}_C\partial_{\tilde F} d_{\alpha\nu})
 \\
 &&L_{\alpha}\check{\rm \Gamma}^{\alpha}_{AC}=0\,\,(\rm{for \,\, the\,\,semisimple \, Lie\, groups,\,\, since}\; L_{\alpha}\bar \rho^{\alpha}_{\mu}=c_{\sigma\alpha}^{\alpha}\bar{\rho}^{\sigma}_{\mu}),
 %\frac12[-N^S_AN^R_CL_{\alpha}\tilde {\mathscr F}^{\alpha}_{SR}-(N^E_AN^p_C-N^E_CN^p_A)L_{\alpha}\tilde %{\mathscr F}^{\alpha}_{Ep}-N^q_AN^p_CL_{\alpha}\tilde {\mathscr F}^{\alpha}_{qp}]=L_{\alpha}\check{\rm \Gamma}^{\alpha}_{AC}
 \\
 &&\check{\rm \Gamma}^{ E}_{ \mu C}\check{\rm \Gamma}^{\mu}_{ AE}=\frac12[N^S_C( h^{ER}\tilde{\mathscr F}^{\sigma}_{SR}+h^{Ep}\tilde{\mathscr F}^{\sigma}_{Sp})+N^q_C(h^{ER}\tilde{\mathscr F}^{\sigma}_{qR}+h^{Ep}\tilde{\mathscr F}^{\sigma}_{qp})]
 \tilde d_{\mu\sigma}
 \\
 &&\;\;\;\;\times
  \frac12[-N^{S'}_AN^{R'}_E\tilde {\mathscr F}^{\mu}_{S'R'}-(N^{R'}_AN^{p'}_E-N^{R'}_EN^{p'}_A)\tilde {\mathscr F}^{\mu}_{R'p'}-N^{q'}_AN^{p'}_E\tilde {\mathscr F}^{\mu}_{q'p'}]
\\
 &&\check{\rm \Gamma}^{ q}_{\mu C}\check{\rm \Gamma}^{\mu}_{A q}=\frac12[N^S_C( h^{qR}\tilde{\mathscr F}^{\sigma}_{SR}+h^{qa}\tilde{\mathscr F}^{\sigma}_{Sa})+N^p_C(h^{qR}\tilde{\mathscr F}^{\sigma}_{pR}+h^{qa}\tilde{\mathscr F}^{\sigma}_{pa})]
 \tilde d_{\mu\sigma}
 \\
 &&\;\;\;\;\;\;\;\;\;\;\;\;\;\times
 \frac12(-N^{R`}_A\tilde {\mathscr F}^{\mu}_{R'q}-N^{p'}_A\tilde {\mathscr F}^{\mu}_{p'q})
\\
 &&\check{\rm \Gamma}^{\mu}_{ BC}\check{\rm \Gamma}^{ B}_{A \mu}=\frac12N^E_B(N^D_C\tilde{\mathscr F}^{\mu}_{DE}+N^p_C\tilde{\mathscr F}^{\mu}_{pE})+\frac12N^p_B(N^E_C\tilde{\mathscr F}^{\mu}_{Ep}+N^q_C\tilde{\mathscr F}^{\mu}_{qp})
 \\
 &&\;\;\;\;\;\;\times\frac12[N^S_A( h^{BR'}\tilde{\mathscr F}^{\sigma}_{SR'}+h^{Bp'}\tilde{\mathscr F}^{\sigma}_{Sp'})+N^{q'}_A(h^{BR'}\tilde{\mathscr F}^{\sigma}_{q'R'}+h^{Bp'}\tilde{\mathscr F}^{\sigma}_{q'p'})]
 \tilde d_{\mu\sigma}
 \\
&&\check{\rm \Gamma}^{ \mu}_{ pC}\check{\rm \Gamma}^{ p}_{A \mu}=-\frac12(-N^R_C\tilde {\mathscr F}^{\mu}_{Rp}-N^q_C\tilde {\mathscr F}^{\mu}_{qp})
\\
&&\;\;\;\;\;\;\;\;\;\;\;\;\times\frac12[N^S_A( h^{pR'}\tilde{\mathscr F}^{\sigma}_{SR'}+h^{pa}\tilde{\mathscr F}^{\sigma}_{Sa})+N^b_A(h^{pR'}\tilde{\mathscr F}^{\sigma}_{bR'}+h^{pa}\tilde{\mathscr F}^{\sigma}_{ba})]
 \tilde d_{\mu\sigma}
\\
&&\check{\rm \Gamma}^{ \mu}_{\nu C}\check{\rm \Gamma}^{\nu}_{A \mu}=\frac12(\tilde d^{\mu\sigma}H_C\tilde d_{\nu\sigma})\,\times\frac12\tilde (d^{\nu\epsilon}H_A\tilde d_{\mu\epsilon})
\\
&&\check{\rm \Gamma}^{ E}_{ AC}\check{\rm \Gamma}^{ \mu}_{\mu E}=N^{\tilde R}_A\,{}^{\rm H}{\Gamma}^E_{C\tilde{R}}\,\times\frac12(\tilde d^{\mu\nu}H_E\tilde d_{\mu\nu})
\\
&&\check{\rm \Gamma}^{ q}_{ AC}\check{\rm \Gamma}^{ \mu}_{\mu q}=N^{\tilde E}_A\,{}^{\rm H}{\Gamma}^q_{C\tilde{E}}\times\frac12(\tilde d^{\mu\nu}H_q\tilde d_{\mu\nu})
\\
&&\check{\rm \Gamma}^{ \epsilon}_{ AC}\check{\rm \Gamma}^{ D}_{ D\epsilon}=\frac12N^E_A(N^{D'}_C\tilde{\mathscr F}^{\epsilon}_{D'E}+N^{p'}_C\tilde{\mathscr F}^{\epsilon}_{p'E})+\frac12N^{p'}_A(N^E_C\tilde{\mathscr F}^{\epsilon}_{Ep'}+N^{q'}_C\tilde{\mathscr F}^{\epsilon}_{q'p'})
\\
&&\;\;\;\;\times\frac12[N^S_D( h^{DR}\tilde{\mathscr F}^{\sigma}_{SR}+h^{Dp}\tilde{\mathscr F}^{\sigma}_{Sp})+N^q_D(h^{DR}\tilde{\mathscr F}^{\sigma}_{qR}+h^{Dp}\tilde{\mathscr F}^{\sigma}_{qp})]
 \tilde d_{\epsilon\sigma}=0
\\
&&\check{\rm \Gamma}^{ \alpha}_{ AC}\check{\rm \Gamma}^{ p}_{p \alpha}=\check{\rm \Gamma}^{ \alpha}_{ AC}
\times [-\frac12( h^{pE}\tilde{\mathscr F}^{\sigma}_{Ep}+h^{pb}\tilde{\mathscr F}^{\sigma}_{bp})\tilde d_{\alpha\sigma}]=0
\\
&&\check{\rm \Gamma}^{ \alpha}_{ AC}\check{\rm \Gamma}^{ \mu}_{\mu \alpha}=0,\; \check{\rm \Gamma}^{ \mu}_{\mu \alpha}=0\;\; (\rm{for\;the\;semisimple\;Lie\; group})\\
\end{eqnarray*}
\begin{eqnarray*}
&&\mathbb C^{\mu}_{AB}\check{\rm \Gamma}^{B}_{\mu C}=[N^E_A(N^D_B\tilde{\mathscr F}^{\mu}_{DE}+N^p_B\tilde{\mathscr F}^{\mu}_{pE})+N^p_A(N^E_B\tilde{\mathscr F}^{\mu}_{Ep}+N^q_B\tilde{\mathscr F}^{\mu}_{qp})]
\\
&&\;\;\;\;\times\frac12[N^S_C( h^{BR}\tilde{\mathscr F}^{\sigma}_{SR}+h^{Bp'}\tilde{\mathscr F}^{\sigma}_{Sp'})+N^{q'}_C(h^{BR}\tilde{\mathscr F}^{\sigma}_{q'R}+h^{Bp'}\tilde{\mathscr F}^{\sigma}_{q'p'})]
 \tilde d_{\mu\sigma}
\\
&&\mathbb C^{\mu}_{Aq}\check{\rm \Gamma}^{q}_{\mu C}=-(N^{R'}_A\tilde{\mathscr F}^{\mu}_{R'q}+N^{p'}_A\tilde{\mathscr F}^{\mu}_{p'q})
\\
&&\;\;\;\;\times\frac12[N^S_C( h^{qR}\tilde{\mathscr F}^{\sigma}_{SR}+h^{qa}\tilde{\mathscr F}^{\sigma}_{Sa})+N^p_C(h^{qR}\tilde{\mathscr F}^{\sigma}_{pR}+h^{qa}\tilde{\mathscr F}^{\sigma}_{pa})]
 \tilde d_{\mu\sigma}\\
%\end{eqnarray*}
%\begin{eqnarray*}
&&\mathbb{C}^{\mu}_{Aq}=-(N^E_A\tilde{\mathscr F}^{\mu}_{Eq}+N^p_A\tilde{\mathscr F}^{\mu}_{pq})\\
&&\mathbb C^{\mu}_{AR}=-N^{\tilde M}_AN^{\tilde T}_R\tilde{\mathscr F}^{\mu}_{\tilde M\tilde T}\\
&&\mathbb C^{\mu}_{AR}\check{\rm \Gamma}^{R}_{\mu C}=
%-N^{\tilde M}_AN^{\tilde T}_R\tilde{\mathscr F}^{\mu}_{\tilde M\tilde T}\times\frac12 {\tilde G}^{RF}N^{\tilde E}_FN^{\tilde S}%_C\tilde{\mathscr F}^{\sigma}_{\tilde{S}\tilde{E} }\tilde d_{\mu\sigma}=
-\frac12({\tilde G}^{RF}N^{\tilde T}_R N^{\tilde E}_F)N^{\tilde M}_A N^{\tilde S}_C \tilde{\mathscr F}^{\mu}_{\tilde M\tilde T}\tilde{\mathscr F}^{\sigma}_{\tilde{S}\tilde{E} }\tilde d_{\mu\sigma}  
\end{eqnarray*}
As a result, we obtain that  \underline{$h^{AC}\check R'_{AC}$} is given by the sum of the following terms:
\begin{align*}
 h^{AC}H_{A}\check{\rm \Gamma}^{\alpha}_{ \alpha C}&=\frac12h^{AC}H_A(\tilde d^{\alpha\nu}H_C\tilde d_{\alpha\nu})
=\frac12h^{AC}N^{\tilde E}_A\partial_{\tilde E}( d^{\alpha\nu}N^{\tilde F}_C\partial_{\tilde F} d_{\alpha\nu})\\
h^{AC}\check{\rm \Gamma}^{E}_{ \mu C}\check{\rm \Gamma}^{\mu}_{ AE}&=-\frac14(h^{TS}h^{FD}\tilde{\mathscr F}^{\mu}_{ DT}\tilde{\mathscr F}^{\sigma}_{FS}+h^{TS}h^{Dq}\tilde{\mathscr F}^{\mu}_{ DT}\tilde{\mathscr F}^{\sigma}_{qS})\tilde d_{\mu\sigma} 
\\
h^{AC}\check{\rm \Gamma}^{q}_{ \mu C}\check{\rm \Gamma}^{\mu}_{ Aq}&=-\frac14(h^{TS}h^{qR}\tilde{\mathscr F}^{\mu}_{ Tq}\tilde{\mathscr F}^{\sigma}_{SR}+h^{TS}h^{qa}\tilde{\mathscr F}^{\mu}_{Tq}\tilde{\mathscr F}^{\sigma}_{Sa})\tilde d_{\mu\sigma} 
\\
h^{AC}\check{\rm \Gamma}^{\mu}_{ BC}\check{\rm \Gamma}^{B}_{ A\mu}&=\frac14(h^{ER}h^{DS}\tilde{\mathscr F}^{\mu}_{ DE}\tilde{\mathscr F}^{\sigma}_{SR}+h^{DS}h^{Ep}\tilde{\mathscr F}^{\mu}_{ DE}\tilde{\mathscr F}^{\sigma}_{Sp})\tilde d_{\mu\sigma} 
\\
 h^{AC}\check{\rm \Gamma}^{\mu}_{ qC}\check{\rm \Gamma}^{q}_{ A\mu}&=\frac14h^{SE}(h^{qR}\tilde{\mathscr F}^{\mu}_{Eq}\tilde{\mathscr F}^{\sigma}_{SR}+h^{qb}\tilde{\mathscr F}^{\mu}_{ Eq}\tilde{\mathscr F}^{\sigma}_{Sb})\tilde d_{\mu\sigma} 
\\
h^{AC}\check{\rm \Gamma}^{\mu}_{ \alpha C}\check{\rm \Gamma}^{\alpha}_{ A\mu}&=\frac14h^{AC}(\tilde d^{\mu\nu} H_C\tilde d_{\nu \alpha})(\tilde d^{\alpha\sigma} H_A\tilde d_{\sigma \mu})
\\
 -h^{AC}\check{\rm \Gamma}^{B}_{ A C}\check{\rm \Gamma}^{\alpha}_{ \alpha B}&=-\frac12h^{EC}(\tilde d^{\alpha\nu} H_B\tilde d_{\nu \alpha})\check{\rm \Gamma}^{B}_{CE}
\\
 -h^{AC}\check{\rm \Gamma}^{q}_{ A C}\check{\rm \Gamma}^{\alpha}_{ \alpha q}&=-\frac12h^{EC}(\tilde d^{\alpha\nu} H_q\tilde d_{\nu \alpha})\check{\rm \Gamma}^{q}_{CE}
\\
-h^{AC}\check{\rm \Gamma}^{\alpha}_{ A C}\check{\rm \Gamma}^{B}_{B \alpha }&=0
\\
-h^{AC}\check{\rm \Gamma}^{\alpha}_{ A C}\check{\rm \Gamma}^{p}_{p \alpha }&=0
\\
 -h^{AC}\check{\rm \Gamma}^{\alpha}_{ A C}\check{\rm \Gamma}^{\mu}_{\mu \alpha }&=0
\\
  -h^{AC}{\mathbb C}^{\mu}_{AB}\check{\rm \Gamma}^{B}_{\mu C}&=\frac12(h^{SE}h^{DR}\tilde{\mathscr F}^{\mu}_{E D}\tilde{\mathscr F}^{\sigma}_{SR}+h^{SE}h^{Dp}\tilde{\mathscr F}^{\mu}_{ ED}\tilde{\mathscr F}^{\sigma}_{Sp})\tilde d_{\mu\sigma} 
 \\
-h^{AC}{\mathbb C}^{\mu}_{Aq}\check{\rm \Gamma}^{q}_{\mu C}&=\frac12(h^{SE}h^{qR}\tilde{\mathscr F}^{\mu}_{E q}\tilde{\mathscr F}^{\sigma}_{SR}+h^{SE}h^{qa}\tilde{\mathscr F}^{\mu}_{ Eq}\tilde{\mathscr F}^{\sigma}_{Sa})\tilde d_{\mu\sigma} 
 \end{align*}

\subsection*{\underline{$h^{Ab}\check R'_{Ab}$}}
\begin{align*}
h^{Ab}H_A \check{\rm \Gamma}^{\alpha}_{ \alpha b}&=\frac12h^{Ab}H_A(\tilde d^{\alpha\nu} H_b\tilde d_{\nu \alpha})=\frac12h^{Eb}\partial_E(d^{\alpha\nu} \partial_b\, d_{\nu \alpha})
\\
-h^{Ab}L_{\alpha}\check{\rm \Gamma}^{\alpha}_{ Ab}&=\frac12h^{Ab}L_{\alpha}(N^E_A\tilde{\mathscr F}^{\alpha}_{Eb}+N^q_A\tilde{\mathscr F}^{\alpha}_{qb})
\\
h^{Ab}\check{\rm \Gamma}^{B}_{\alpha b}\check{\rm \Gamma}^{\alpha}_{AB}&=-\frac14h^{Eb}(h^{DT}\tilde{\mathscr F}^{\sigma}_{Tb}\tilde{\mathscr F}^{\alpha}_{DE}+h^{Dq}\tilde{\mathscr F}^{\sigma}_{qb}\tilde{\mathscr F}^{\alpha}_{DE})\tilde d_{\alpha\sigma}
\\
h^{Ab}\check{\rm \Gamma}^{q}_{\alpha b}\check{\rm \Gamma}^{\alpha}_{Aq}&=-\frac14(h^{Db}h^{qE}\tilde{\mathscr F}^{\alpha}_{qD}\tilde{\mathscr F}^{\sigma}_{Eb}+h^{Db}h^{qa}\tilde{\mathscr F}^{\alpha}_{qD}\tilde{\mathscr F}^{\sigma}_{ab})\tilde d_{\alpha\sigma}
\\
h^{Ab}\check{\rm \Gamma}^{\alpha}_{ C b}\check{\rm \Gamma}^C_{A\alpha}&=\frac14(h^{Sb}h^{ER}\tilde{\mathscr F}^{\alpha}_{bE}\tilde{\mathscr F}^{\sigma}_{SR}+h^{Sb}h^{Ep}\tilde{\mathscr F}^{\alpha}_{bE}\tilde{\mathscr F}^{\sigma}_{Sp})\tilde d_{\alpha\sigma}
\\
h^{Ab}\check{\rm \Gamma}^{\alpha}_{ p b}\check{\rm \Gamma}^p_{A\alpha}&=\frac14(h^{Sb}h^{pR}\tilde{\mathscr F}^{\alpha}_{bp}\tilde{\mathscr F}^{\sigma}_{SR}+h^{Sb}h^{pa}\tilde{\mathscr F}^{\alpha}_{bp}\tilde{\mathscr F}^{\sigma}_{Sa})\tilde d_{\alpha\sigma}
\\
h^{Ab}\check{\rm \Gamma}^{\alpha}_{\mu b}\check{\rm \Gamma}^{\mu}_{A\alpha}&=\frac14h^{Ab}(\tilde d^{\alpha\nu} H_b\tilde d_{\nu \mu})(\tilde d^{\mu\sigma} H_A\tilde d_{\sigma \alpha})
\\
-h^{Ab}\check{\rm \Gamma}^{E}_{A b}\check{\rm \Gamma}^{\alpha}_{\alpha E}&=-\frac12h^{Db}\,{}^{\rm H}\Gamma^E_{bD}(\tilde d^{\alpha\nu}H_E\tilde d_{\nu\alpha})
\\
-h^{Ab}\check{\rm \Gamma}^{q}_{A b}\check{\rm \Gamma}^{\alpha}_{\alpha q}&=-\frac12h^{Eb}\,{}^{\rm H}\Gamma^q_{bE}(\tilde d^{\alpha\nu}H_q\tilde d_{\nu\alpha})
\\
-h^{Ab}\check{\rm \Gamma}^{\alpha}_{A b}\check{\rm \Gamma}^{B}_{B\alpha }&=0
\\
-h^{Ab}\check{\rm \Gamma}^{\alpha}_{A b}\check{\rm \Gamma}^{p}_{p\alpha }&=0
\\
-h^{Ab}\check{\rm \Gamma}^{\alpha}_{A b}\check{\rm \Gamma}^{\mu}_{\mu\alpha }&=0\\
-h^{Ab}\mathbb{C}^{\alpha}_{AB}\check{\rm \Gamma}^{B}_{\alpha b}&=\frac12(h^{Eb}h^{DT}\tilde{\mathscr F}^{\alpha}_{DE}\tilde{\mathscr F}^{\sigma}_{Tb}+h^{Eb}h^{Da}\tilde{\mathscr F}^{\alpha}_{DE}\tilde{\mathscr F}^{\sigma}_{ab})\tilde d_{\alpha\sigma}
\\
-h^{Ab}\mathbb{C}^{\alpha}_{Ap}\check{\rm \Gamma}^{p}_{\alpha b}&=\frac12(h^{Eb}h^{pR}\tilde{\mathscr F}^{\alpha}_{pE}\tilde{\mathscr F}^{\sigma}_{Rb}+h^{Eb}h^{pa}\tilde{\mathscr F}^{\alpha}_{pE}\tilde{\mathscr F}^{\sigma}_{ab})\tilde d_{\alpha\sigma}
\end{align*}

\subsection*{\underline{$h^{bA}\check R'_{bA}$}}
\begin{align*}
h^{bA}H_b \check{\rm \Gamma}^{\alpha}_{ \alpha A}&=\frac12h^{bA}H_b(\tilde d^{\alpha\nu} H_A\tilde d_{\nu \alpha})\\
%\;\;\;\;\;\;\;\;\;\;\;\;\;\;\;\;
&=\frac12[h^{bE}\partial_b(d^{\alpha\nu} \partial_E d_{\nu \alpha})+h^{bA}\partial_b(N^p_A)(d^{\alpha\nu} \partial_p d_{\nu \alpha})]
\\
-h^{bA}L_{\alpha}\check{\rm \Gamma}^{\alpha}_{ bA}&=-h^{bA}L_{\alpha}(-\check{\rm \Gamma}^{\alpha}_{ Ab})=-\frac12h^{bA}L_{\alpha}(N^E_A\tilde{\mathscr F}^{\alpha}_{Eb}+N^q_A\tilde{\mathscr F}^{\alpha}_{qb})
\\
h^{bA}\check{\rm \Gamma}^{B}_{\alpha A}\check{\rm \Gamma}^{\alpha}_{bB}&=-\frac14(h^{bS}h^{ER}\tilde{\mathscr F}^{\sigma}_{RS}\tilde{\mathscr F}^{\alpha}_{Eb}+h^{bS}h^{Ep}\tilde{\mathscr F}^{\sigma}_{pS}\tilde{\mathscr F}^{\alpha}_{Eb})\tilde d_{\alpha\sigma}
\\
h^{bA}\check{\rm \Gamma}^{q}_{\alpha A}\check{\rm \Gamma}^{\alpha}_{bq}&=-\frac14(h^{Sb}h^{qR}\tilde{\mathscr F}^{\alpha}_{bq}\tilde{\mathscr F}^{\sigma}_{SR}+h^{Sb}h^{qa}\tilde{\mathscr F}^{\alpha}_{bq}\tilde{\mathscr F}^{\sigma}_{Sa})\tilde d_{\alpha\sigma}
\\
h^{bA}\check{\rm \Gamma}^{\alpha}_{CA}\check{\rm \Gamma}^{C}_{b\alpha }&=\frac14(h^{Db}h^{ET}\tilde{\mathscr F}^{\alpha}_{ED}\tilde{\mathscr F}^{\sigma}_{Tb}+h^{Db}h^{Ea}\tilde{\mathscr F}^{\alpha}_{ED}\tilde{\mathscr F}^{\sigma}_{ab})\tilde d_{\alpha\sigma}
\\
h^{bA}\check{\rm \Gamma}^{\alpha}_{pA}\check{\rm \Gamma}^{p}_{b\alpha }&=\frac14(h^{Eb}h^{pR}\tilde{\mathscr F}^{\alpha}_{pE}\tilde{\mathscr F}^{\sigma}_{Rb}+h^{Eb}h^{pa}\tilde{\mathscr F}^{\alpha}_{pE}\tilde{\mathscr F}^{\sigma}_{ab})\tilde d_{\alpha\sigma}
\\
h^{bA}\check{\rm \Gamma}^{\alpha}_{\mu A}\check{\rm \Gamma}^{\mu}_{b\alpha }&=\frac14h^{bA}(\tilde d^{\alpha\nu}H_A\tilde d_{\nu\mu})(\tilde d^{\mu\sigma}H_b\tilde d_{\sigma\alpha})
\\
-h^{bA}\check{\rm \Gamma}^{\alpha}_{\alpha E}\check{\rm \Gamma}^{E}_{ba}&=-\frac12h^{Db}(\tilde d^{\alpha\nu}H_E\tilde d_{\nu\alpha})\,{}^{\rm H}\Gamma^E_{bD}
\\
-h^{bA}\check{\rm \Gamma}^{\alpha}_{\alpha q}\check{\rm \Gamma}^{q}_{bA}&=-\frac12h^{bA}(\tilde d^{\alpha\nu}H_q\tilde d_{\nu\alpha})\,{}^{\rm H}\Gamma^q_{bA}
\\
-h^{bA}\check{\rm \Gamma}^{\alpha}_{bA}\check{\rm \Gamma}^{B}_{B\alpha}&=0
\\
-h^{bA}\check{\rm \Gamma}^{\alpha}_{bA}\check{\rm \Gamma}^{p}_{p\alpha}&=0
\\
-h^{bA}\check{\rm \Gamma}^{\alpha}_{bA}\check{\rm \Gamma}^{\mu}_{\mu\alpha}&=0
\\
-h^{bA}\mathbb{C}^{\alpha}_{bB}\check{\rm \Gamma}^{B}_{\alpha A}
%=-h^{bA}(-\mathbb{C}^{\alpha}_{Bb})\check{\rm \Gamma}^{B}_{\alpha A}
&=\frac12(h^{ER}h^{bS}\tilde{\mathscr F}^{\alpha}_{bE}\tilde{\mathscr F}^{\sigma}_{SR}+h^{Ep}h^{bS}\tilde{\mathscr F}^{\alpha}_{bE}\tilde{\mathscr F}^{\sigma}_{Sp})\tilde d_{\alpha\sigma}
\\
-h^{bA}\mathbb{C}^{\alpha}_{bp}\check{\rm \Gamma}^{p}_{\alpha A}&=\frac12(h^{bS}h^{pR}\tilde{\mathscr F}^{\alpha}_{bp}\tilde{\mathscr F}^{\sigma}_{SR}+h^{bS}h^{pa}\tilde{\mathscr F}^{\alpha}_{bp}\tilde{\mathscr F}^{\sigma}_{Sa})\tilde d_{\alpha\sigma}
\end{align*}

\subsection*{\underline{$h^{ab}\check R'_{ab}$}}
\begin{align*}
 h^{ab}H_a \check{\rm \Gamma}^{\alpha}_{ \alpha b}&=\frac12h^{ab}H_a(\tilde d^{\alpha\nu} H_b\tilde d_{\nu \alpha})=\frac12h^{ab}\partial_a( d^{\alpha\nu} \partial_b d_{\nu \alpha})
\\
-h^{ab}L_{\alpha}\check{\rm \Gamma}^{\alpha}_{ ab}&=-h^{ab}L_{\alpha}(-\frac12\tilde{\mathscr F}^{\alpha}_{ab})=0
 \\
 h^{ab}\check{\rm \Gamma}^{B}_{\alpha b}\check{\rm \Gamma}^{\alpha}_{aB}&=-\frac14h^{ab}(h^{ET}\tilde{\mathscr F}^{\sigma}_{Tb}\tilde{\mathscr F}^{\alpha}_{Ea}+h^{Ep}\tilde{\mathscr F}^{\sigma}_{pb}\tilde{\mathscr F}^{\alpha}_{Ea})\tilde d_{\alpha\sigma}
 \\
 h^{ab}\check{\rm \Gamma}^{p}_{\alpha b}\check{\rm \Gamma}^{\alpha}_{ap}&=-\frac14h^{ab}(h^{pE}\tilde{\mathscr F}^{\sigma}_{Eb}\tilde{\mathscr F}^{\alpha}_{pa}+h^{pq}\tilde{\mathscr F}^{\sigma}_{qb}\tilde{\mathscr F}^{\alpha}_{pa})\tilde d_{\alpha\sigma}
 \\
 h^{ab}\check{\rm \Gamma}^{\alpha}_{B b}\check{\rm \Gamma}^{B}_{a\alpha}&=\frac14h^{ab}(h^{ET}\tilde{\mathscr F}^{\alpha}_{Eb}\tilde{\mathscr F}^{\sigma}_{Ta}+h^{Ep}\tilde{\mathscr F}^{\alpha}_{Eb}\tilde{\mathscr F}^{\sigma}_{pa})\tilde d_{\alpha\sigma}
 \\
 h^{ab}\check{\rm \Gamma}^{\alpha}_{p b}\check{\rm \Gamma}^{p}_{a\alpha}&=\frac14h^{ab}(h^{pE}\tilde{\mathscr F}^{\alpha}_{pb}\tilde{\mathscr F}^{\sigma}_{Ea}+h^{pq}\tilde{\mathscr F}^{\alpha}_{pb}\tilde{\mathscr F}^{\sigma}_{qa})\tilde d_{\alpha\sigma}
 \\
  h^{ab}\check{\rm \Gamma}^{\alpha}_{\mu b}\check{\rm \Gamma}^{\mu}_{a\alpha}&=\frac14h^{ab}(\tilde d^{\alpha\nu} H_b\tilde d_{\nu \mu})(\tilde d^{\mu\sigma} H_a\tilde d_{\sigma \alpha})
 \\
 -h^{ab}\check{\rm \Gamma}^{R}_{a b}\check{\rm \Gamma}^{\alpha}_{\alpha R}&=-\frac12h^{ab}(\tilde d^{\alpha\nu} H_R\tilde d_{\nu \alpha})\,{}^{\rm H}{\rm \Gamma}^R_{ab},
 \\
 -h^{ab}\check{\rm \Gamma}^{p}_{a b}\check{\rm \Gamma}^{\alpha}_{\alpha p}&=-\frac12h^{ab}(\tilde d^{\alpha\nu} H_p\tilde d_{\nu \alpha})\,{}^{\rm H}{\rm \Gamma}^p_{ab}
 \\
 -h^{ab}\check{\rm \Gamma}^{\alpha}_{a b}\check{\rm \Gamma}^{R}_{R\alpha }&=0
 \\
 -h^{ab}\check{\rm \Gamma}^{\alpha}_{a b}\check{\rm \Gamma}^{p}_{p\alpha }&=0
 \\
 -h^{ab}\check{\rm \Gamma}^{\alpha}_{a b}\check{\rm \Gamma}^{\mu}_{\mu\alpha }&=0
 \\
 -h^{ab}\mathbb{C}^{\alpha}_{aB}\check{\rm \Gamma}^{B}_{\alpha b}&=\frac12h^{ab}(h^{ET}\tilde{\mathscr F}^{\sigma}_{Tb}\tilde{\mathscr F}^{\alpha}_{Ea}+h^{Ep}\tilde{\mathscr F}^{\sigma}_{pb}\tilde{\mathscr F}^{\alpha}_{Ea})\tilde d_{\alpha\sigma}
 \\
 -h^{ab}\mathbb{C}^{\alpha}_{ap}\check{\rm \Gamma}^{p}_{\alpha b}&=\frac12h^{ab}(h^{pE}\tilde{\mathscr F}^{\sigma}_{Eb}\tilde{\mathscr F}^{\alpha}_{pa}+h^{pq}\tilde{\mathscr F}^{\sigma}_{qb}\tilde{\mathscr F}^{\alpha}_{pa})\tilde d_{\alpha\sigma}
 \end{align*}

\subsection*{\underline{$\tilde{d}^{\alpha\beta}\check R'_{\alpha \beta}$}}
\begin{align*}
 \tilde d^{\alpha\beta}L_{\alpha}\check{\rm \Gamma}^{B}_{B\beta}&=0,\;\;\;\tilde d^{\alpha\beta}L_{\alpha}\check{\rm \Gamma}^{q}_{q\beta}=0\\
 %&&\tilde d^{\alpha\beta}L_{\alpha}\check{\rm \Gamma}^{q}_{q\beta}=0\\
 -\tilde d^{\alpha\beta}H_B\check{\rm \Gamma}^{B}_{\alpha\beta}&=-\tilde d^{\alpha\beta}H_B(-\frac12h^{BC}H_C\tilde d_{\alpha\beta}-\frac12h^{Bb}H_b\tilde d_{\alpha\beta})
 \\
 -\tilde d^{\alpha\beta}H_q\check{\rm \Gamma}^{q}_{\alpha\beta}&=-\tilde d^{\alpha\beta}H_q(-\frac12h^{qC}H_C\tilde d_{\alpha\beta}-\frac12h^{qb}H_b\tilde d_{\alpha\beta})
 \\
 \tilde d^{\alpha\beta}\check{\rm \Gamma}^{B}_{D \beta}\check{\rm \Gamma}^{D}_{\alpha B}&=-\frac14(h^{TR}h^{SE}\tilde{\mathscr F}^{\sigma}_{SR}\tilde{\mathscr F}^{\epsilon}_{ET}+h^{Tp}h^{SE}\tilde{\mathscr F}^{\sigma}_{Sp}\tilde{\mathscr F}^{\epsilon}_{ET}\\
 %&&\;\;\;\;\;\;\;\;\;\;\;\;\;\;\;\;\;\;\;\;\;\;
 &\;\;\;\;+h^{TR}h^{Sa}\tilde{\mathscr F}^{\sigma}_{SR}\tilde{\mathscr F}^{\epsilon}_{aT}+h^{Tp}h^{Sa}\tilde{\mathscr F}^{\sigma}_{Sp}\tilde{\mathscr F}^{\epsilon}_{aT})\tilde d_{\epsilon\sigma}
 \\
\tilde d^{\alpha\beta}\check{\rm \Gamma}^{B}_{q \beta}\check{\rm \Gamma}^{q}_{\alpha B}&=-\frac14(h^{ST}h^{qR}\tilde{\mathscr F}^{\sigma}_{Tq}\tilde{\mathscr F}^{\epsilon}_{SR}+h^{Sp}h^{qR}\tilde{\mathscr F}^{\sigma}_{pq}\tilde{\mathscr F}^{\epsilon}_{SR}\\
 %&&\;\;\;\;\;\;\;\;\;\;\;\;\;\;\;\;\;\;\;\;\;\;
 &\;\;\;\;+h^{ST}h^{qa}\tilde{\mathscr F}^{\sigma}_{Tq}\tilde{\mathscr F}^{\epsilon}_{Sa}+h^{Sp}h^{qa}\tilde{\mathscr F}^{\sigma}_{pq}\tilde{\mathscr F}^{\epsilon}_{Sa})\tilde d_{\epsilon\sigma},
 \\ 
 \tilde d^{\alpha\beta}\check{\rm \Gamma}^{B}_{\mu \beta}\check{\rm \Gamma}^{\mu}_{\alpha B}&=-\frac14[h^{BC}(\tilde d^{\alpha\beta}H_C\tilde d_{\mu\beta})+h^{Bb}(\tilde d^{\alpha\beta}H_b\tilde d_{\mu\beta})](\tilde d^{\mu\nu}H_B\tilde d_{\nu\alpha})
 \\
 \tilde d^{\alpha\beta}\check{\rm \Gamma}^{q}_{R \beta}\check{\rm \Gamma}^{R}_{\alpha q}&=-\frac14(h^{qD}h^{ST}\tilde{\mathscr F}^{\sigma}_{SD}\tilde{\mathscr F}^{\epsilon}_{Tq}+h^{qa}h^{ST}\tilde{\mathscr F}^{\sigma}_{Sa}\tilde{\mathscr F}^{\epsilon}_{Tq}\\
 %&&\;\;\;\;\;\;\;\;\;\;\;\;\;\;\;\;\;\;\;\;\;\;
 &\;\;\;\;+h^{qD}h^{Sb}\tilde{\mathscr F}^{\sigma}_{SD}\tilde{\mathscr F}^{\epsilon}_{bq}+h^{qa}h^{Sb}\tilde{\mathscr F}^{\sigma}_{Sa}\tilde{\mathscr F}^{\epsilon}_{bq})\tilde d_{\epsilon\sigma}
 \\
 \tilde d^{\alpha\beta}\check{\rm \Gamma}^{q}_{p \beta}\check{\rm \Gamma}^{p}_{\alpha q}&=-\frac14(h^{qE}h^{pR}\tilde{\mathscr F}^{\sigma}_{Ep}\tilde{\mathscr F}^{\epsilon}_{qR}+h^{qb}h^{pR}\tilde{\mathscr F}^{\sigma}_{bp}\tilde{\mathscr F}^{\epsilon}_{qR}\\
 %&&\;\;\;\;\;\;\;\;\;\;\;\;\;\;\;\;\;\;\;\;\;\;
 &\;\;\;\;+h^{qE}h^{pa}\tilde{\mathscr F}^{\sigma}_{Ep}\tilde{\mathscr F}^{\epsilon}_{qa}+h^{qb}h^{pa}\tilde{\mathscr F}^{\sigma}_{bp}\tilde{\mathscr F}^{\epsilon}_{qa})\tilde d_{\epsilon\sigma},
 \\
 \tilde d^{\alpha\beta}\check{\rm \Gamma}^{q}_{\mu \beta}\check{\rm \Gamma}^{\mu}_{\alpha q}&=-\frac14[h^{qC}(\tilde d^{\alpha\beta}H_C\tilde d_{\mu\beta})+h^{qb}(\tilde d^{\alpha\beta}H_b\tilde d_{\mu\beta})](\tilde d^{\mu\nu}H_q\tilde d_{\nu\alpha})
 \\
 \tilde d^{\alpha\beta}\check{\rm \Gamma}^{\mu}_{B \beta}\check{\rm \Gamma}^{B}_{\alpha \mu}&=-\frac14[h^{BC}(\tilde d^{\alpha\beta}H_C\tilde d_{\alpha\mu})+h^{Bb}(\tilde d^{\alpha\beta}H_b\tilde d_{\alpha\mu})](\tilde d^{\mu\nu}H_B\tilde d_{\nu\beta})
 \\
 \tilde d^{\alpha\beta}\check{\rm \Gamma}^{\mu}_{q \beta}\check{\rm \Gamma}^{q}_{\alpha \mu}&=-\frac14[h^{qC}(\tilde d^{\alpha\beta}H_C\tilde d_{\alpha\mu})+h^{qb}(\tilde d^{\alpha\beta}H_b\tilde d_{\alpha\mu})](\tilde d^{\mu\nu}H_q\tilde d_{\nu\beta})
 \\
 -\tilde d^{\alpha\beta}\check{\rm \Gamma}^{E}_{\alpha \beta}\check{\rm \Gamma}^{R}_{RE}&=\frac12[h^{EB}(\tilde d^{\alpha\beta}H_B\tilde d_{\alpha\beta})+h^{Eb}(\tilde d^{\alpha\beta}H_b\tilde d_{\alpha\beta})]N^{\tilde D}_R\,{}^{\rm H}\Gamma^R_{E\tilde D}
 \\
 -\tilde d^{\alpha\beta}\check{\rm \Gamma}^{E}_{\alpha \beta}\check{\rm \Gamma}^{q}_{qE}&=\frac12[h^{EB}(\tilde d^{\alpha\beta}H_B\tilde d_{\alpha\beta})+h^{Eb}(\tilde d^{\alpha\beta}H_b\tilde d_{\alpha\beta})]\,{}^{\rm H}\Gamma^q_{qE}
\\
-\tilde d^{\alpha\beta}\check{\rm \Gamma}^{E}_{\alpha \beta}\check{\rm \Gamma}^{\mu}_{\mu E}&=\frac14[h^{EB}(\tilde d^{\alpha\beta}H_B\tilde d_{\alpha\beta})+h^{Eb}(\tilde d^{\alpha\beta}H_b\tilde d_{\alpha\beta})](\tilde d^{\mu\nu}H_E\tilde d_{\mu\nu})
\\
-\tilde d^{\alpha\beta}\check{\rm \Gamma}^{q}_{\alpha \beta}\check{\rm \Gamma}^{R}_{Rq}&=\frac12[h^{qB}(\tilde d^{\alpha\beta}H_B\tilde d_{\alpha\beta})+h^{qb}(\tilde d^{\alpha\beta}H_b\tilde d_{\alpha\beta})]N^{\tilde E}_R\,{}^{\rm H}\Gamma^R_{q\tilde E}
\\
-\tilde d^{\alpha\beta}\check{\rm \Gamma}^{q}_{\alpha \beta}\check{\rm \Gamma}^{p}_{pq}&=\frac12[h^{qB}(\tilde d^{\alpha\beta}H_B\tilde d_{\alpha\beta})+h^{qb}(\tilde d^{\alpha\beta}H_b\tilde d_{\alpha\beta})]\,{}^{\rm H}\Gamma^p_{pq}
\\
-\tilde d^{\alpha\beta}\check{\rm \Gamma}^{q}_{\alpha \beta}\check{\rm \Gamma}^{\mu}_{\mu q}&=\frac14[h^{qB}(\tilde d^{\alpha\beta}H_B\tilde d_{\alpha\beta})+h^{qb}(\tilde d^{\alpha\beta}H_b\tilde d_{\alpha\beta})](\tilde d^{\mu\nu}H_q\tilde d_{\mu\nu})
\\
-\tilde d^{\alpha\beta}\check{\rm \Gamma}^{\mu}_{\alpha \beta}\check{\rm \Gamma}^{R}_{R\mu }&=0,\;\;
 %\\
 %&&
 -\tilde d^{\alpha\beta}\check{\rm \Gamma}^{\mu}_{\alpha \beta}\check{\rm \Gamma}^{p}_{p\mu }=0,\;\;
 %\\
 %&&
 -\tilde d^{\alpha\beta}\check{\rm \Gamma}^{\mu}_{\alpha \beta}\check{\rm \Gamma}^{\nu}_{\nu\mu }=0
  \end{align*}
 
\appendix
\section*{Appendix C}
\section*{The identity $N^A_C\sigma_A+N^a_C\sigma_a=\sigma_C$}
 %\underline{Identity $N^A_C\sigma_A+N^a_C\sigma_a=\sigma_C$}

Consider the following expression
\[
 K^A_{\alpha}\frac{\partial}{\partial Q^{\ast A}}d_{\mu\nu}(Q^{\ast},\tilde f)+K^a_{\alpha}\frac{\partial}{\partial \tilde f^a}d_{\mu\nu}(Q^{\ast},\tilde f).
\]
 %\[
 %\displaystyle
 %K^A_{\alpha}\partial_A d_{\mu\nu}(Q^{\ast},\tilde f)+K^a_{\alpha}\partial_a d_{\mu\nu}(Q^{\ast},\tilde f).
%\]
Since $d_{\mu\nu}$ is given by
\[
 d_{\mu\nu}=K^C_{\mu}(Q^{\ast})G_{CD}(Q^{\ast})K^D_{\nu}(Q^{\ast})+K^p_{\mu}(\tilde f)G_{pq}K^q_{\nu}(\tilde f),
 \]
we have
 \begin{eqnarray*}
 &&K^A_{\alpha}K^C_{\mu,A}G_{CD}K^D_{\nu}+K^A_{\alpha}K^C_{\mu}G_{CD,A}K^D_{\nu}+K^A_{\alpha}K^C_{\mu}G_{CD}K^D_{\nu,A}\\
  &&\;\;\;+K^a_{\alpha}K^p_{\mu,a}G_{pq}K^q_{\nu}+K^a_{\alpha}K^p_{\mu}G_{pq}K^q_{\nu,a}.
\end{eqnarray*}
Using the Killing relation 
\[
 K^A_{\alpha}G_{CD,A}=-G_{CR}K^R_{\alpha,D}-G_{RD}K^R_{\alpha,C}
\]
in the first part of the previous expression for $d_{\mu\nu}$, we get
\begin{eqnarray*}
 &&K^A_{\alpha}K^{C'}_{\mu,A}G_{C'D}K^D_{\nu}-K^C_{\mu}K^D_{\nu}(G_{CD'}K^{D'}_{\alpha,D}+G_{C'D}K^{C'}_{\alpha C})+K^A_{\alpha}K^C_{\mu}G_{CD}K^D_{\nu,A}\\
&&
 =G_{C'D}K^D_{\nu}(K^A_{\alpha}K^{C'}_{\mu,A}-K^C_{\mu}K^{C'}_{\alpha,C})+G_{CD'}K^{C}_{\mu}(K^A_{\alpha}K^{D'}_{\nu,A}-K^D_{\nu}K^{D'}_{\alpha,D})\\
&&
 =G_{C'D'}K^{D'}_{\nu}c^{\sigma}_{\alpha\mu}K^{C'}_{\sigma}+G_{C'D'}K^{C'}_{\mu}c^{\sigma}_{\alpha\nu}K^{D'}_{\sigma}={\gamma}_{\nu \sigma}c^{\sigma}_{\alpha\mu}+{\gamma}_{\mu \sigma}c^{\sigma}_{\alpha\nu}.
\end{eqnarray*}
In the second part of the expression for $d_{\mu\nu}$ we use the following Killing relation:
\[
 -G_{qb'}K^{b'}_{\alpha, a}-G_{ab'}K^{b'}_{\alpha,q}=0.
\]
Then we have
\begin{eqnarray*}
 &&K^a_{\alpha}K^p_{\mu,a}G_{pq}K^q_{\nu}+K^a_{\alpha}K^p_{\mu}G_{pq}K^q_{\nu,a}-K^q_{\nu}G_{qb'}K^{b'}_{\alpha, a}K^a_{\mu}-K^q_{\nu}G_{ab'}K^{b'}_{\alpha,q}K^a_{\mu}\\
&&
 =K^p_{\mu}G_{pq}(K^a_{\alpha}K^q_{\nu,a}-K^b_{\nu}K^q_{\alpha,b})+K^q_{\nu}G_{pq}(K^a_{\alpha}K^p_{\mu,a}-K^a_{\mu}K^p_{\alpha,a})
 \\
 &&={\gamma}'_{\nu\sigma}c^{\sigma}_{\alpha\mu}+{\gamma}'_{\mu \sigma}c^{\sigma}_{\alpha\nu}.
\end{eqnarray*}
Therefore, the sum of the resulting expressions leads to
\[
 \displaystyle
 K^A_{\alpha}\partial_A d_{\mu\nu}(Q^{\ast},\tilde f)+K^a_{\alpha}\partial_a d_{\mu\nu}(Q^{\ast},\tilde f)=d_{\nu\sigma}c^{\sigma}_{\alpha\mu}+d_{\mu \sigma}c^{\sigma}_{\alpha\nu}.
\]
Multiplying both sides of the previous equality by $d^{\mu\nu}$, we obtain that in the case of a semisimple Lie group the following holds:
\[
 d^{\mu\nu}(K^A_{\alpha}\partial_A d_{\mu\nu}+K^a_{\alpha}\partial_a d_{\mu\nu})=c^{\mu}_{\alpha\mu}+c^{\nu}_{\alpha\nu}=0.
\]
Hence it follows that $\;\;\;N^A_C\sigma_A+N^a_C\sigma_a=\sigma_C$.

As a consequence, one can get
\[
h^{BM}N^A_{B,M}\sigma _A+h^{BM}N^a_{B,M}\sigma _a=0,
\]
since $N^A_{B,M}=-K^A_{\alpha,M}\Lambda^{\alpha}_B-K^A_{\alpha}\Lambda^{\alpha}_{B,M}$, $N^a_{B,M}=-K^a_{\alpha}\Lambda^{\alpha}_{B,M}$, and given that $h^{BM}=G^{EF}N^B_EN^M_F$ and $N^B_E\Lambda^{\alpha}_B=0$.

 \end{document}